\documentclass[aps,prb,groupedaddress,nofootinbib,twocolumn]{revtex4-2}
\usepackage[overload]{empheq} 
\usepackage{cases} 
\usepackage{hyperref}
\hypersetup{colorlinks=true,linkcolor=blue,anchorcolor=blue,citecolor=blue,filecolor=blue,urlcolor=blue,bookmarksnumbered=true,pdfview=FitB}
\usepackage[normalem]{ulem}
\usepackage[english]{babel}
\usepackage{bookmark}
\usepackage{xcolor} 
\usepackage{amsmath}
\usepackage{amssymb} 
\usepackage[retainorgcmds]{IEEEtrantools} 
\usepackage{hyperref}
\usepackage{bbold}
\usepackage{graphicx}
\usepackage[export]{adjustbox}
\usepackage{changepage}
\usepackage{calc}
\usepackage{braket}
\usepackage{comment}

\usepackage[caption=false]{subfig}
\usepackage[normalem]{ulem}

\usepackage{svg}
\usepackage[off]{svg-extract}
\svgsetup{clean=true}

\usepackage{relsize}
\usepackage[T1]{fontenc}
\usepackage{etoolbox}

\newcommand{\be}{\begin{IEEEeqnarray}{lll}}
\newcommand{\ee}{\end{IEEEeqnarray}}
\DeclareMathOperator{\tr}{Tr}
\newcommand{\dm}[1]{\textcolor{blue}{#1}}
\newcommand{\ag}[1]{\textcolor{red}{#1}}

\newcommand{\bea}{\begin{eqnarray}}
\newcommand{\eea}{\end{eqnarray}}
\newcommand{\ve}{\varepsilon}

\newcommand{\vf}{v_{\mathrm{F}}}
\newcommand{\kf}{k_\text{F}}

\newcommand{\bk}{{\textbf k}}

\newcommand{\bp}{{\textbf p}}
\newcommand{\bq}{{\textbf q}}

\newcommand{\nn}{\nonumber}
\newcommand{\bwt}{\begin{widetext}}
\newcommand{\ewt}{\end{widetext}}

\newcommand{\bse}{\begin{subequations}}
\newcommand{\ese}{\end{subequations}}

\newcommand{\bv}{{\textbf v}}

\newcommand{\e}{\epsilon}
\newcommand{\vd}{v_{\mathrm{D}}}
\newcommand{\oi}{\omega_{\mathrm{I}}}
\newcommand{\od}{\omega_{\mathrm{D}}}
\newcommand{\fer}{\text{F}}
\newcommand{\EF}{E_\text{F}}
\newcommand{\bos}{n_\text{B}}

\newcommand{\ef}{E_\text{F}}
\newcommand{\op}{\omega_{\text{p}d}}
\newcommand{\opthree}{\omega_{\text{p}3}}
\newcommand{\optwo}{\omega_{\text{p}2}}

\newcommand{\bkq}{{\mathbf{k+q}}}

\newcommand{\bpq}{{\mathbf{p+q}}}

\newcommand{\dee}{\text{d}}

\begin{document}

\title{Intrinsic optical absorption in Dirac metals}

\author{Adamya P. Goyal}
\author{Prachi Sharma}\thanks{Current address: School of Physics and Astronomy, University of Minnesota, Minneapolis, MN 55455, USA} \author{ Dmitrii L. Maslov}
\affiliation{
Department of Physics, University of Florida, P. O. Box 118440, Gainesville, FL 32611-8440, USA}

\date{\today}

\begin{abstract}
 A Dirac metal is a doped (gated) Dirac material  with the Fermi energy ($\EF$) lying either in the conduction or valence bands. In the non-interacting picture, optical absorption in gapless Dirac metals occurs only if the frequency of incident photons ($\Omega$) exceeds the direct (Pauli) frequency threshold, equal to $2\EF$. 
 In this work,  we study, both analytically and numerically, the role of electron-electron (\emph{ee}) and electron-hole (\emph{eh}) interactions in optical absorption of two-dimensional (2D) and three-dimensional (3D) Dirac metals in the entire interval of frequencies below $2\EF$. We show that, for $\Omega\ll \EF$, the optical conductivity, $\Re\sigma(\Omega)$, arising from the combination of \emph{ee} and certain \emph{eh} scattering processes, scales as $\Omega^2\ln\Omega$ in 2D and as $\Omega^2$ in 3D, respectively, both for short-range (Hubbard) and long-range (screened Coulomb) interactions. Another type of \emph{eh} processes, similar to Auger-Meitner (AM) processes in atomic physics, starts to contribute for $\Omega$ above the direct threshold, equal to $\EF$.  Similar to the case of doped semiconductors with parabolic bands studied in prior literature, the AM contribution to $\Re\sigma(\Omega)$ in Dirac metals is manifested by a threshold singularity, $\Re\sigma(\Omega)\propto (\Omega-\EF)^{d+2}$, where $d$ is the spatial dimensionality and $0<\Omega-\EF\ll \EF$. In contrast to doped semiconductors, however, the AM contribution in Dirac metals is completely overshadowed by the \emph{ee} and other \emph{eh} contributions. Numerically, $\Re\sigma(\Omega)$ 
 happens to be small in almost the entire range of $\Omega<2\EF$. This finding may have important consequences for collective modes in Dirac metals lying below $2\EF$.
\end{abstract}

\maketitle

\section{Introduction}
\label{sec:Intro}

The characteristic feature of Dirac 
materials is
the presence of symmetry-protected band-touching points which, in certain cases, 
is accompanied by the eponymous Dirac dispersion near these points. Realizations of these systems include monolayer
graphene~\cite{neto:2009}
and the surface state of a three-dimensional topological insulator~\cite{hasan:2010}
in two dimensions (2D), and Weyl/Dirac semi-metals~\cite{vafek:2014,Burkov:2018,Armitage:2018}
in three dimensions (3D).\footnote{For the purposes of present discussion, the topological distinction between Weyl and Dirac materials is irrelevant, and we will be referring to both of the them as to ``Dirac materials''.}
Owing to zero band gap, these materials exhibit semi-metallic behavior at charge neutrality.

At the level of non-interacting (NI) electrons, 
a pristine 2D Dirac material is 
characterized by a frequency-independent and universal optical conductivity \cite{neto:2009}
\bea
\Re\sigma_{\text{NI}2}(\Omega)=\frac{Ne^2}{16\hbar},
\label{2Dfree0}
\eea
whereas the conductivity of a pristine 3D Dirac material scales linear with frequency \cite{Hosur:2012,Ashby:2014}
\bea\Re\sigma_{\text{NI}3}(\Omega)=\frac{Ne^2\Omega}{24\pi\hbar \vd},\label{3Dfree0}
\eea
 where $N$ is the total (spin times valley) degeneracy, $\vd$ is the Dirac velocity.\footnote{Throughout the paper, we set $\hbar=1$ in the intermediate results but display it in the final results for the conductivity. Also, without loss of generality, we take $\Omega>0$ and assume that the Fermi energy lies in the conduction band.}
 These predictions were corroborated by multiple experiments, see for e.g.,  reviews Ref.~\cite{review_Peres2010,DasSarma:2011,kotov:2012,Hosur:2013,vafek:2014,Burkov:2018,Armitage:2018}.

The effect of electron-electron (\emph{ee}) interactions on the optical conductivity of Dirac materials  was studied extensively in 2D, see, e.g., reviews \cite{review_Peres2010,DasSarma:2011,kotov:2012} and references therein, and also in 3D \cite{Rosenstein:2013,Roy:2018}. As Coulomb interaction is marginally irrelevant both in 2D and 3D, it leads to a logarithmic renormalization of the Dirac velocity and thus of the coupling constant, $e^2/\vd$
\cite{kotov:2012,Abrikosov:1971}. 
Consequently, $\Re\sigma_{\text{NI}2}(\Omega)$ acquires a multiplicative renormalization factor,  which varies with $\Omega$ logarithmically and, at $\Omega\to 0$, approaches a constant equal to $1$ \cite{kotov:2012} or $1+1/(N+1)$ \cite{Roy:2018} in 2D and 3D, respectively. Note that this renormalization starts already at first order in the bare Coulomb potential, which implies that it does not involve collisions between particles in the intermediate states (the latter start at second order). On the other hand, a short-range (Hubbard) interaction is irrelevant in both 2D and 3D.

\begin{figure}
    \centering
         \includegraphics[width=0.5\linewidth
        ,center]
        {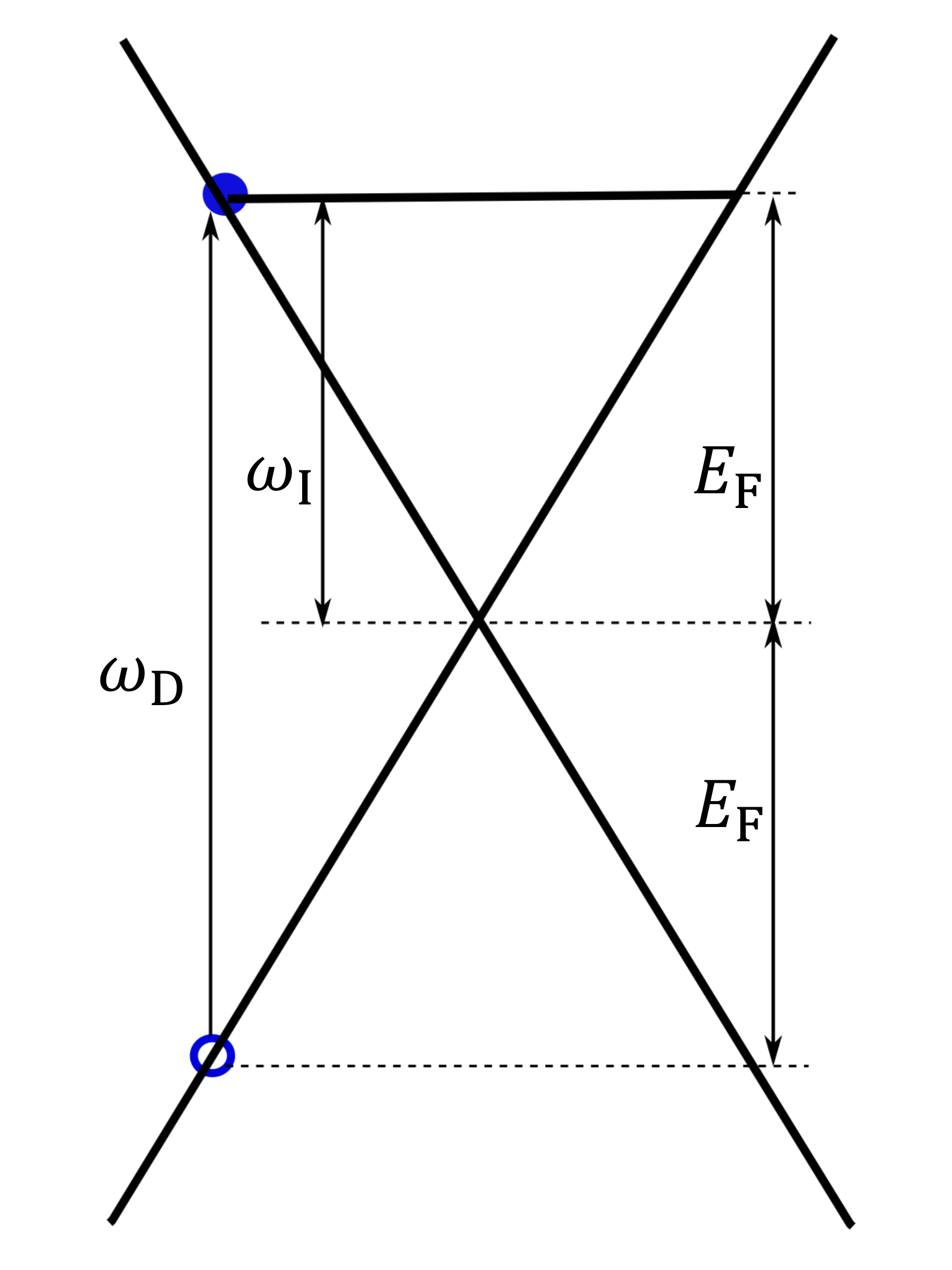}
       \caption{Band diagram of Dirac metal  symmetric conduction and valence bands showing the direct (Pauli) threshold $\od=2\EF$ for single-particle inter-band transitions. Also shown is the indirect threshold for many-body Auger-Meitner transitions, $\oi=\EF$.
           \label{fig:BandDiagram}}
\end{figure}

In a typical experiment, Dirac 
materials
are 
doped (gated) away from charge neutrality, either intentionally or unintentionally. From now on, we will be referring to such systems as  ``Dirac metals''. 
In this case,
the Pauli principle dictates that the optical conductivity of an ideal Dirac metal is strictly zero below
the ``direct'' (or Pauli)
threshold, 
\bea
\od=2\EF,
\label{omegaD}
\eea
where $\EF$ is the Fermi energy, measured from the Dirac point.
Experimentally, however,  one observes significant absorption for frequencies above the Drude tail but below $\od$ \cite{li_dirac_2008,Mak:2008,Horng:2011,Mak:2012,Drew:2016} and significant Raman response in the same frequency range \cite{riccardi_gate-dependent_2016}, both of which indicate a deviation from the single-particle picture. 
Absorption below the 
Pauli threshold in doped graphene 
due to a combined effect of disorder, electron-phonon and electron-electron interaction
has also been addressed theoretically
in Refs.~\cite{peres2007,stauber:2008,peres:2008,peres:2010PRL}. 
In this paper, we focus on intrinsic absorption due to \emph{ee} and electron-hole (\emph{eh}) interactions for $\Omega<\od$.

Absorption due to \emph{ee} interaction in a Dirac metal was studied in Refs.~\cite{Principi:2013,Sharma:2021}. For $\Omega\ll \EF$, the conductivity was found to scale as $\Omega^2\ln\Omega$  and $\Omega^2$ in 2D and 3D, respectively \cite{Sharma:2021}.\footnote{An earlier result of Ref.~\cite{Principi:2013} was missing a logarithmic factor in the 2D case.} A quadratic scaling of the conductivity can be understood as the  consequence of partially broken Galilean invariance in a Dirac-Fermi liquid (DFL). Indeed,
the optical conductivity can be cast into a Drude-like form
\bea\Re\sigma(\Omega)\propto \frac{1}{\Omega^2\tau_j(\Omega)},\label{Drude}
\eea
where $\tau_j(\Omega)$ is the current relaxation time. If Galilean invariance is broken completely, e.g., by umklapp scattering, $\tau_j(\Omega)$ is of the same order as the quasiparticle lifetime in a Fermi liquid (FL): $\tau_j(\Omega)\sim \tau_{\text{qp}}(\Omega)\propto \Omega^{-2}$. 
In this case, Eq.~\eqref{Drude} produces a familiar ``FL foot": $\Re\sigma(\Omega)=\text{const}$. On the other hand, if Galilean invariance is intact, current cannot be relaxed in \emph{ee} collisions: although $\tau_{\text{qp}}(\Omega)$ is finite, $\tau_j(\Omega)=\infty$ and thus $\Re\sigma(\Omega)=0$. A DFL 
occupies an intermediate niche between the two limits described above. On one hand, its non-parabolic spectrum allows for current relaxation; on the other hand, the spectrum is still isotropic (at low doping) and current relaxation is impossible for electrons right on the Fermi surface (FS) \cite{Sharma:2021}. For states away from the FS, the current relaxation time is finite but long, $\tau_j(\Omega)\propto \Omega^{-4}$ (modulo a $\ln\Omega$ factor in 2D), while $\tau_{\text{qp}}(\Omega)$ still 
scales in a FL way, i.e., as
$\Omega^{-2}$. 
According to Eq.~\eqref{Drude}, the quartic scaling of $1/\tau_j(\Omega)$ translates into the quadratic scaling of the conductivity.

In this paper,  we extend the results of Ref.~\cite{Sharma:2021} to the entire interval of frequencies below $\od$. 
Such an extension necessarily requires to account for both \emph{ee} and \emph{eh} interaction processes. We consider 
2D and 3D Dirac metals
with two types of interaction: Hubbard and Coulomb.
Our analytic results follow from the analysis of the Kubo formula and are applicable in two regions: i) for $\Omega\ll\oi$, where
\bea
\omega_{\text{I}}=
E_\text{F},\label{omegaI}
\eea
is the ``indirect''  threshold and ii)  just above the indirect threshold, i.e, for $\Omega\gtrapprox\oi$.
In the rest of the interval $0<\Omega<\od$, the conductivity is calculated numerically, but only for a Dirac metal with Hubbard interaction.
For $\Omega\ll\oi$,  we show that the \emph{eh} contribution to the conductivity scales as $\Omega^2$, i.e., it is comparable to the \emph{ee} one found in Ref.~\cite{Sharma:2021} in 3D and  is subleading to the \emph{ee} one in 2D, but  only in the leading logarithm sense.

This $\Omega^2$-scaling of the \emph{eh} contribution to the conductivity can also be understood in terms of the Drude formula \eqref{Drude}. Current relaxation due to \emph{eh} scattering is not limited by (partially broken) Galilean invariance, so that $\tau_j(\Omega)\sim \tau_{\text{qp}}(\Omega)\propto\Omega^{-2}$. [Unlike $\tau_{\text{qp}}(\Omega)$,  $\tau_j(\Omega)$ does not have an extra logarithmic factor in 2D.] However, the energies of electrons and holes differ now by $\EF$ rather than $\Omega$; therefore, the factor of $\Omega^2$ in Eq.~\eqref{Drude} is replaced by $\EF^2$, and the conductivity scales as $\Omega^2$.

Another channel of absorption due to \emph{eh} interaction opens up when $\Omega$ exceeds the indirect threshold $\oi$ [Eq.~\eqref{omegaI}]. 
Since the seminal 1969 paper by Gavoret et al.~\cite{gavoret_optical_1969}, absorption of light by degenerate semiconductors due to a particular type of \emph{eh} interaction processes,
similar to Auger-Meitner (AM) processes in atomic physics
\cite{meitner:1922,Auger:1923,PhysicsToday:2019},
have been studied by a large number of researchers, see, e.g., Refs.~\cite{ruckenstein_many-body_1987,Sham:1990,Hawrylak:1991,Pimenov:2017}. 
Although we consider only gapless systems, 
our result for the AM contribution just above $\oi$ exhibits a threshold singularity of the same type as found for a gapped spectrum \cite{gavoret_optical_1969,ruckenstein_many-body_1987,Sham:1990,Hawrylak:1991,Pimenov:2017}, i.e.,
\bea
\Re\sigma(\Omega)\propto \theta(\delta\Omega)\delta\Omega^{\beta_{\text{A}}},\label{Auger}
\eea
where
$\delta\Omega\equiv\Omega-\oi\ll \oi$ and $\beta_{\text{A}}=d+2$ with $d$ being the spatial dimensionality and $\theta(x)$ is the Heaviside step function. 
Equation~\eqref{Auger} can be obtained by estimating the conductivity as $\Re\sigma(\Omega)\propto \delta\Omega \mathcal{N}(\delta\Omega)/\tau_{\text{qp}}(\delta\Omega)$, where $\mathcal{N}(\epsilon)\propto \epsilon^{d
-1}$ is the density of states of a gapless Dirac metal and $1/\tau_{\text{qp}}(\epsilon)\propto \epsilon^2$.

More important, however, is the fact that for a non-parabolic spectrum the AM contribution occurs at the background of \emph{ee} and other \emph{eh} contributions, which start at the lowest frequencies (as $\Omega^2$ and $\Omega^2\ln\Omega$ in 3D and 2D, respectively)  and are still present both near and above $\oi$.  Therefore, the AM threshold singularity is masked by these other contributions. These competing contributions were not taken into account in the previous work on AM processes  \cite{gavoret_optical_1969,ruckenstein_many-body_1987,Sham:1990,Hawrylak:1991,Pimenov:2017}, which  considered two strictly  parabolic bands separated by a gap ($2\Delta$).
 To clarify the difference in absorption by materials with parabolic and Dirac bands, we invoke temporarily a gapped Dirac spectrum, $\epsilon_\bk=\pm\sqrt{\vd^2k^2+\Delta^2}$. A gapped semiconductor with parabolic conduction and valence band
can be viewed as the $\Delta\to\infty$ limit of this spectrum.  
In this case, intra-band \emph{ee} interaction does not affect the conductivity due to Galilean invariance, as we already discussed above. Moreover, inter-band absorption accompanied by electron-hole conversion processes, i.e., processes that do not conserve the numbers of electrons and holes separately, 
is also forbidden in the parabolic limit, because the corresponding eigenstates are either purely electron-like or purely hole-like, with zero overlap between the two. Therefore, the interaction part of the corresponding Hamiltonian conserves the numbers of electrons and holes separately, and absorption is absent for $\Omega<\oi$. 
For a strongly non-parabolic, e.g., gapless Dirac spectrum, the \emph{ee} contribution is not suppressed by Galilean invariance,  while electron-hole conversion 
processes are generically as important as other processes.

As far as the interval of $\Omega>\od$ is concerned,
Abedinpour et al.~\cite{abedinpour_drude_2011} showed that the conductivity of doped graphene (a 2D Dirac metal, in our terminology) with Coulomb interaction exhibits a logarithmic renormalization which, for $\Omega\gg \od$, is reduced to the well-studied case of undoped graphene, and is logarithmically enhanced for $\Omega\gtrapprox\od$ both for Coulomb and Hubbard interactions Both of these effects arise already at first order in the corresponding interaction and reflect renormalization of the Dirac velocity and, consequently, of the coupling constant. To the best of our knowledge, the interval of $\Omega\gg\od$ has not been studied for a 3D Dirac metal but, in analogy with the results for the undoped 3D case \cite{Rosenstein:2013,Roy:2018}, we would also expect a logarithmic renormalization starting at first order. 
On the other hand, absorption processes studied in our paper correspond to real collisions between electrons, and between electrons and holes, which occur starting from the second order in the interaction. Therefore, these processes are subleading to the first-order effects described above  studied in Ref.~\cite{abedinpour_drude_2011}, and we will not extend our results above $\od$.  

Our numerical results agree with analytic ones, where applicable, and allow one to trace the behavior of the conductivity for almost entire frequency range of interest, $0<\Omega<\od$,
except for a narrow interval of width 
$\mathcal{O}(\alpha_{\text{H,C}}^2E_\text{F})$ around $\omega_{\text{D}}$, 
where $\alpha_{\text{H,C}}\ll 1$ is the dimensionless coupling constant of Hubbard and Coulomb interactions, respectively. 
In this interval, our perturbative expansion breaks down and one needs to re-sum the diagrammatic series.

The rest the paper is organized as follows.  In Sec.~\ref{sec:model}, we set up the model Hamiltonians for 2D and 3D Dirac metals. 
In Sec.~\ref{sec:Formalism}, we 
outline the formalism for calculating  the optical conductivity via the Kubo formula. In Sec.~\ref{sec:processes}, we identify the \emph{ee} and \emph{eh} scattering processes that contribute to the conductivity in a given frequency range.   In section~\ref{sec:archetype-conductivity}, we analyze the general structure of the contributions to the conductivity from the self-energy and vertex diagrams, which serve as archetypes for other contributions. 
In sections~\ref{sec:3D} and~\ref{sec:2Dsystems},  we present our analytical and numerical results for the optical conductivity of 3D and 2D Dirac metals, respectively. Our conclusions are given in Sec.~\ref{sec:Conclusions}.

\section{Model Hamiltonians of Dirac metals}
\label{sec:model}

In this section we define our model Hamiltonians for 2D and 3D Dirac metals.

\subsection{3D Hamiltonian\label{sec:3DHamiltonian}}

We model a 3D Dirac metal
by a $4\times4$ low-energy Hamiltonian with two orbital degrees of freedom per spin which describes a single Dirac point~\cite{Burkov:2011,Koshino:2016,RevModPhys.90.015001} 
\bse
\bea
\hat{\mathcal{H}}_{\text{3D}}&=&\hat{\mathcal{H}}_0+\hat{\mathcal{H}}_{\text{int}},
\\
\hat{\mathcal{H}}_0
&=&\sum\limits_{\mathbf{k}}\Psi^{\dagger}_{\mathbf{k}}\left[v_{\mathrm{D}}\hat{\sigma}_x\otimes(\boldsymbol{\hat{\varsigma}}\cdot\mathbf{k}) -\EF\hat{\sigma}_0\otimes \hat{\varsigma}_0
\right]\Psi^{\phantom{}}_{\mathbf{k}},
\label{eq:3DKineticPartHamiltonian}\IEEEeqnarraynumspace\\
\hat{\mathcal{H}}_{\text{int}}&=&\frac{1}{2\mathcal{V}}\sum\limits_{\mathbf{q}}V_{\text{3D}}(\mathbf{q})\hat{n}_\mathbf{q}\hat{n}_\mathbf{-q},
\label{Hint3D}
\eea
\ese
where $v_{\mathrm{D}}$ is the Dirac velocity, 
$\Psi_\mathbf{k}$ is the $4\times1$ Dirac spinor, Pauli matrices $\boldsymbol{\hat{\varsigma}}=(\hat{\varsigma}_x, \hat{\varsigma}_y,\hat{\varsigma}_z)$  and $\boldsymbol{\hat{\sigma}}=(\hat{\sigma}_x,\hat{\sigma}_y,\hat{\sigma}_z)$ represent (real) spin  and pseudospin, respectively, $\hat{\varsigma}_0$ and $\hat{\sigma}_0$ are the identity matrices in the corresponding subspaces, $\hat{n}_\mathbf{q}=\sum\limits_\mathbf{k}\Psi^\dagger_\mathbf{k}\Psi_\mathbf{k+q}$ is the density operator, $V_{\text{3D}}(\mathbf{q})$ is the interaction potential, and $\mathcal{V}$ is the system volume. In general, we assume that there are $N$ identical Dirac points. 

The  eigenvalues and orthonormal eigenfunctions of $\hat{\mathcal{H}}_0$ in Eq.~(\ref{eq:3DKineticPartHamiltonian}) 
are given by
\bea
\xi_\bk^s=s\epsilon_\mathbf{k}-\EF,\;
\epsilon_\bk=v_{\mathrm{D}}k
\label{energy}
\eea
and
\be
\ket{\mathbf{k},+}=
\frac{1}{\sqrt{2}}
\begin{bmatrix}
    \psi_1\\
\left(\boldsymbol{\hat{\varsigma}}\cdot\hat{k}\right)
    \psi_1
\end{bmatrix}
,
\ket{\mathbf{k},-}=
\frac{1}{\sqrt{2}}\begin{bmatrix}
    -\left(\boldsymbol{\hat{\varsigma}}\cdot\hat{k}\right)
    \psi_2\\
    \psi_2
\end{bmatrix}\IEEEeqnarraynumspace,\nn\\
\label{3Dstates}
\ee
respectively. Here  $\hat k=\bk/k$, $s=\pm 1$ is the helicity index,
and  $\psi_{1,2}$ are the $2\times1$ spinor states such that $\psi_{1,2}^\dagger\psi_{1,2}=1$. We choose $\psi_{1}=\psi_{2}=(0,1)^{T}$. 
The Green's function of $\hat{\mathcal{H}}_0$ is given by
\bse
\be
\hat{G}(\mathbf{k},i\omega)=\frac{1}{2}\sum\limits_{s=\pm}\hat{M}^s_{\bk}g_s(\mathbf{k},i\omega),\label{Eq:3DfullGFdefinition}\\
\hat{M}^s_\bk=\hat{\sigma}_0\otimes \hat{\varsigma}_0+s\left(
\hat{\sigma}_x\otimes(\boldsymbol{\hat{\varsigma}}\cdot \hat{k})
\right),\label{eq:3DMatrixPartDefinition}\IEEEeqnarraynumspace\\
g_s(\mathbf{k},i\omega)=\frac{1}{i\omega-\xi_\mathbf{k}^s}.
\label{eq:3DScalarGreensFunction}
\ee
\ese
For the sake of brevity, we will be omitting index $n$ in Matsubara frequencies, which will be distinguished 
from real ones by a factor of the imaginary unit, $i$. For example, $\omega$ in Eq.~\eqref{eq:3DScalarGreensFunction} stands for a Matsubara frequency.
We will be referring
to the bands with helicity $s=\pm 1$ as the ``conduction'' and ``valence'' bands, respectively. 
The density of states at the Fermi level per spin per valley is equal to $\mathcal{N}_{\fer,3}=
\EF^2/2\pi^2 v_{\mathrm{D}}^3$. 

The velocity operator corresponding to $\hat{\mathcal{H}}_0$ in (\ref{eq:3DKineticPartHamiltonian}) is
\be
\hat{\mathbf{v}}
=\vd\hat{\sigma}_x\otimes\boldsymbol{\hat{\varsigma}}\label{v3D}
\ee
with matrix elements
\bea
\bv_\bk^{s,s'}=\bra{\mathbf{k},s}\hat{\mathbf{v}}\ket{\mathbf{k},s'}.\label{vssp}
\eea
In what follows, we will need explicit expressions for the intra- and inter-band matrix elements of the velocity operator, which are given by
\be
\mathbf{v}^{s,s}_\mathbf{k}=
\bra{\mathbf{k},s}\hat{\mathbf{v}}\ket{\mathbf{k},s}=s
v_\text{D}\hat{k}
\label{vintra}
\ee
and
\bea
\mathbf{v}^{+,-}_{\mathbf{k}}&=&\left(\mathbf{v}^{-,+}_{\mathbf{k}}\right)^*=\bra{\mathbf{k},+}\hat{\mathbf{v}}\ket{\mathbf{k},-}\nn\\
&=&\vd\psi_1^\dagger\left[\boldsymbol{\hat{\varsigma}}
-\left(\boldsymbol{\hat{\varsigma}}\cdot\hat{k}\right)\boldsymbol{\hat{\varsigma}}\left(\boldsymbol{\hat{\varsigma}}\cdot\hat{k}\right)
\right]\psi_2,
\label{vinter3D}
\eea
respectively.

We now turn to the interaction part of the Hamiltonian. In what follows, we will consider two models for the interaction $V_{\text{3D}}(\mathbf{q})$:
\begin{subequations}
    \begin{align}[left = {V_{\text{3D}}(\mathbf{q})=\empheqlbrace\,}]
      & \lambda_{3},&\text{ (3D, Hubbard)}\label{eq:3DHubbard} \\
      & \frac{4\pi e^2}{q^2},&\text{ (3D, Coulomb)}\label{eq:3DCoulombInteraction}
    \end{align}
\end{subequations}
where $\lambda_{3}>0$ is a constant and $e$ is the magnitude of electron charge.
We focus on the case of low doping, when $\kf$ is much smaller than the distance between the nearby Dirac points, $b$. By ``Hubbard interaction'' we then mean an interaction that is constant for $q$ less or comparable to $\kf$ and falls off rapidly in the interval $\kf\ll q\ll b$. In that case, one can neglect scattering processes that swap electrons between the Dirac points.
The Hubbard model, though not completely realistic, captures the essential physics and allows one to obtain both analytic results for the optical conductivity in certain frequency regimes and numerical results for all frequencies. Thus, we focus most of our discussion on the Hubbard model. The Coulomb model allows one to obtain analytic results in certain frequency regimes but is very expensive computationally for arbitrary frequencies, and we will restrict our analysis of this model to analytic results only. 
We discuss both Hubbard and Coulomb interactions in more detail in Section~\ref{sec:listofdiagrams}.

Note that in  the basis of electron and hole creation/annihilation operators, which diagonalizes $\hat{\mathcal{H}}_0$, the Hamiltonian \eqref{Hint3D} accounts for {\em all} possible interaction processes, including those that do not conserve the number of electrons and holes. As mentioned in Sec.~\ref{sec:Intro}, our approach is more general in this regard than the one in 
prior studies
of optical absorption in doped semiconductors
\cite{gavoret_optical_1969,ruckenstein_many-body_1987,Sham:1990,Hawrylak:1991,Pimenov:2017}. 
These studies considered a model Hamiltonian, which allows only for the density-density interaction between electrons and holes
\bea
\hat{\mathcal{H}}'_{\text{int}}=\frac{1}{2\mathcal{V}}\sum_{\substack{\bk,\bp,\bq,\\\varsigma=\pm,s=\pm}
} V_\text{int}(\bq) \hat d^\dagger_{\bk+\bq,\varsigma,s}\hat d^\dagger_{\bp-\bq,\varsigma',-s}\hat d^{\phantom{\dagger}}_{\bp,\varsigma'-s}\hat d_{\bk,\varsigma,s}^{\phantom{\dagger}},\nn\\
\label{prev}
\eea
where $\hat d^\dagger_{\bk,\varsigma,\pm}$ is the operator creating an electron/hole with momentum $\bk$ and spin $\varsigma$. Such a Hamiltonian is correct for a parabolic spectrum, in which case intra-band absorption is forbidden by Galilean invariance while processes of electron-hole conversion are absent due to the vanishing overlap of the electron and hole states. 
However, it is not applicable to the gapless Dirac spectrum studied in this paper.

\subsection{2D Hamiltonian\label{sec:2DHamiltonian}}
As an example of a 2D Dirac metal, we consider monolayer graphene
described by the 
standard
Hamiltonian~\cite{neto:2009}:
\bse
\bea
\hat{\mathcal{H}}_{\text{2D}}&=&\hat{\mathcal{H}}_0+\hat{\mathcal{H}}_{\text{int}},\\
\hat{\mathcal{H}}_0
&=&\sum\limits_{\mathbf{k}}\Psi^{\dagger}_{\mathbf{k}}\left[\vd\left(
\tau_z
\hat{\sigma}_xk_x+\hat{\sigma}_yk_y\right)-\hat\sigma_0 \EF
\right]
\Psi_{\mathbf{k}},\IEEEeqnarraynumspace
\label{eq:2DKineticPartHamiltonian}\\
\hat{\mathcal{H}}_{\text{int}}&=&\frac{1}{2\mathcal{V}}
\sum\limits_{\mathbf{q}}V_{\text{2D}}(\mathbf{q})\hat{n}_\mathbf{q}\hat{n}_\mathbf{-q},\label{Hint}
\eea
\ese
where 
$\tau_z=
\pm
1$,
$\Psi_\mathbf{k}$ is a $2\times 1$ Dirac spinor,
the set of Pauli matrices $\hat{\boldsymbol{\sigma}}=(\hat{\sigma}_x,\hat{\sigma}_y,\hat{\sigma}_z)$ describes pseudospin, $\hat{\sigma}_0$ is the identity matrix in the same subspace,
$\hat{n}_\mathbf{q}=\sum\limits_\mathbf{k}\Psi^\dagger_\mathbf{k}\Psi_\mathbf{k+q}$ is the density operator, and $\mathcal{V}$ is the system area. To use the large-$N$ approximation afterwards, we assume that fermions carry spin $\varsigma$, such that the total degeneracy is $N=2(2\varsigma+1)$.

The  eigenvalues of $\hat{\mathcal{H}}_0$ in Eq.~(\ref{eq:2DKineticPartHamiltonian}) are the same as in Eq.~\eqref{energy}, while its orthonormal eigenfunctions 
are given by
\be
\ket{\mathbf{k},+}=
\frac{1}{\sqrt{2}}\begin{bmatrix}
    1\\
    \tau_ze^{i\tau_z\phi_\mathbf{k}}
\end{bmatrix}
,
\ket{\mathbf{k},-}=
\frac{1}{\sqrt{2}}
\begin{bmatrix}
    -\tau_z e^{-i\tau_z\phi_\mathbf{k}},
    \\
    1
\end{bmatrix}
\nn\\
\label{states}
\ee
where
$\phi_\bk$ is the azimuthal angle of $\bk$.
The Green's function of $\hat{\mathcal{H}}_0$ is given by
\bse
\be
\hat{G}(\mathbf{k},i\omega)=\frac{1}{2}\sum\limits_{s=\pm}\hat{M}^s_{\bk}g_s(\mathbf{k},i\omega),\label{Eq:2DfullGFdefinition}\\
\hat{M}^s_{\bk}=\hat{\sigma}_0+s\left(\vd\frac{\hat{\sigma}_x\tau_z k_x+\hat{\sigma}_yk_y}{\epsilon_\bk}
\right),\label{eq:2DMatrixPartDefinition}
\ee
\ese
where $g_s(\bk,i\omega)$ is the same as in Eq.~\eqref{eq:3DScalarGreensFunction}. 
The density of states at the Fermi level per spin per valley is equal to   $\mathcal{N}_{\fer,2}=\EF/2\pi v_{\mathrm{D}}^2$.

The velocity operator corresponding to $\hat{\mathcal{H}}_0$ is
\be
\hat{\mathbf{v}}
=\vd\left(\tau_z\hat{\sigma}_x,\hat{\sigma}_y\right),\label{v2D}
\ee
with its intra-band matrix element being the same as in Eq.~\eqref{vintra},
while the inter-band matrix element is given by
\be
\mathbf{v}^{+,-}_{\mathbf{k}}&=\left(\mathbf{v}^{-,+}_{\mathbf{k}}\right)^*=\bra{\mathbf{k},+}\hat{\mathbf{v}}\ket{\mathbf{k},-}
\nonumber\\
&=iv_\text{D}
e^{-i\tau_z\phi_\mathbf{k}}
\left(\hat{k}\times\hat{z}\right),
\label{vinter2D}
\ee
where
$(\hat{x},\hat{y},\hat{z})$ are the Cartesian unit vectors. 
As in 3D, 
the intra- and inter-band velocities are orthogonal to each other.

Lastly, similar to the 3D case, we consider two models of the interaction
\begin{subequations}
    \begin{align}[left = {V_{\text{2D}}(\mathbf{q})=\empheqlbrace\,}]
      & \lambda_{2},&\text{ (2D, Hubbard)}\label{eq:2DHubbard} \\
      & \frac{2\pi e^2}{q},&\text{ (2D, Coulomb)}\label{eq:2DCoulombInteraction}
    \end{align}
\end{subequations}
where $\lambda_{2}>0$ is a constant. As in 3D, by ``Hubbard'' interaction we mean the interaction with radius shorter than the Fermi wavelength but longer that the lattice constant, which
cannot transfer electrons between the valleys. As in 3D, we will present both the analytical and numerical results for the Hubbard case, and only the analytical results for the Coulomb case.

\section{Optical conductivity: general formalism
\label{sec:Formalism}}
\subsection{Kubo formula}
In linear response, the real part of the optical conductivity is given by the Kubo formula
\be
\Re\sigma_{\alpha\beta}(
\Omega)=
-\frac{1}{\Omega}\Im\Pi_{\alpha\beta\text{,R}}(\mathbf{Q=0},\Omega),\label{eq:ReConducDefi}
\ee
where $\alpha,\beta=x,y (z)$ in 2D and 3D, respectively, and $\Pi_{\alpha\beta\text{,R}}(\mathbf{Q=0},\Omega)$ is the retarded current-current correlation function (denoted by subscript ``R"), which is obtained by analytic continuation of its Matsubara counterpart:
\be
\Pi_{\alpha\beta\text{,R}}(\mathbf{Q},\Omega)=\Pi_{\alpha\beta}(\mathbf{Q},i\Omega\rightarrow\Omega+i0^+),\nn\\
\Pi_{\alpha\beta}(\mathbf{Q},i\Omega)=-\frac{1}{\mathcal{V}}\int\limits_{0}^{1/k_\text{B}T}d\tau e^{i\Omega\tau}\left<T_{\tau}\hat{j}_{\alpha}^\dagger(\mathbf{Q},\tau)\hat{j}_{\beta}(\mathbf{Q},0)\right>.\nn\\
\label{retcurrcurr}
\ee
In the basis of conduction/valence bands, the current operator is written as 
\be
\hat{\boldsymbol{j}}(\mathbf{Q},\tau)=-e\sum\limits_{\mathbf{k},s,s'}\bv^{s,s'}_{\mathbf{k}}\quad \hat{d}^{\dagger}_{\mathbf{k}-\frac{\mathbf{Q}}{2},s}(\tau)\hat{d}^{\phantom{}}_{\mathbf{k}+\frac{\mathbf{Q}}{2},s'}(\tau),
\ee
where $\bv_{\bk}^{s,s'}$ is given by Eq.~\eqref{vssp}.

For isotropic systems, considered in this paper, the conductivity tensor is diagonal and symmetric.
In this case, we define 
\bea
\Pi(\mathbf{Q},i\Omega)&\equiv&\frac{1}{d}\sum\limits_{\alpha
}
\Pi_{\alpha\alpha}(\mathbf{Q},i\Omega),\nn\\
\Pi_\text{R}(\mathbf{Q},\Omega)&\equiv &
\frac{1}{d}\sum\limits_{\alpha
}\Pi_{\alpha\alpha,\text{R}}(\mathbf{Q},\Omega),\nn\\
\Re\sigma(\Omega)&=& -\frac{1}{\Omega}\Im\Pi_\text{R}(\mathbf{Q=0},\Omega).
\eea

We also assume that temperature is much smaller than any other energy scale of the problem and consider only the $T=0$ limit.

\begin{figure}
        \includegraphics[width=0.9\linewidth,
        left]{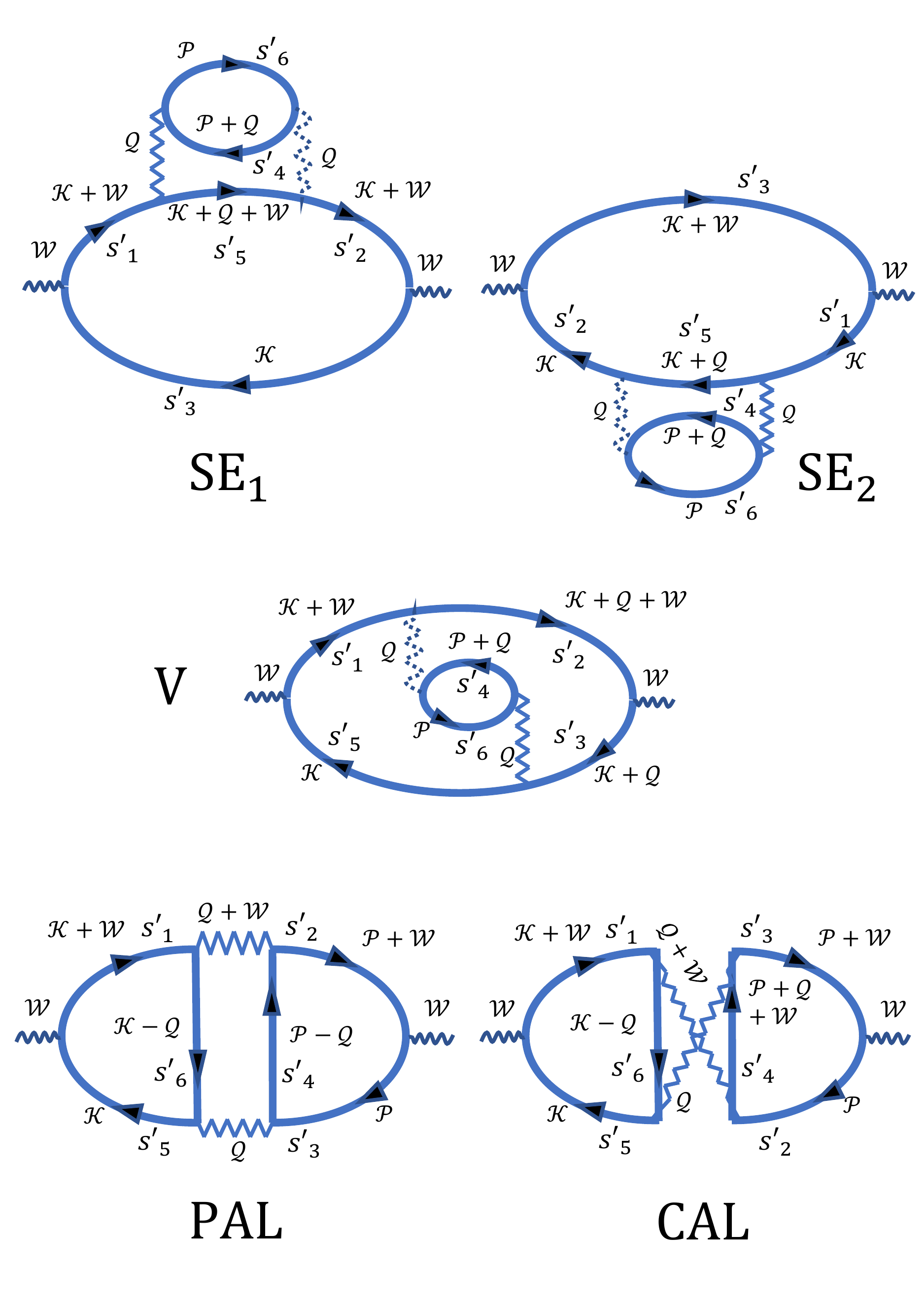}       
        \caption{Leading-order diagrams for the current-current correlation function. Thick solid lines depict the matrix Green's functions, given by Eqs.~\eqref{Eq:3DfullGFdefinition} and \eqref{Eq:2DfullGFdefinition} in 3D and 2D, respectively. For Hubbard interaction, the solid and broken wavy lines are identical and depict the Hubbard interaction $\lambda_d=\text{const}$ in $d$ dimensions, and the displayed diagrams are the leading ones in the large $N$-approximation. 
        For Coulomb interaction, the solid and broken wavy lines depict the dynamically and statically screened Coulomb potentials, respectively [Eqs.~\eqref{eq:VfullRPA},~\eqref{eq:3DCoulombstaticscree},~\eqref{eq:2DCoulombstaticscree}], and the displayed diagrams are the leading ones within the  random-phase approximation.
        The external momentum has only the frequency component: $\mathcal{W}=({\bf 0}, i\Omega)$.    From top to bottom: self-energy (SE$_1$ and SE$_2$), vertex (V), parallel 
        (PAL) and crossed  Aslamazov-Larkin (CAL) diagrams.  
        Indices $s'_1\dots s'_6=\pm 1$ indicate helicities that are being summed over.
        \label{fig:AllDiagrams}}
\end{figure}

\subsection{Relevant diagrams\label{sec:listofdiagrams}}
In Dirac metals,
optical absorption occurs already for non-interacting particles, if  the frequency of incident light exceeds the direct threshold, $\od
=2\EF$.  
The main focus of this paper is the range of $0<\Omega<2\EF$, where absorption occurs only if electrons interact with other degrees in freedom, in particular, both among themselves and with holes. Dissipation occurs only if the interaction is dynamic, i.e., if the bare interaction, either Hubbard or Coulomb,  
is dressed by particle-hole pairs. 
Diagrammatically, this corresponds to renormalizing the interaction lines either by 
particle-hole polarization bubbles or ``Aslamazov-Larkin triangles'' (cf. Fig.~\ref{fig:AllDiagrams}).

\subsubsection{Hubbard interaction}
\label{sec:Hubbard}
To make the analysis tractable, we assume that the number of identical Dirac points is large ($N\gg 1$) and 
also adopt the  
weak-coupling approximation, i.e, we assume that 
$\alpha_{\text{H}}
N\ll 1$,
where
\bea
\alpha_{\text{H}}
=\lambda_{d}\mathcal{N}_{\fer,d}\label{eq:alphadef}
\eea 
is the dimensionless coupling constant.
The first assumption allows us to retain only diagrams with the highest number of fermion loops, while the second one allows us to keep the lowest order in the interaction at which dissipation occurs, to wit: the second.  The relevant diagrams for the current-current correlation function are shown in Fig.~\ref{fig:AllDiagrams}. For the Hubbard case, the solid and broken interaction lines are identical and denote the Hubbard coupling $\lambda_{d}$.

\subsubsection{Coulomb interaction\label{sec:CoulombModelDefi}}
Within the random-phase approximation (RPA), the dynamically screened Coulomb interaction is given by
\be
V(\mathbf{q},i\nu)=\frac{1}{V^{-1}_0(\mathbf{q})+\pi_0(\mathbf{q},i\nu)}\label{eq:VfullRPA}
\ee
where 
\bea
\pi_0(\bq,i\nu)=-\int_\mathcal{K}\tr\left[\hat{G}(\bk+\bq,i\omega+i\nu)\hat{G}(\bk,i\omega)\right],
\label{eq:pi0Definition}
\eea
is the polarization bubble, $\int_{\mathcal{K}}$ is a short-hand for $(2\pi)^{-(d+1)}\int \dee^dk\int \dee\omega$, and $\hat G$ is the free-electron Green's function given by Eqs.~\eqref{Eq:3DfullGFdefinition} and \eqref{Eq:2DfullGFdefinition} in 3D and 2D, respectively. 
Since only the dynamic interaction contributes to dissipation, it is convenient to subtract off the static part of the interaction
and treat the remaining dynamic part as the effective interaction. The dynamic part is given by
\bea
V_\text{dyn}(\mathbf{q},i\nu)&\equiv& V(\bq,i\nu)-V(\bq,0)\nn\\
&=&-V(\mathbf{q},i\nu)
V(\mathbf{q},0)
\pi_{0,\text{dyn}}(\mathbf{q},i\nu),\label{eq:VdynRPA}
\eea
where
\be\pi_{0,\text{dyn}}(\mathbf{q},i\nu)=\pi_{0}(\mathbf{q},i\nu)-\pi_{0}(\mathbf{q},0)\label{pi_dyn}\ee is the dynamic part of the polarization bubble.
The lowest two-loop order diagrams in $V_\text{dyn}(\mathbf{q},i\nu)$ are shown in Fig.~\ref{fig:AllDiagrams}, where now the solid and broken wavy lines depict the dynamic and static parts of the interaction, respectively.

As opposed to the Hubbard case, the Coulomb one has an additional energy scale, \be\omega_{\text{p}d}=
\vd\kappa_d,\label{oplasma}\ee
where 
\bse
\bea
\kappa_3=\left(4\pi e^2 N\mathcal{N}_{\fer,3}\right)^{1/2}\label{kapp3}
\eea 
and \bea
\kappa_{2}=2\pi e^2 N\mathcal{N}_{\fer,2}\label{kappa2}\eea 
\ese
are the inverse screening radii in 3D and 2D, respectively. For $d=3$, $\omega_{\text{p}3}$ is on the order of the plasmon frequency at $q=0$. For $d=2$, $\omega_{\text{p}2}$ is on the order of the plasmon dispersion evaluated  at $q\sim \kappa_2$. The condition for the Coulomb interaction to be treated via  within RPA
is $\kappa_d\ll \kf$, which implies that $\omega_{\text{p}d}\ll \ef$. Correspondingly, the frequency region $0<\Omega\ll \ef$ is divided into two subregions: $0<\Omega\ll \op$ and $\op\ll \Omega\ll \ef$.
In the first subregion, a
typical energy transfer, $\nu$, is on the order of $\Omega$, while a typical momentum transfer, $q$, is on the order of $\kappa_d$. Therefore, $\nu\ll 
\vd q\sim\omega_{\text{p}d}$.
In this case, one can set $\nu=0$ in the first factor on the RHS of Eq.~\eqref{eq:VdynRPA} with the result
\be
V_\text{dyn}(\mathbf{q},i\nu)\approx 
-V^2(\mathbf{q},0)
\pi_{0,\text{dyn}}(\mathbf{q},i\nu).\label{eq:Vdynexpanded}
\ee
Diagrammatically, this amounts to replacing all the solid wavy lines by broken wavy ones in Fig.~\ref{fig:AllDiagrams}. 
Because $q\ll \kf$, the static screened potential is described by the usual Thomas-Fermi form:
\begin{subequations}
    \begin{align}[left = {V(\mathbf{q},0)=\empheqlbrace\,}]
      & \frac{4\pi e^2}{q^2+\kappa_3^2},&\text{for 3D}\label{eq:3DCoulombstaticscree} \\
      & \frac{2\pi e^2}{q+\kappa_{2}},&\text{for 2D}.\label{eq:2DCoulombstaticscree} 
    \end{align}
\end{subequations}

In the second subregion ($\omega_{\text{p}d}\ll\Omega\ll E_\fer$), typical energy and momentum transfers are $\nu\sim\vd q\sim\Omega\gg \omega_{\text{p}d}$. In this case, screening is irrelevant and the effective dynamic interaction is given by
\bea
V_{\text{dyn}}(\bq,i\nu)=-V^2_0(\bq) \pi_{0,\text{dyn}}(\bq,i\nu),\label{Cdyn}
\eea
where $V_0(\bq)$ is the bare Coulomb potential.

Note that we do not need to use the large-$N$ approximation for the Coulomb case, it is enough to require that the dimensional coupling constant of the Coulomb interaction
\bea\alpha_{\text{C}}=\frac{\vd\kappa_{d}}{\EF}\label{tkappa}
\eea
is small, which is the condition for the validity of RPA. For the Coulomb case, therefore, we will restrict our analysis to the actual value of $N$ for a specific system.

\subsection{Current-current correlation function on  the Matsubara axis}
In this section, we describe the general structure of the diagrams for the current-current correlation function.
The set of diagrams in Fig.~\ref{fig:AllDiagrams} includes two self-energy (SE) diagrams, SE$_1$ and SE$_2$, a vertex correction diagram (V),  and two Aslamazov-Larkin (AL) diagrams in the particle-particle and particle-hole channels, labelled as PAL (``parallel AL'') and CAL (``crossed AL''), respectively.
The contributions of individual diagrams to the current-current correlation function at the external $d+1$ momentum 
$\mathcal{W}\equiv({\bf 0},i\Omega)$ are given by
\bse
\be
\Pi^{\text{SE}_1}(\mathcal{W})=\frac{1}{d}\int_\mathcal{K}\tr\left[\hat{\bv}
\hat{S}(\mathcal{K+W})\cdot\hat{\bv}
\hat{G}(\mathcal{K})\right],\label{eq:SEPi1}\\
\Pi^{\text{SE}_2}(\mathcal{W})=\frac{1}{d}\int_\mathcal{K}\tr\left[\hat{\bv}
\hat{G}(\mathcal{K+W})\cdot\hat{\bv}
\hat{S}(\mathcal{K})\right],\label{eq:SEPi2}\\
\Pi^{\text{V}}(\mathcal{W})=\frac{1}{d}\int_\mathcal{K'}\tr\left[\hat{\boldsymbol{\Gamma}}
\left(\mathcal{K';W}\right)\hat{G}(\mathcal{K'+W})\cdot\hat{\bv}
\hat{G}(\mathcal{K'})\right],\nn\\
\label{PiV}\\
\Pi^{\text{\text{PAL}}}(\mathcal{W})=-\frac{1}{d}\int_\mathcal{Q}V^2_{\text{st}}(
\bq)
{\bf A}(\mathcal{Q},\mathcal{W})\cdot{\bf B}(\mathcal{Q},\mathcal{W}),\\
\Pi^{\text{CAL}}(\mathcal{W})=-\frac{1}{d}\int_\mathcal{Q}V^2_{\text{st}}
(\bq
)
{\bf A}(\mathcal{Q},\mathcal{W})\cdot{\bf C}(\mathcal{Q},\mathcal{W}),
\label{eq:CALPi}
\ee
\ese
where
\bse
\be
\hat{S}(\mathcal{L})=\hat{G}(\mathcal{L})\hat{\Sigma}(\mathcal{L})\hat{G}(\mathcal{L}),\label{curlyS}\\
\hat{\Sigma}(\mathcal{L})=-\int_\mathcal{Q}\tilde{V}(\mathcal{Q})\hat{G}(\mathcal{L+Q}),\label{Sigma}\\
\tilde{V}(\mathcal{Q})=-V_\text{st}^2(\mathbf{q})\pi_0(\mathcal{Q}),\label{curlyV}
\\
\hat{\boldsymbol{\Gamma}}
\left(\mathcal{K';W}\right)=-\int_{\mathcal{K}}\tilde{V}(\mathcal{K'-K})\hat{G}(\mathcal{K})\hat{\bv}
\hat{G}(\mathcal{K+W}),\\
{\bf A}
(\mathcal{Q},\mathcal{W})=-\int_\mathcal{K}\tr\left[\hat{G}(\mathcal{K})\hat{\bv}
\hat{G}(\mathcal{K+W})\hat{G}(\mathcal{K-Q})\right], 
\\
{\bf B}
(\mathcal{Q},\mathcal{W})=-\int_\mathcal{P}\tr\left[\hat{G}(\mathcal{P+W})\hat{\bv}
\hat{G}(\mathcal{P})\hat{G}(\mathcal{P-Q})\right], 
\\
{\bf C}
(\mathcal{Q},\mathcal{W})=-\int_\mathcal{P}\tr\left[\hat{G}(\mathcal{P})\hat{G}(\mathcal{P+W+Q})\hat{G}(\mathcal{P+W})\hat{\bv}
\right],\nn
\\
\ee
\ese
$\mathcal{K}\equiv(\mathbf{k},i\omega)$, $\mathcal{K'}\equiv(\mathbf{k'},i\omega)$, $\mathcal{P}\equiv(\mathbf{p},i\omega)$, $\mathcal{Q}\equiv(\mathbf{q},i\nu)$, 
$\pi_0(\mathcal{Q})$ is defined by Eq.~\eqref{eq:pi0Definition},
and
$V_\text{st}(\mathbf{q})$
is the static part of the interaction, equal to $V(\mathbf{q},0)$ [Eqs.~\eqref{eq:3DCoulombstaticscree} and \eqref{eq:2DCoulombstaticscree}] and to $\lambda_d$ for the Coulomb and Hubbard cases, respectively. 
The expressions above are valid for Coulomb interaction 
at the lowest frequencies ($\Omega\ll\EF$)  and for any frequency for Hubbard interaction. 
Using the free rather than dressed Green's functions is justified for any frequency except for a narrow region near the direct threshold (a precise condition will be formulated later, cf. Sec.~\ref{sec:IF3D}).
The total current-current correlation function is the sum of all the contributions displayed above:
\be
\Pi(\mathcal{W})=\sum\limits_{J}\Pi^J(\mathcal{W}),\label{eq:cuur-curr-corr-sum-J}
\ee
where 
$J\in\{\text{SE},\text{V},\text{PAL},\text{CAL}\}$, and ``SE'' refers to both the self-energy diagrams collectively.

Equations \eqref{eq:SEPi1}-\eqref{eq:CALPi} 
become more transparent if written in the electron-hole basis, in which $\hat{\mathcal{H}}_0$ is diagonal.
Indeed, any diagram contains six Green's functions,
each being the sum 
of an electron and hole parts with helicities $s'=\pm 1$, respectively. 
This gives rise to a set of six helicities $\mathcal{S}'
=\{
s'_1\dots s'_6\}$ that are to be summed over. Thus, each  diagram is the sum of $2^6=64$ terms     \be
\Pi^{J}(\mathcal{W})=\sum\limits_{\mathcal{S}'
        }\Pi^{J}_{\mathcal{S}'
                }(\mathcal{W}),\label{eq:curr-currMatsu}
\ee
where summation goes over all 64 configurations of $\mathcal{S}'$.
Each $\Pi^{J}_{\mathcal{S}'
        }(\mathcal{W})$ term in the sum contains a product of two integrals over the frequency
\be
\int d\omega\prod\limits_{l=1}^{L}g_{s_l}(\mathbf{k}_l,i\omega+i\nu_l)
\int d\omega'\prod\limits_{l'=1}^{L'}g_{s_{l'}}(\mathbf{k}_{l'},i\omega'+i\nu_{l'}),\nn\\
\label{conv}
\ee
where $L=4$ for all diagrams, $L'=2$ for SE$_{1,2}$ and V diagrams,   and $L=L'=3$ for PAL and CAL diagrams, $s_l,s_{l'}\in\mathcal{S}'
$, and $g_{s}(\mathbf{k},i\omega)$
is the Green's function in the diagonal basis, defined by Eq.~\eqref{eq:3DScalarGreensFunction}.
The integrals in Eq.~\eqref{conv} vanish if the poles of the integrands are located in the same halves of the complex plane. 
Because $\xi_{\mathbf{k}_l}^{s_l}<0$ for $s_l=-1$, at least one of the helicities in each of the integrals in Eq.~\eqref{conv} must be positive for a non-zero result. 
Thus, instead of  $2^6=64$ terms we would, in general, have only $2^4=16$ terms in the sum of helicities in Eq.~\eqref{eq:curr-currMatsu}.

\subsection{Retarded 
current-current correlation function
\label{sec:AnalyticContinuation}}
Upon analytic continuation, 
the imaginary part of the retarded current-current correlation function                     can be written 
                as a sum over         the new terms, $\mathcal{R}^{J}_{\mathcal{S}}(\Omega)$:
\be
\Im\Pi^{J}_{\text{R}}(\Omega)=\sum\limits_{\mathcal{S}'
        }\Im\Pi^{J}_{\mathcal{S}'
            ,\text{R}}(\Omega)=
    \sum\limits_{\mathcal{S}}\mathcal{R}^{J}_{\mathcal{S}}(\Omega),
    \label{RJ}
\ee
where $\mathcal{S}\in\{s_1\ldots s_6\}$ is another set of helicities, which is different from $\mathcal{S}'$, and the subscript ``R'' stands for ``retarded''. Note that while the equality between the sums in Eq.~\eqref{RJ} is always valid, there is, in general, no one-to-one correspondence between the individual terms of the two sums.\footnote{
The rationale 
behind transitioning from 
$\Im\Pi^J_{\mathcal{S'},\text{R}}(\Omega)$ to $\mathcal{R}^{J}_{\mathcal{S}}(\Omega)$, which differ only in labeling of the helicities, is mere convenience.
Namely, it allows one 
to systematically collect contributions with similar behaviors
into
$\mathcal{R}^{J
}_{\mathcal{S}}(\Omega)$.
\label{ftnt:PivsRexplanation}} 
Looking ahead, it will be convenient to represent not only the self-energy but also all other diagrams 
as sums of two terms, which we will distinguish by assigning a label 
$u=1,2$ to the diagram index $J$, i.e., 
\be
\mathcal{R}^{J}_{\mathcal{S}}(\Omega) = \sum\limits_{u=1,2}\mathcal{R}^{J_u}_{\mathcal{S}}(\Omega),
\ee
 where $J_{1,2}\in\{\text{SE}_{1,2},\text{V}_{1,2},\text{PAL}_{1,2},\text{CAL}_{1,2}\}$. Note that whereas the subscript $u$ refers to 
two topologically distinct diagrams 
for the SE case, its 
meaning for the V, PAL and CAL contributions
is purely algebraic. For example, the contribution of the vertex diagram is represented by a sum of two terms in Eq.~\eqref{ImV}, and similarly for the AL diagrams.

We remind the reader that we chose $\Omega>0$. With this choice, as shown in Appendix~\ref{app:genstructure}, 
any of the $\mathcal{R}^{J_u
}_{\mathcal{S}}(\Omega)$ terms 
has the following structure 
\begin{widetext}
\be
\mathcal{R}^{J_u
}_{\mathcal{S}}(\Omega)=K^{J_u}\int_{\mathbf{k,p,q}}\int_\nu &V^2_{\text{st}}(\mathbf{q})
\mathcal{T}^{J_u}_\mathcal{S}\left(\bk,\bp,\bq\right)\mathcal{G}^{J_u}_\mathcal{S}(\bk,\bp,\bq,\Omega)\nn\\
&\times 
\theta(\Omega+\xi^{s_3}_\mathbf{k})
\theta(-\xi^{s_3}_\mathbf{k})\theta(\xi^{s_5}_\mathbf{k+q})\theta(-\xi^{s_4}_\mathbf{p})\theta(\xi^{s_6}_\mathbf{p+q})\delta(\Omega+\nu+\xi^{s_3}_\mathbf{k}-\xi^{s_5}_\mathbf{k+q})\delta(\nu+\xi^{s_6}_\mathbf{p+q}-\xi^{s_4}_\mathbf{p}).\label{Eq:generalzeroTexpression}
\ee
\end{widetext}
[A rather complicated form 
of Eq.~\eqref{Eq:generalzeroTexpression} 
will be clarified later 
by an example of the $\text{SE}_1$ diagram; see Eq.~\eqref{eq:Im-curr-curr-total} and Sec.~\ref{sec:archetype-conductivity}.]
Here, $\int_{{\bf n}}$ is a shorthand for $\int \dee^dn/(2\pi)^d$, $\int_\nu$ stands for $\int^\infty_{-\infty} \dee\nu/2\pi$, and
\bea 
K^{J_u}=
\left\{
\begin{array}{ccc}
-\pi^2/32,\;\text{for}\; J_{u}=\text{SE}_{1,2},\text{CAL}_{1,2},\\
\pi^2/32,\;\text{for}\; J_{u}=\text{V}_{1,2},\text{PAL}_{1,2}.
\end{array}
\right.
\label{KJu}
\eea
Further, $\mathcal{T}^{J_u}_\mathcal{S}$ denote the trace of matrix products coming from the spinor wavefunctions
and 
$\mathcal{G}^{J_u}_\mathcal{S}$ are the products of the real parts of the Green's functions, given by
\be
\mathcal{T}^{\text{SE}_1}_\mathcal{S}=&\frac{1}{d}\tr\left(\hat{\bv}
\hat{M}_\mathbf{k}^{s_1}\hat{M}_{\mathbf{k+q}}^{s_5}\hat{M}_{\mathbf{k}}^{s_2}\cdot\hat{\bv}
\hat{M}_{\mathbf{k}}^{s_3}\right)\nn\\
&\times\tr\left(\hat{M}_\mathbf{-p-q}^{s_6}\hat{M}_\mathbf{-p}^{s_4}\right),
\nonumber\\
\mathcal{T}^{\text{SE}_2}_\mathcal{S}=&\frac{1}{d}\tr\left(\hat{\bv}
\hat{M}_\mathbf{-k-q}^{s_1}\hat{M}_{\mathbf{-k}}^{s_3}\hat{M}_{\mathbf{-k-q}}^{s_2}\cdot\hat{\bv}
\hat{M}_{\mathbf{-k-q}}^{s_5}\right)\nonumber\\
&\times\tr\left(\hat{M}_\mathbf{p}^{s_4}\hat{M}_\mathbf{p+q}^{s_6}\right),\nonumber\\
\mathcal{G}^{\text{SE}_1}_\mathcal{S}=&\frac{1}{\Omega
-\xi^{s_1}_\mathbf{k}+\xi^{s_3}_\mathbf{k}}\frac{1}{\Omega
-\xi^{s_2}_\mathbf{k}+\xi^{s_3}_\mathbf{k}},\nn
\\
\mathcal{G}^{\text{SE}_2}_\mathcal{S}=&\frac{1}{\Omega-\xi^{s_5}_\mathbf{k+q}+\xi^{s_1}_\mathbf{k+q}}\frac{1}{\Omega-\xi^{s_5}_\mathbf{k+q}+\xi^{s_2}_\mathbf{k+q}},\label{eq:firsteqofresultlist}
\ee
\be
\mathcal{T}^{\text{V}_1}_\mathcal{S}=&\frac{1}{d}\tr\left(\hat{\bv}
\hat{M}_\mathbf{k}^{s_1}\hat{M}_{\mathbf{k+q}}^{s_5}\cdot\hat{\bv}
\hat{M}_{\mathbf{k+q}}^{s_2}\hat{M}_{\mathbf{k}}^{s_3}\right)
\nonumber\\
&\times
\text{Tr}\left(\hat{M}_{\mathbf{-p-q}}^{s_6}\hat{M}_{\mathbf{-p}}^{s_4}\right),
\nonumber\\
\mathcal{T}^{\text{V}_2}_\mathcal{S}=&\frac{1}{d}\tr\left(\hat{\bv}
\hat{M}_{\mathbf{-k-q}}^{s_5}\hat{M}_{\mathbf{-k}}^{s_1}\cdot\hat{\bv}
\hat{M}_{\mathbf{-k}}^{s_3}\hat{M}_{\mathbf{-k-q}}^{s_2}\right)
\nonumber\\
&\times
\text{Tr}\left(\hat{M}_\mathbf{p}^{s_4}\hat{M}_{\mathbf{p+q}}^{s_6}\right),\nonumber\\
\mathcal{G}^{\text{V}_1}_\mathcal{S}=&
\mathcal{G}^{\text{V}_2}_\mathcal{S}
=\frac{1}{\Omega-\xi^{s_1}_\mathbf{k}+\xi^{s_3}_\mathbf{k}}\frac{1}{\Omega-\xi^{s_5}_\mathbf{k+q}+\xi^{s_2}_\mathbf{k+q}},\label{eq:VertexTGresults}
\ee
\be
\mathcal{T}^{\text{PAL}_1}_\mathcal{S}=&\frac{1}{d}\text{Tr}\left(\hat{\bv}
\hat{M}_{\mathbf{-k}}^{s_1}\hat{M}_{\mathbf{-k-q}}^{s_5}\hat{M}_{\mathbf{-k}}^{s_3}\right)
\nonumber\\
&
\times
\cdot
\text{Tr}\left(\hat{\bv}
\hat{M}^{s_2}_{\mathbf{p+q}}\hat{M}_{\mathbf{p}}^{s_4}\hat{M}_{\mathbf{p+q}}^{s_6}\right),\nonumber\\
\mathcal{T}^{\text{PAL}_2}_\mathcal{S}=&\frac{1}{d}\text{Tr}\left(\hat{\bv}
\hat{M}^{s_3}_{\mathbf{-k}}\hat{M}_{\mathbf{-k-q}}^{s_5}\hat{M}_{\mathbf{-k}}^{s_1}\right)
\nonumber\\
&
 \times
 \cdot
\text{Tr}\left(\hat{\bv}
\hat{M}_{\mathbf{p+q}}^{s_6}\hat{M}_{\mathbf{p}}^{s_4}\hat{M}_{\mathbf{p+q}}^{s_2}\right),\nonumber\\
\mathcal{G}^{\text{PAL}_1}_\mathcal{S}=&\mathcal{G}^{\text{PAL}_2}_\mathcal{S}=\frac{1}{\Omega-\xi^{s_1}_{\mathbf{k}}+\xi^{s_3}_{\mathbf{k}}}\frac{1}{\Omega-\xi^{s_6}_{\mathbf{p+q}}+\xi^{s_2}_{\mathbf{p+q}}},\IEEEeqnarraynumspace\label{eq:PALTGresults}
\ee
\be
\mathcal{T}^{\text{CAL}_1}_\mathcal{S}=&\frac{1}{d}\text{Tr}\left(\hat{\bv}
\hat{M}_{\mathbf{-k}}^{s_1}\hat{M}_{\mathbf{-k-q}}^{s_5}\hat{M}_{\mathbf{-k}}^{s_3}\right)
\nonumber\\
&
 \times
 \cdot
\text{Tr}\left(\hat{\bv}
\hat{M}_\mathbf{p}^{s_4}\hat{M}_{\mathbf{p+q}}^{s_6}\hat{M}_\mathbf{p}^{s_2}\right),\nonumber\\
\mathcal{T}^{\text{CAL}_2}_\mathcal{S}=&\frac{1}{d}\text{Tr}\left(\hat{\bv}
\hat{M}_{\mathbf{k+q}}^{s_5}\hat{M}_{\mathbf{k}}^{s_3}\hat{M}_{\mathbf{k+q}}^{s_1}\right)
\nonumber\\
&
 \times
 \cdot
\text{Tr}\left(\hat{\bv}
\hat{M}_{\mathbf{-p-q}}^{s_2}\hat{M}_{\mathbf{-p}}^{s_4}\hat{M}_{\mathbf{-p-q}}^{s_6}\right),\nonumber\\
\mathcal{G}^{\text{CAL}_1}_\mathcal{S}=&\frac{1}{\Omega-\xi^{s_2}_{\mathbf{p}}+\xi^{s_4}_{\mathbf{p}}}\frac{1}{\Omega-\xi^{s_1}_{\mathbf{k}}+\xi^{s_3}_{\mathbf{k}}},\nonumber\\
\mathcal{G}^{\text{CAL}_2}_\mathcal{S}=&\frac{1}{\Omega-\xi^{s_6}_{\mathbf{p+q}}+\xi^{s_2}_{\mathbf{p+q}}}\frac{1}{\Omega-\xi^{s_5}_{\mathbf{k+q}}+\xi^{s_1}_{\mathbf{k+q}}}.
\label{eq:CALexactexpressions}
\ee
Here, $\hat M^t_{\bf{l}}$ is the matrix part of the Green's function given by Eqs.~\eqref{eq:3DMatrixPartDefinition} and 
\eqref{eq:2DMatrixPartDefinition} in 3D and 2D, respectively.
Note that 
for 
all diagrams
$\mathcal{T}^{J_u}_\mathcal{S}$ are separable functions of the momenta $\mathbf{k}$ and $\mathbf{p}$, i.e.,
\be
\mathcal{T}^{J_u}_\mathcal{S}(\bk,\bp,\bq)=\mathcal{T}_1^{J_u}(\mathbf{k,q})\mathcal{T}^{J_u}_2(\mathbf{p,q}).
\ee
From the 
$\theta$-functions in Eq.~\eqref{Eq:generalzeroTexpression}, we see that the result is non-zero only if $\xi^{s_5}_\mathbf{k+q}>0$ and $\xi^{s_6}_\mathbf{p+q}>0$, which implies that
\be
s_5=s_6=+1,
\label{eq:PositivehelicitiesConstraint}
\ee
i.e., the corresponding solid lines in diagrams describe electrons in the conduction band. This is a particular instance of the general constraint 
discussed after Eq.~\eqref{conv},
thanks  to which the sum over helicities contains now only  $2^4=16$ instead of  $2^6=64$ terms.
The remaining 
helicities belong to the subset 
\be
\mathcal{S}_A\in\{s_1,s_2,s_3,s_4\}.\label{eq:SAdefi}
\ee
Therefore,  the contribution of diagram $J$ to the sum in Eq.~\eqref{RJ} 
is given by the sum of $16$ terms of the type $\mathcal{R}^{J}_{\mathcal{S}_A}(\Omega)$:
\be
\Im\Pi^{J}_{\text{R}}(\Omega)=\sum\limits_{\mathcal{S}_A}\mathcal{R}^{J}_{\mathcal{S}_A}(\Omega)=\sum\limits_{\mathcal{S}_A,u}\mathcal{R}^{J_u}_{\mathcal{S}_A}(\Omega).\label{eq:RJnewSum}
\ee
Thus, the total retarded current-current correlation function  $\Im\Pi_\text{R}(
\Omega)$,  is given by [cf. Eqs.~\eqref{eq:cuur-curr-corr-sum-J} and \eqref{RJ}]:
\be
\Im\Pi_\text{R}(\Omega)
=\sum\limits_{J}\Im\Pi^{J}_{\text{R}}(\Omega)
 =\sum\limits_{\mathcal{S}_A,J}\mathcal{R}^{J}_{\mathcal{S}_A}(\Omega)
=\sum\limits_{\mathcal{S}_A,J_u}\mathcal{R}^{J_u}_{\mathcal{S}_A}(\Omega)\nn\\
=\sum_{
s_3,s_4}\int_{\mathbf{k,p,q},\nu}\mathcal{D}(\bk,\bp,\bq,\nu,\Omega)
\nn\\
\qquad\qquad\times\sum_{J_u,
s_1,s_2}K^{J_u}\mathcal{T}^{J_u
}_{\mathcal{S}_A}(\bk,\bp,\bq)\mathcal{G}^{J_u
}_{\mathcal{S}_A}(\bk,\bp,\bq,\Omega),
\label{eq:Im-curr-curr-total}
\ee
where
\be
\mathcal{D}(\bk,\bp,&\bq,\nu,\Omega)=
\theta(\Omega+\xi^{s_3}_\mathbf{k})
\theta(-\xi^{s_3}_\mathbf{k})
\theta(\xi^{+}_\mathbf{k+q})
\theta(-\xi^{s_4}_\mathbf{p})
\theta(\xi^{+}_\mathbf{p+q})
\nn\\
&\times\delta(\Omega+\nu+\xi^{s_3}_\mathbf{k}-\xi^{
+
}_\mathbf{k+q})\delta(\nu+\xi^{
+
}_\mathbf{p+q}-\xi^{s_4}_\mathbf{p})\IEEEeqnarraynumspace 
\label{Ddefinition}
\ee
is the block of kinematic constraints represented by the theta- and delta-functions with Eq.~\eqref{eq:PositivehelicitiesConstraint} implemented.
The $\theta$-functions reflect the Pauli principle, while the $\delta$-functions manifest the energy conservation.
Note that $\mathcal{D}(\mathbf{k,p,q},\nu)$ 
depends only on helicities $s_3,s_4$ and is the same for all diagrams; therefore, 
it can be pulled out of the sum over $J_u,s_1$, and $s_2$. From now onward, we assume that the constraint Eq.~\eqref{eq:PositivehelicitiesConstraint} has already been implemented.

\section{
Scattering processes}
\label{sec:processes}
\subsection{Frequency thresholds}
\label{sec:Frequencyregimes}
Different terms in Eq.~\eqref{eq:RJnewSum} start to contribute at frequencies above certain thresholds.
These thresholds can be deduced from the kinematic constraints 
in Eq.~\eqref{Ddefinition}, which depend only on the helicities $s_3,s_4$ and are the same for all diagram types.
(For the reader's convenience, helicity sets corresponding to different scattering processes are summarized in Table~\ref{table:helicitiesSummary}.)

Equation~\eqref{Ddefinition} gives rise to the following kinematic constraints: 
\bse
\bea
&&\xi^{+}_\mathbf{k+q}=\epsilon_{\bk+\bq}-\EF>0,\;\xi^{+}_\mathbf{p+q}=\epsilon_{\bp+\bq}-\EF>0,\nn\\\label{c0}\\
&&\xi^{s_3}_\mathbf{k}=s_3\epsilon_\bk-\EF<0,\;\xi^{s_4}_\mathbf{p}=s_4\epsilon_\bp-\EF<0,\label{ca}\\
&&\Omega+\nu+\xi^{s_3}_\mathbf{k}=\xi^+_\mathbf{k+q};\;\xi^{s_4}_\mathbf{p}-\nu=\xi^{+}_\mathbf{p+q}.\label{cb}
\eea
\ese
The inequalities \eqref{c0} and \eqref{cb} imply that 
\be
\EF-\Omega-s_3\epsilon_\bk<\nu<s_4\epsilon_\bp-\EF,
\ee
which, in 
\dm{its} 
turn, leads to
\be
s_4\epsilon_\bp+s_3\epsilon_\bk>2\EF-\Omega.\label{c1}
\ee
\begin{figure}
    \centering
        \includegraphics[width=0.9\linewidth
                ,left]
      {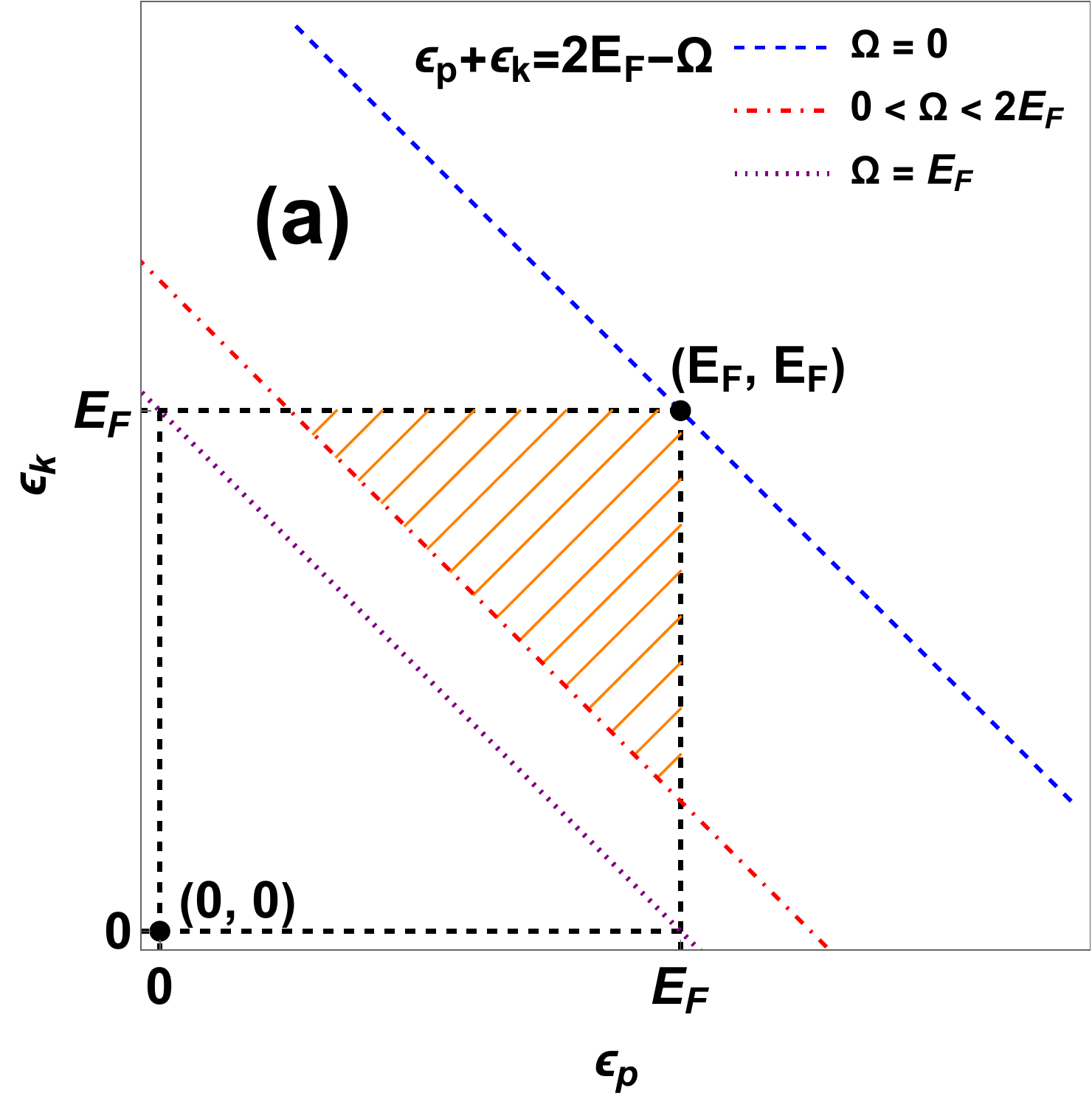}
        \includegraphics[width=0.9\linewidth
                ,left]
        {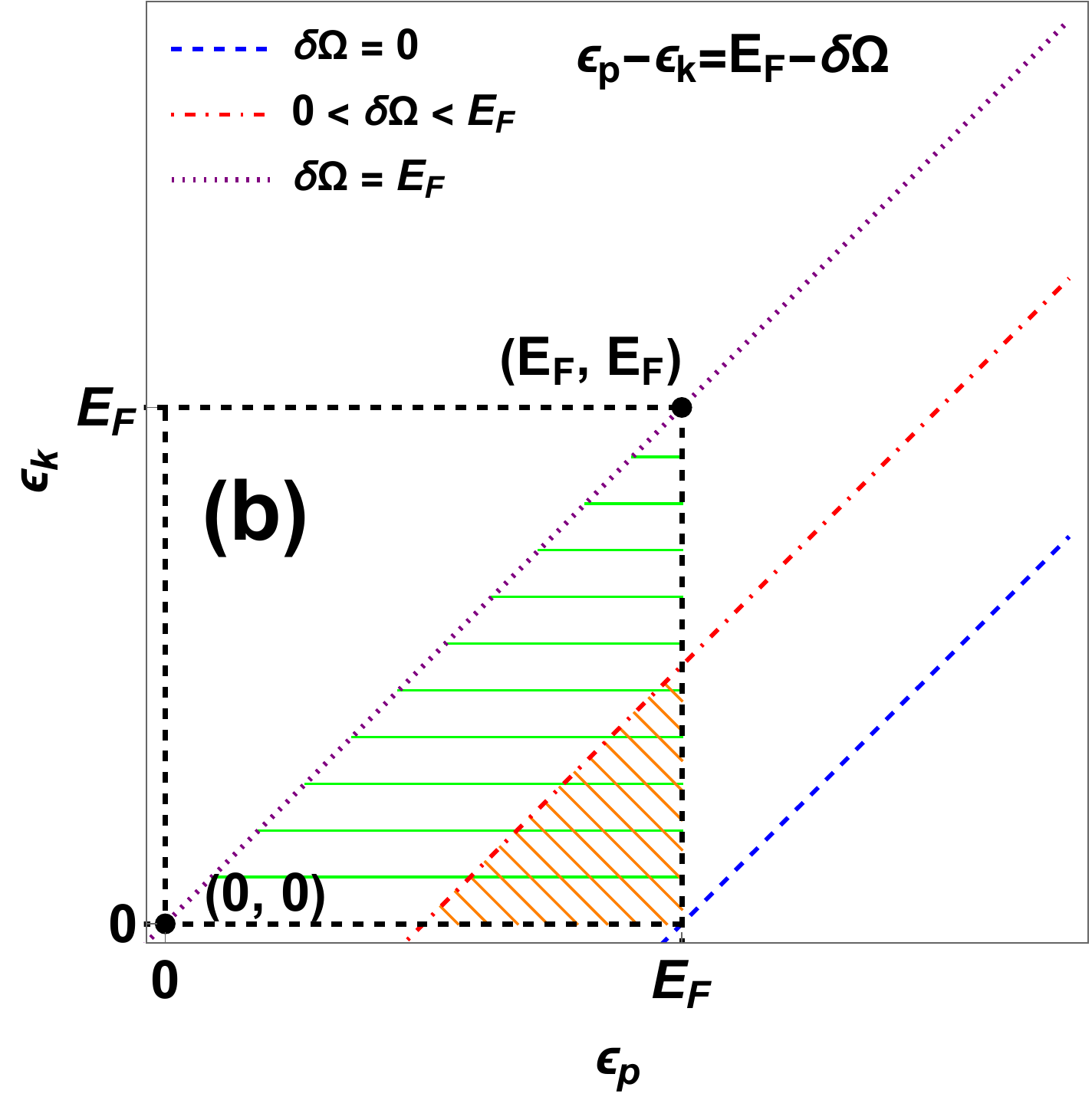}
                           \caption{
           a) Region in the $(\e_\bk,\e_\bp)$ plane contributing to absorption due to electron-electron and electron-hole scattering processes, described in Sec.~\ref{sec:LFregime}, for frequencies in the range $0<\Omega<\od=2\EF$. Straight lines are the equations $\epsilon_\bp+\epsilon_\bk=2\EF-\Omega$, for different choices of $\Omega$, as specified in the legend.
                     b) Region in the $(\e_\bk,\e_\bp)$ plane contributing to absorption due to AM-like processes, described in Sec.~\ref{sec:IFregime}, which occur in the range $\oi=\EF<\Omega<2\EF
                           $. 
              Straight lines are the equations $\epsilon_\bp-\epsilon_\bk=\EF-\delta\Omega$, for different choices of $\Omega$, as specified in the legend.
       \label{fig:LFIFkp}}
\end{figure}
Making all possible choices 
of
$s_4=\pm 1$ and $s_3=\pm 1$, we obtain the frequency thresholds which delineate three
frequency regimes, as described in the following sections.
\bse
\begin{table*}
    \caption{Summary of the helicity sets for different scattering processes. Here $\od=2\EF$, $\oi=\EF$, while \emph{ee}, \emph{eh1}, \emph{eh2} stand for absorption processes involving electrons only, one hole, and two holes, respectively.  Note that for Auger-Meitner (AM) processes, the choices of $s_1=\pm 1$ and $s_2=\pm 1$ are not correlated either to each other or to the choices of $s_3$ and $s_4$, while the choices of $s_3$ and $s_4$ are correlated to each other.
    Examples of diagrams involving \emph{eh1} and \emph{eh2} processes are shown in Fig.~\ref{fig:1h2hdiags}; examples of diagrams involving  AM processes are shown in Fig.~\ref{fig:Augerdiags}. 
 \label{table:helicitiesSummary}}
    \begin{ruledtabular}
    \begin{tabular}{c|c|c|c|c|c|c|c}
Frequency range & Type  & $s_1$
& $s_2$ & $s_3$ & $s_4$ & $s_5$ & $s_6$ 
\\
\hline

$ 0<\Omega<\od$ & \emph{ee}& $+1$
& $+1$ & $+1$ & $+1$ & $+1$ & $+1$ 
  \\
    \hline  
 $0<\Omega<\od$ & \emph{eh1}
    &   $\pm1$
& $\mp1$ & $+1$ & $+1$ & $+1$ & $+1$ 
  \\
    \hline  
 $0<\Omega<\od$ & \emph{eh2}
    &   $-1$
& $-1$ & $+1$ & $+1$ & $+1$ & $+1$ 
  \\
    \hline  
 $\oi<\Omega<\od$ & AM
        &   $\pm 1$
& $\pm 1$ & $\pm 1$ & $\mp 1$ & $+1$ & $+1$ 
         \end{tabular}
    \end{ruledtabular}
\end{table*}
\ese

\subsubsection{All frequencies: $0<\Omega< \od$\label{sec:LFregimedefinition}}
\label{sec:LFregime}

The choice of $s_3=s_4=+1$ corresponds to processes  whose contributions start right at $\Omega>0$ and continue up to $\od=2\EF$ (and beyond). 
      Combining Eqs.~\eqref{ca} and \eqref{c1},  we see that dispersions $\epsilon_\bk$ and $\epsilon_\bp$ are constrained by the following inequalities: 
    \bea
    0<\e_\bk<\EF,\;0<\e_\bp<\EF,\;\e_\bk+\e_\bp>2\EF-\Omega.
    \eea
    Geometrically, these constraints are shown in Fig.~\ref{fig:LFIFkp}$a$. At $\Omega=0$, the slanted line $\e_\bk+\e_\bp=2\EF-\Omega$ touches the corner
        of the square, which formed by the horizontal line $\e_\bk=\EF$ and vertical line $\e_\bp=\EF$.
    For $0<\Omega<2\EF$, the  slanted line $\e_\bk+\e_\bp=2\EF$ cuts 
through
the square, such that the allowed values of $\e_\bk$ and $\e_\bp$ lie in the diagonally hatched region.
    
   This 
      regime includes processes of pure intra-band absorption ($s_1=s_2=+1$), 
      when all the six states are in the conduction band, and dissipation occurs in the same way as 
      in a DFL~\cite{Sharma:2021}. 
                  In addition, this regime includes scattering processes between electrons and holes. With four out of six helicities chosen positive ($s_3=s_4=s_5=s_6=+1$), 
   either one of helicities $s_1$ and $s_2$  or both of them    can be negative. Therefore, such scattering processes involve up to two states
      in the valence band, while the numbers of electrons and holes are not conserved separately.
  
As we discussed in Sec.~\ref{sec:Intro}, 
absorption due to all processes described above is absent within the model of a gapped semiconductor with parabolic bands and the interaction Hamiltonian given in Eq.~\eqref{prev},
which was considered in
Refs.~\cite{gavoret_optical_1969,ruckenstein_many-body_1987,Sham:1990,Hawrylak:1991,Pimenov:2017}.

    \subsubsection{Intermediate frequencies: $\omega_{\text{I}}\leq\Omega<\omega_{\text{D}}$}
    \label{sec:IFregime}
    In addition to still active \emph{ee} and \emph{eh} processes, described in the previous section, another type of \emph{eh} processes contributes to the conductivity in  the intermediate-frequency regime, defined as $\omega_{\text{I}}\leq \Omega < \omega_{\text{D}}$.
      This regime corresponds to the following helicity choices: 1)  $s_3=-1,\,s_4=+1$ and 2) $s_3=+1,\,s_4=-1$.
   For the first choice, Eqs.~\eqref{ca} and \eqref{c1} imply that
\bea
&&0<\epsilon_\bk<\infty,\; 0<\epsilon_\bp<\EF,\;\nn\\
&&\epsilon_\bp-\epsilon_\bk>2\EF-\Omega=\EF-\delta\Omega,
\eea
where $\delta\Omega\equiv\Omega-\EF$.
Geometrically, these constraints are depicted in Fig.~\ref{fig:LFIFkp}$b$. The constraints are satisfied if the line 
$\epsilon_\bp-\epsilon_\bk=\EF-\delta\Omega$
cuts across the semi-infinite 
band, defined by the inequalities $0<\epsilon_\bk<\infty$ and $0<\epsilon_\bp<\EF$,  which is only possible if $\Omega>\EF= \omega_{\text{I}}$.
The 
contribution from the
second choice, $s_4=-1,s_3=+1$, can be re-written in terms of the first one 
via an appropriate re-labelling of helicities,
and thus this case does not need to be analyzed separately.

\begin{figure}
                                               \includegraphics[width=\linewidth
                ,left]
        {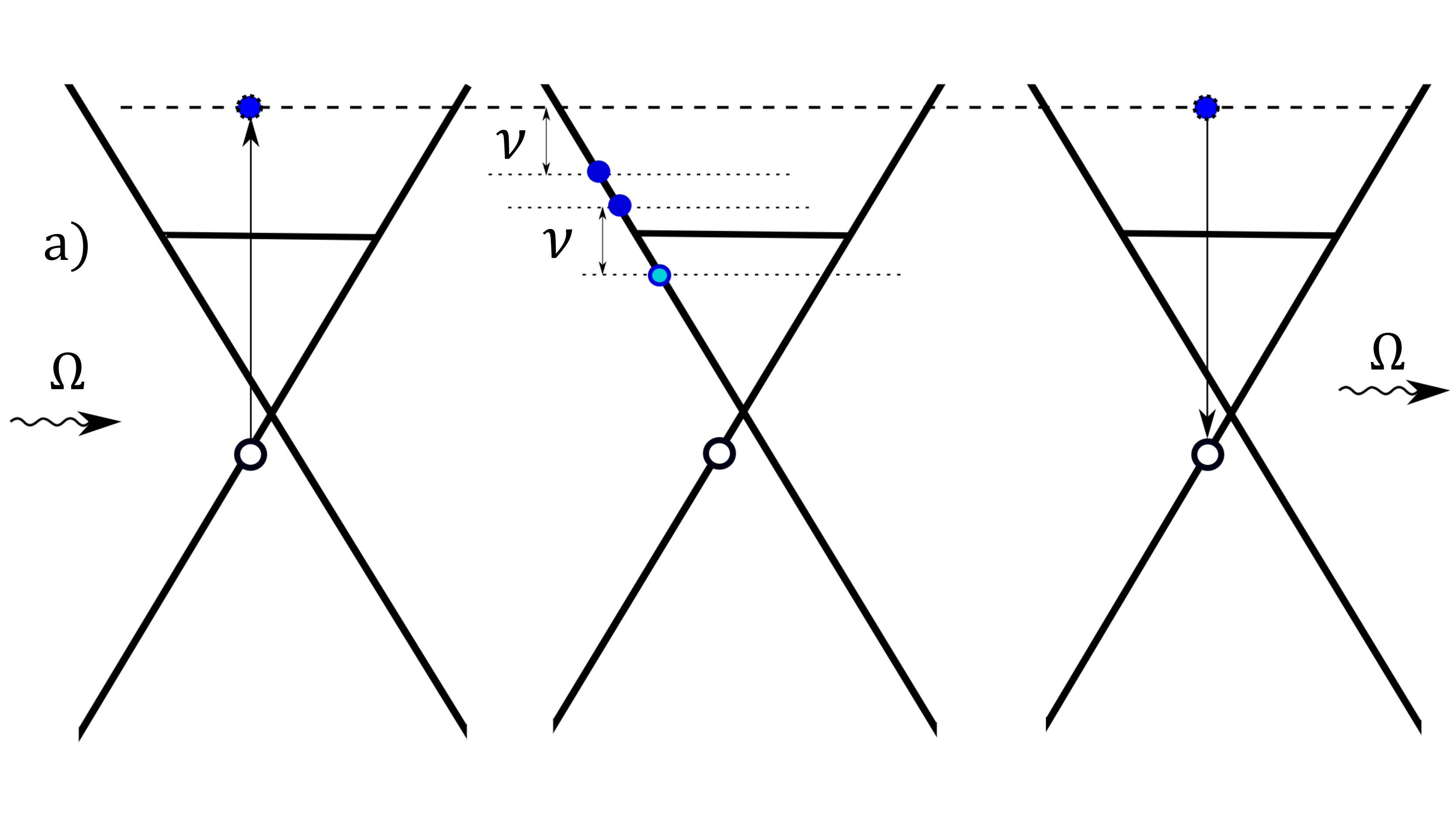}
                                                                \includegraphics[width=\linewidth
                ,left]
        {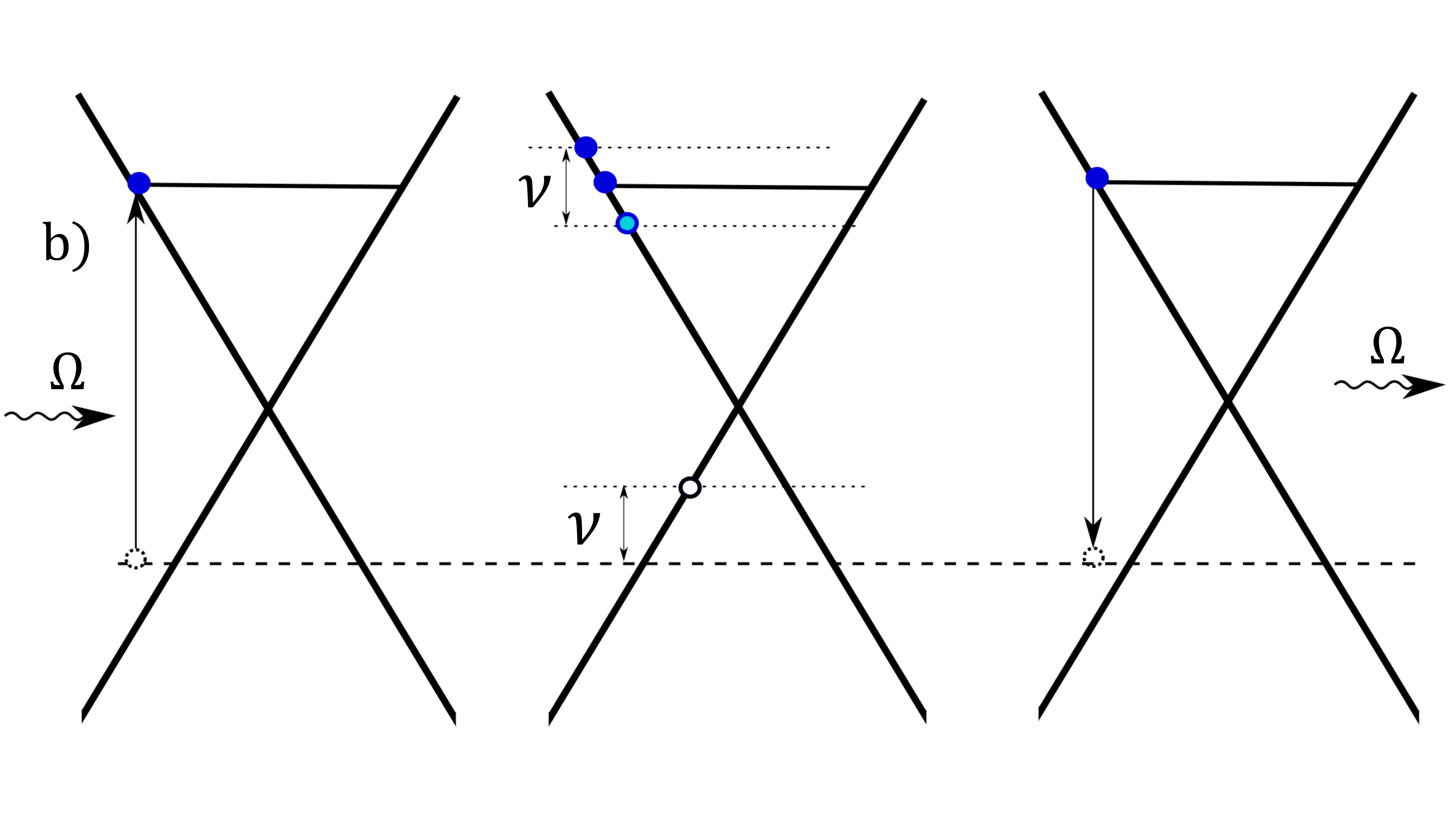}
            
        \caption{Examples of Auger-Meitner--like processes  corresponding to two different helicity sets: 
        $s_3=-1,s_4=+1$ (panel \emph{a})  and $s_3=+1,s_4=-1$
        (panel \emph{b}).
                        The state on the horizontal dashed line is a virtual (off-shell) one.
            \label{fig:AugerProcessCartoon}}
   \end{figure}

The threshold $\Omega=\omega_{\text{I}}$ demarcates the onset of AM-like processes, first introduced in the context of doped semi-conductors in Ref.~\cite{gavoret_optical_1969} and further studied in Refs.~\cite{ruckenstein_many-body_1987,Sham:1990, Hawrylak:1991,Pimenov:2017}. 
Figure~\ref{fig:AugerProcessCartoon} 
depicts two kinds of AM processes that occur for $s_3=-s_4=-1$ (panel \emph{a}) and $s_3=-s_4=+1$ (panel \emph{b}). In Fig.~\ref{fig:AugerProcessCartoon}\emph{a}, 
an incoming photon of energy $\Omega>\omega_\text{I}$ 
creates a hole state and a virtual state at the same momentum. The virtual state decays into an electron and a particle-hole pair, formed by two electron states with energy $\nu$. The particle-hole pair and the  electron then 
decay into another virtual state, which annihilates the hole,
and the photon is emitted  back. 
In Fig.~\ref{fig:AugerProcessCartoon}\emph{b},
an incoming photon 
creates an 
electron and 
a virtual state.
The virtual state decays into a real electron and a particle-hole pair, formed by the electron in the conduction band and hole in the valence band. Finally the virtual state annihilates the electron, and the photon is emitted back.

\subsubsection{High frequencies: $\Omega>\od$}
As we said before, absorption for $\Omega>\od$ occurs even in the absence of electron-electron interaction. The corresponding optical conductivity is plateaued at the universal value in 2D and increases linearly with frequency in 3D. Electron-electron interaction gives rise to logarithmic renormalizations of the velocity and coupling constant~\cite{neto:2009,kotov:2012,Hosur:2013}, which occur already to first order in the static interaction.
            Dissipative processes, considered in this paper, contribute only to second  order in the interaction (cf. Fig.~\ref{fig:AllDiagrams}) and thus can be neglected in this frequency range.
     
As the final remark for this section, we note that,
in addition to being independent of the diagram type,
 the 
  frequency thresholds are also independent of a particular form of the dispersion and dimensionality.
\bse
\begin{table*}
    \caption{Summary of the analytic results for the optical conductivity of 2D and 3D 
        Dirac metals, $\Re\sigma(\Omega)$, with Hubbard interaction.
        Here, $\eta_d=\Re\sigma(\Omega)/\sigma_{0d}\alpha_{\text{H}}^2N^2$, 
    $\sigma_{0d}=e^2\kf^{d-2}/\hbar$, 
    $\kf$ is the Fermi momentum,
        $d=2,3$ is the spatial dimensionality,
    $\alpha_{\text{H}}$ is the dimensionless coupling constant of Hubbard interaction [Eq.~\eqref{eq:alphadef}],  and
        $N$ is number of flavors.
    The results are valid in two regions: at the lowest frequencies (first row), 
    $\Omega\ll \EF$,
        and just above $\oi=\EF$ (second row),     where AM process start to contribute, i.e., for $0\leq \delta\Omega\equiv \Omega-\oi
        \ll\EF$, where $\oi=\EF$ is the indirect threshold.
 Equation numbers after the formulas refer to their locations in the text.
 \label{table:ResultsSummary}}
    \begin{ruledtabular}
    \begin{tabular}{ccc}
Frequency range  & $\eta_3$
& $\eta_2$
\\
\hline
$ \Omega\ll E_\fer$ &  
          $\frac{4}{175\pi}
    \left(\frac{\Omega}{\EF}\right)^2$ 
    ;~\eqref{3D_sigma_Hubbard}
        & $
  \left(\frac{1} {80\pi^2}\ln\frac{E_\text{F}}{\Omega}+
   \frac{5\ln 2+4}{200\pi^2}
   \right)
  \left(\frac{\Omega}{\EF}\right)^2$
  ;~\eqref{2D_sigma_Hubbard}
    \\
    \hline 
 $0<\delta\Omega\ll E_\fer$
        &  $
    \frac{71}{3240\pi}\left(\frac{\delta\Omega}{
        E_\fer}\right)^{5}$
    ;~\eqref{3D:IFT}
        & $\frac{5}{108\sqrt{3}\pi}
    \left(\frac{\delta\Omega}{
        E_\fer}\right)^{4}$
    ;~\eqref{2D:IFT}
        \end{tabular}
    \end{ruledtabular}
\end{table*}
\begin{table*}
    \caption{Summary of the analytic results for the optical conductivity of 
        2D and 3D Dirac metals, 
        $\Re\sigma^\text{C}
    (\Omega)$, with Coulomb interaction.
        Here, 
$\eta_d^{\text{C}}=\Re\sigma^\text{C}(\Omega)/\sigma_{0d}$,
$\alpha_{\text{C}}$ 
is the dimensionless coupling constant of the Coulomb interaction [Eq.~\eqref{tkappa}], and $\omega_{\text{p}d}$ is the effective plasma frequency in $d$ dimensions, defined in Eq.~\eqref{oplasma}.  The rest of notations are the same as Table \ref{table:ResultsSummary}. Equation numbers after the formulas refer to their location in the text. 
  \label{table:ResultsSummaryCoulomb}}
    \begin{ruledtabular}
    \begin{tabular}{ccc}
Frequency range  & $\eta_3^\text{C}$
& $\eta_2^\text{C}
$
 \\
\hline
$ \Omega\ll 
\omega_{\text{p}d}
\ll E_\fer$ &  
    $    \frac{1}{480}\alpha_{\text{C}}
        \left(\frac{\Omega}{\EF}\right)^2$ 
    ;~\eqref{Cee1}
        & $
\left( 
    \frac{1} {80\pi^2}\ln\frac{\vd\kappa_{2}}{\Omega}+
    \frac{5\ln 2-2}{16\pi^2}
  \right)\left(\frac{\Omega}{\ef}\right)^2$
  ;~\eqref{Cee2D} 
         \\
\hline
$ 
\omega_{\text{p}d}
\ll\Omega\ll E_\fer$ &  
    $
    \frac{(3-4\ln 2)}{24\pi}\alpha_{\text{C}}^4
\left(\frac{\EF}{\Omega}\right)$
;~\eqref{Cee2}
    & $\frac{5}{576\pi^2}\frac{e^2}{\vd}$
    ;~\eqref{Cee2_2D} 
       \\
    \hline
      $0<\delta\Omega\ll E_\fer$
        &  $
        \frac{71}{3240\pi}\alpha_{\text{C}}^4
        \left(\frac{\delta\Omega}{
        E_\fer}\right)^{
        5}$
    ;~\eqref{3D:IFTC}
        & $
        \frac{5}{108\sqrt{3}\pi}\alpha_{\text{C}}^2
        \left(\frac{\delta\Omega}{
        E_\fer}\right)^
        {4}$
    ;~\eqref{2D:IFTC}
        \end{tabular}
    \end{ruledtabular}
\end{table*}
\ese

\subsection{
        Archetypal contributions to the optical conductivity 
    \label{sec:archetype-conductivity}}

We now analyze the structure of $\mathcal{R}_{\mathcal{S}_A}^{J_u}$ [Eq.~\eqref{eq:Im-curr-curr-total}], using one of the  self-energy diagram, namely, $\text{SE}_1$ in Fig.~\ref{fig:AllDiagrams},
as an example. 
As follows from Eq.~\eqref{eq:Im-curr-curr-total}, 
the contribution of this diagram
can be written as
\bwt
\bea
\sum_{\mathcal{S}_A
}\mathcal{R}_{\mathcal{S}_A}^{\text{SE}_1}(\Omega)
&=&-\frac{\pi^2}{32}\int_{\mathbf{k},\bp,\bq}
\theta(-\xi^{s_3}_\mathbf{k})
\theta(\Omega+\xi^{s_3}_\mathbf{k})
\sum_{\mathcal{S}_A}
\frac{
\mathcal{T}^{\text{SE}_1}_{\mathcal{S}_A}(\bk,\bp,\bq)}{(\Omega-\xi^{s_1}_{\mathbf{k}}+\xi^{s_3}_{\mathbf{k}})(\Omega-\xi^{s_2}_\mathbf{{k}}+\xi^{s_3}_{\mathbf{k}})}
\mathcal{U}
(\bk,\bp,\bq,\Omega+\xi^{s_3}_{\mathbf{k}}),
\label{sigmaSE}
\eea
where 
we used the third line of Eq.~\eqref{eq:firsteqofresultlist} 
for $\mathcal{G}^{\text{SE}_1}_{\mathcal{S}_A}$,
$\mathcal{T}^{\text{SE}_1}_{\mathcal{S}_A}$ is given by first line of the same equation, 
and
\be
\mathcal{U}&(\mathbf{k},\bp,\bq,\omega)=
\int_\nu V^2_{\text{st}}(\mathbf{q})
\theta(\xi^{+}_\mathbf{k+q})\theta(-\xi^{s_4}_\mathbf{p})\theta(\xi^{+}_\mathbf{p+q})
\delta(\omega+\nu-\xi^{+}_\mathbf{k+q})\delta(\nu+\xi^{+}_\mathbf{p+q}-\xi^{s_4}_\mathbf{p})\label{Sbar}.
\ee
\ewt
The structure of the expressions above can be understood by comparing them to their counterparts for the scalar case, when the trace part in Eq.~\eqref{sigmaSE}
is equal to unity. 
For our choice of $\Omega>0$, the theta-functions in Eq.~\eqref{sigmaSE} come from the difference of the Fermi functions in the current-current bubble of the SE$_1$ diagram,
where the imaginary part of the Green's function at the bottom of SE$_1$ diagram was replaced by the $\delta$-function, and the ensuing constraint on the frequency ($\omega=\xi^{s_3}_\bk$) was resolved.
Next, with the trace part replaced by unity, the integral of 
$\mathcal{U}(\bk,\bp,\bq,\Omega+\xi^{s_3}_{\mathbf{k}})$  in Eq.~\eqref{Sbar} over $\bp$ and $\bq$ gives the imaginary part of the self-energy at momentum $\bk$ and frequency $\omega=\Omega+\xi^{s_3}_{\mathbf{k}}$.
The denominator of the integrand in Eq.~\eqref{sigmaSE} comes from the product of the real parts of two Green's functions adjacent to the self-energy block.

As noted earlier, the theta- and delta-function constraints are the same for all diagrams with the only difference being the scalar factors $K^{J_u}$, the trace factors $\mathcal{T}^{J_u}_\mathcal{S}$ and the products $\mathcal{G}^{J_u}_\mathcal{S}$. Thus, similarly, the contribution from V$_1$ diagram is given by 
\bwt
\bea
\sum_{\mathcal{S}_A}\mathcal{R}_{\mathcal{S}_A}^{\text{V}_1}(\Omega)
&=&\frac{\pi^2}{32}\int_{\mathbf{k},\bp,\bq}
\theta(-\xi^{s_3}_\mathbf{k})
\theta(\Omega+\xi^{s_3}_\mathbf{k})
\sum_{\mathcal{S}_A}
\frac{
\mathcal{T}^{\text{V}_1}_{\mathcal{S}_A}(\bk,\bp,\bq)}{(\Omega-\xi^{s_1}_\mathbf{k}+\xi^{s_3}_\mathbf{k})(\Omega-\xi^{+}_\mathbf{k+q}+\xi^{s_2}_\mathbf{k+q})}
\mathcal{U}(\bk,\bp,\bq,\Omega+\xi^{s_3}_{\mathbf{k}}),
\label{eq:V1}
\eea
\ewt
and so on for other values of $J_u$.

For the reader's convenience, the analytic results 
for the optical conductivity
are summarized in Tables~\ref{table:ResultsSummary} and~\ref{table:ResultsSummaryCoulomb} for Hubbard and Coulomb interactions, respectively.

\section{Optical conductivity of a 3D Dirac metal}
\label{sec:3D}
In this section, we derive the analytic results for  the optical conductivity of a 3D Dirac metal.
\subsection{Lowest frequencies: \texorpdfstring{$\Omega\ll E_\text{F}$}{Omega<<EF}} 
\label{sec:lowOmega}

This is the case with $s_3=s_4=+1$ (cf. Sec.~\ref{sec:LFregimedefinition}). 
With $s_3$ and $s_4$ being fixed, the only free helicities remaining are $s_1$ and $s_2$.
The case $s_1=s_2=+1$ corresponds to a purely intra-band absorption, with all states being in the conduction band.
The cases of $s_1=-s_2=\pm 1$ and $s_1=s_2=-1$ correspond to absorption due to scattering processes which involve up to two holes.

\subsubsection
{Intra-band absorption due to electron-electron interaction \label{sec:ee3D}}
We start with purely intra-band absorption due to electron-electron
(\emph{ee})
interaction, when all the helicities are positive:
$s_i=+1$, $i=1\dots 6$.
Because the hole states in this case are totally passive, one can view the system as a FL, which is isotropic yet not Galilean-invariant due to a non-parabolicity of the electron spectrum, i.e., as a DFL. The absorption probability in this case is severely restricted by momentum conservation. 
In Refs.~\cite{rosch:2005,rosch:2006,maslov_optical_2016,Sharma:2021} it was shown
  that, for the single-band case,  momentum conservation brings in a factor of
the ``velocity imbalance'', $(\Delta\bv)^2$,
 to the integrand of the expression of the conductivity. Here, $\Delta \bv$ is the difference between the velocities of the initial and final states of an \emph{ee} scattering process.
The same factor appears in our case as well. To see this, we first note that in the \emph{ee} case  the denominators of the fractions in Eqs.~\eqref{sigmaSE} and \eqref{eq:V1} are reduced to a factor of $\Omega^2$ (and the same is true for other contributions). 
Next, as shown in Appendix~\ref{appen:CombingLFeediagrams},
the sum of the trace parts 
of all diagrams 
in Fig.~\ref{fig:AllDiagrams}
is given by 
\bea
\mathcal{T}_{\mathcal{S}_+}&=&\sum_{J_u} K^{J_u}\mathcal{T}^{J_u}_{\mathcal{S}_+}(\bk,\bp,\bq)\nn\\
&=&-\frac{\pi^2}{
64
}
(\Delta\bv)^2
\left|\Phi^{+,+}_\mathbf{p,p+q}\right|^2\left|\Phi^{+,+}_\mathbf{k,k+q}\right|^2,\nn\\
\label{sumTmain}
\eea
where 
$K^{J_u}$ is defined in Eq.~\eqref{KJu}, $\mathcal{S}_+$ denotes the set
$\{s_1=+1,s_2=+1\dots s_6=+1\}$, $J_{1,2}\in\{\text{SE}_{1,2},\text{V}_{1,2},\text{PAL}_{1,2},\text{CAL}_{1,2}\}$,
\bea
\Delta\bv=
\mathbf{v}^{+,+}_{\mathbf{k}}+\mathbf{v}^{+,+}_{\mathbf{p+q}}-\mathbf{v}^{+,+}_\mathbf{k+q}-\mathbf{v}^{+,+}_{\mathbf{p}},
\label{Deltav}
\eea
$\bv^{+,+}_\bk$ is the matrix element of the velocity operator between electron-like states, given by Eq.~\eqref{vintra}, 
and $\Phi^{+,+}_{\bk,\bk'}=\braket{\bk,+\vert\bk',+}$ is the matrix element of two electron-like states.\footnote{
For $\bk\to \bk'$, $\Phi^{+,+}_{\bk,\bk'}\to 1$, and Eq.~\eqref{sumTmain} is reduced to the result of  Ref.~\cite{Sharma:2021}, which considered a Dirac metal with long-range Coulomb interaction.}
$\Delta\bv$ in Eq.~\eqref{Deltav} is the change in the total velocity (proportional to the current) due a to collision between two electrons 
with initial momenta 
$\bk$ and $\bpq$, 
and final momenta 
$\bkq$ and $\bp$,
respectively. In a Galilean-invariant system, momentum-conserving electron-electron scattering does not lead to current relaxation and thus does not affect the conductivity. Indeed, we see that $\Delta\bv=0$ if $\bv^{+,+}_{\bf k}={\bf k}/m$ with $m$ being the electron mass. A Dirac metal has finite conductivity only inasmuch as  it violates Galilean invariance. Furthermore, even if the system is not Galilean-invariant but isotropic, $\Delta\bv$ vanishes if all the momenta in Eq.~\eqref{Deltav} are projected onto the Fermi surface and, to get a finite conductivity, one needs to expand $\Delta\bv$ near the Fermi surface. For $\Omega\ll E_\text{F}$, a typical deviation of the quasiparticle energy from the Fermi energy is on the order of $\Omega$. Then $(\Delta \bv)^2$ can be estimated as
\bea
(\Delta \bv)^2\sim w^2\left(\frac{\Omega}{k_\text{F}}\right)^2,\label{Delta_v_est}
\eea
where the ``non-parabolicity coefficient''
\bea
w=1-\frac 12 \frac{\dee^2\epsilon_\bk}{\dee k^2}\frac{\dee(k^2)}{\dee\epsilon_\bk}\Big\vert_{k=k_\text{F}}\label{w}
\eea 
quantifies a deviation from Galilean invariance~\cite{Sharma:2021}. Introducing a gapped Dirac spectrum, $\epsilon_\bk=\sqrt{\vd^2k^2+\Delta^2}$, for a moment, 
we get
\be
w=1-\frac{\Delta^2}{(\Delta+\EF)^2}.
\ee
For $E_\text{F}\gg\Delta$,
  the Dirac spectrum is almost linear, and thus the deviation from the Galilean-invariant case is the strongest. In this case, $w=1-\Delta^2/\EF^2\approx 1$. 
  For $E_\text{F}\ll \Delta$,
  the gapped Dirac spectrum is almost parabolic and, correspondingly, $w$ is small: $w\approx 2E_\text{F}/\Delta\ll 1$. 

To obtain an order-of-magnitude  estimate for the conductivity due to \emph{ee} interaction,
one can replace 
the trace part 
of
Eq.~\eqref{sigmaSE} by $(\Delta \bv)^2
$, and use Eq.~\eqref{Delta_v_est} with $w=1$ for $(\Delta\bv)^2$ (gapless Dirac spectrum).
This yields the following estimate for the conductivity
\bwt
\bea
\Re\sigma_{\text{ee}}(\Omega)\sim\frac{e^2}{\Omega^3}
\int_{\mathbf{k}}\theta(-\xi^{+}_\mathbf{k})\theta(\Omega+\xi^{+}_\mathbf{k})
|\Im\Sigma_{\text{ee}}(\bk,\Omega+\xi^{+}_{\mathbf{k}})|(\Delta\bv)^2.
\label{sigmaSE2}
\eea
As discussed just below Eq.~\eqref{Sbar}, the theta-function constraints in the equation above come from the current-current bubble with the choice of $\Omega>0$. Furthermore,
\bea
\Im\Sigma_{\text{ee}}(\mathbf{k},\omega)
\sim- 
\int_\nu \int_{\bp,\bq}
V^2_{\text{st}}(\mathbf{q})
\theta(\xi^{+}_\mathbf{k+q})\theta(-\xi^{+}_\mathbf{p})\theta(\xi^{+}_\mathbf{p+q})
\delta(\omega+\nu-\xi^{+}_\mathbf{k+q})\delta(\nu+\xi^{+}_\mathbf{p+q}-\xi^{+}_\mathbf{p})\label{Sigma_full}
\eea
\ewt
is the imaginary part of the self-energy due to {\em ee} interaction.
As long as $\Omega\ll \EF$, typical electronic momenta are close to $\kf$, therefore, $\xi_\bk^+\sim \Omega$,
and the integral over $\bk$ gives a factor of $\mathcal{N}_{\fer,3}\Omega$.
Therefore,
\bea
\Re\sigma_{\text{ee}}(\Omega)\sim e^2 \mathcal{N}_{F,3}\frac{|\Im\Sigma_{\text{ee}}(\Omega)|}{\Omega^2}
\left(\frac{\Omega}{\kf}\right)^2
,\label{power_count}
\eea
where $\Sigma_{\text{ee}}(\Omega)\equiv \Sigma_{\text{ee}}(\kf,\Omega)$.

For the Hubbard case, 
the self-energy is of the usual FL form,
\bea
\Im\Sigma_{\text{ee}}(
\Omega)
\sim -(N\alpha_{\text{H}})^2 \frac{\Omega^2}{E_\text{F}},
\label{imsigma3D}
\eea
and thus
\bea
\Re\sigma_{\text{ee}}(\Omega)
\sim
e^2k_\text{F}(N\alpha_{\text{H}})^2\left(\frac{\Omega}{E_\text{F}}\right)^2.\nn\\
\label{sigma_ee_est}
\eea
A detailed  calculation
presented in Appendix~\ref{appen:asymp_eval} 
gives
\bea
\Re\sigma_{\text{ee}}(\Omega)=\frac{38}{4725\pi}\frac{e^2\kf}{\hbar
}(N\alpha_{\text{H}})^2 \left(\frac{\Omega}{\EF}\right)^2,\label{3D_sigma_ee}
\eea
 which agrees with the estimate \eqref{sigma_ee_est}.

The Coulomb case for 
$\Omega\ll\omega_{\text{p}3}\ll\EF$ is similar to the Hubbard one, in a sense that the self-energy is also of the canonical FL form, except for a different coupling constant:
\be
\Im\Sigma_{\text{ee}}(\Omega)\sim -\frac {\kappa_3}{\kf} \frac{\Omega^2}{E_\fer}.\label{ImSOm2}
\ee
Consequently, the conductivity is obtained by replacing $(N\alpha_{\text{H}})^2$ with $\kappa_3/\kf$ in Eq.~\eqref{sigma_ee_est},
\bea
\Re\sigma^{\text{C}}_{\text{ee}1}(\Omega)\sim e^2\kappa_3\left(\frac{\Omega}{E_\text{F}}\right)^2.\label{Cest1}
\eea
The actual calculation gives 
\bea
\Re\sigma^{\text{C}}_{\text{ee}1}(\Omega)=\frac{1}{480}\frac{e^2\kf}{\hbar}\alpha_{\text{C}} \left(\frac{\Omega}{\EF}\right)^2,\label{Cee1}
\eea
which agrees with the estimate in Eq.~\eqref{Cest1}.\footnote{
We are using this opportunity to point out that the numerical coefficient in the result for the same quantity in Ref.~\cite{Sharma:2021} by a subset of current authors (PS and DLM)  is incorrect.
}

In the range of frequencies $\omega_{\text{p}3}\ll\Omega\ll\EF$, electrons interact with their own  plasmon modes. In this regime, we can replace the screened Coulomb potential with the bare one, as specified in Eq.~\eqref{Cdyn}.  Recalling also that the imaginary part of the retarded polarization bubble behaves as  $\Im\pi_{0,\text{R}}(\bq,\omega)\sim \mathcal{N}_{\fer,3}\omega/\vf q$ for $|\omega|/\vf\leq q\ll \kf$, we obtain the following estimate for the imaginary part of the self-energy 
\bea
\Im\Sigma_{\text{ee}}(\Omega)&\sim&-
\frac{\kappa^4_3}{\mathcal{N}_{\fer,3}\vf^2}\int^\Omega_0 \dee\nu \nu \int^{\infty}_{\max\{\Omega,\Omega-\nu\}/\vf} \frac{\dee q}{q^4}\nn\\
&\sim&-\frac{\kappa_3^2}{\kf^2}\frac{\omega^2_{\text{p}3}}{\Omega}.\label{ImS1Om}
\eea
A crossover between Eqs.~\eqref{ImSOm2} and \eqref{ImS1Om} occurs at $\Omega\sim \omega_{\text{p}3}$, as it should.
Substituting Eq.~\eqref{ImS1Om} into \eqref{power_count},
we obtain
\bea
\Re\sigma^{\text{C}}_{\text{ee}2}(\Omega)\sim e^2\frac{\kappa^4_3}{\kf^3} 
\frac{\EF}{\Omega}
\sim \frac{e^6\kf^2}{\hbar \vd\Omega},\label{plasmon_est}
\eea
and
the actual calculation gives 
\be
\Re\sigma^{\text{C}}_{\text{ee}2}(\Omega)=
\frac{(3-4\ln 2)}{24\pi}\frac{e^2\kf}{\hbar}\alpha_{\text{C}}^4
\left(\frac{\EF}{\Omega}\right),
\label{Cee2}
\ee
which matches the estimate \eqref{plasmon_est}.
Equations ~\eqref{Cee1} and \eqref{Cee2}
imply that the conductivity exhibits a maximum at $\Omega\sim\omega_{\text{p}3}$.

\begin{figure}
    \includegraphics[width=\linewidth
                ,left]
        {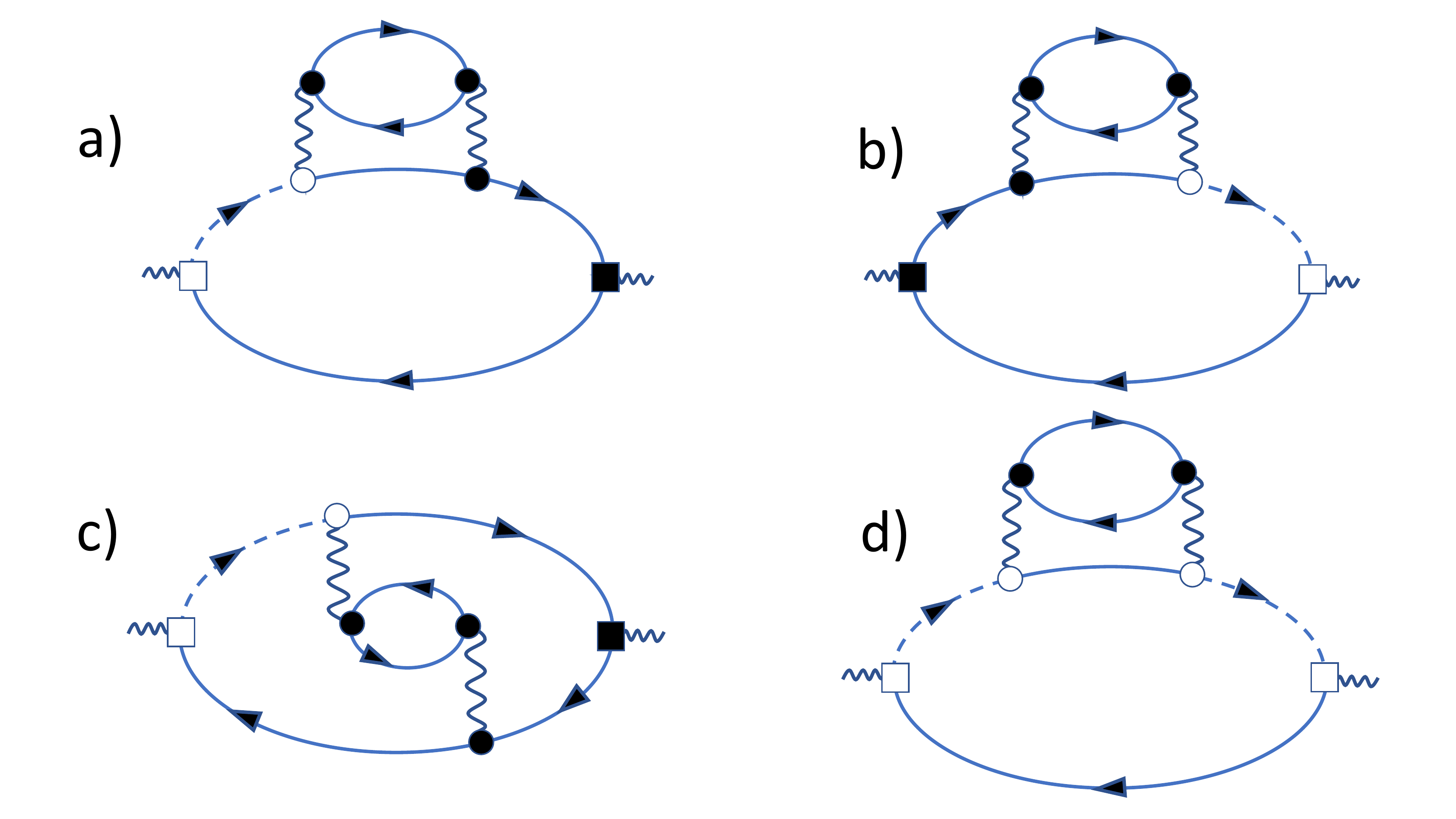}

        \caption{
            Examples of single-hole (\emph{a}-\emph{c}) and two-hole (\emph{d}) diagrams. Solid (dashed) lines depict the Green's functions in the diagonal basis [Eq.~\eqref{eq:3DScalarGreensFunction}] for positive (negative) helicities. The filled and blank circles denote matrix elements of $ee$ and $eh$ interactions, respectively. The filled and blank squares denote the intra- and inter-band current vertices, respectively.
        \label{fig:1h2hdiags}}
\end{figure}

\subsubsection{ Absorption processes involving up to two holes
\label{sec:eh3D}}
We now turn to absorption processes that involve holes. There are two types of such processes:  with one hole ({\it eh1}) 
and with two holes ({\it eh2}).  Recalling that $s_3=s_4=+1$  in the 
all-frequencies regime (cf. Sec.~\ref{sec:LFregime}), we have two choices: 
$s_1=-s_2=\pm 1$, which corresponds to {\it eh1}, and  $s_1=s_2=-1$, which corresponds to {\it eh2}. Examples of \emph{eh1} and \emph{eh2} 
diagrams are shown in Fig.~\ref{fig:1h2hdiags}, where the solid and dashed lines depict the electron and hole Green's functions, respectively, given by Eq.~\eqref{eq:3DScalarGreensFunction} with $s=\pm 1$.

We first look at the {\it eh1} case, when the sum over helicities in the second line of Eq.~\eqref{eq:V1}
contains two terms: one with $s_1=+1,s_2=-1$ and another one with $s_1=-1,s_2=+1$.
In \emph{eh1} diagrams (Fig.~\ref{fig:1h2hdiags}\emph{a}-\emph{c}), 
one of the current vertices is of the intra-band type while another one is of the inter-band type.  In self-energy diagrams \emph{a} and \emph{b}, the current vertices enter at the same momenta and are thus orthogonal to each other, see Eqs.~\eqref{vintra} and~\eqref{vinter3D}.
Therefore, the \emph{eh1} self-energy diagrams vanish. On the other hand, vertex-type diagrams, e.g., diagram \emph{c} in Fig.~\ref{fig:1h2hdiags}, contain current vertices at different momenta, which are not orthogonal to each other, and thus the vertex contribution is finite. In what follows, we will analyze the \emph{eh1} vertex diagrams, whose general algebraic structure is given by Eq.~\eqref{eq:V1}.

As soon as a scattering process involves at least one hole, constraints due to momentum conservation are lifted, and the factor of $(\Delta\bv)^2$ does not bring an additional smallness to the result. However, in contrast to the \emph{ee} case, typical energies involved are now on the order of $\EF$ rather than $\Omega$, and the \emph{eh1} contribution to the conductivity still scales as $\Omega^2$.
Indeed, the sum over $s_1=-s_2=\pm 1$ in Eq.~\eqref{eq:V1} 
gives a factor of 
$1/(\Omega-2\epsilon_{\bk+\bq})(\Omega+2\epsilon_\bk)$ which, for $\Omega\ll\EF$ and $\epsilon_{\bk},\epsilon_{\bk+\bq}\approx\EF$, is of order $1/\EF^2$, 
as opposed to $1/\Omega^2$ for the \emph{ee} case, cf. Eq.~\eqref{power_count}. 
Next, the intra- and inter-band matrix elements of the velocity 
can be estimated as $\vd$.
Finally, a joint between the dashed and solid lines brings in an inter-band matrix element, $\langle\bar\bk,+|\bar\bk',-\rangle$, where $\bar\bk$ and $\bar\bk'$ are the typical electron momenta. 
With all of the above taken into account, the \emph{eh1} contribution to the conductivity can be estimated as
\bea
\Re\sigma_{\text{eh1}}\sim e^2 \mathcal{N}_{\fer,3}
\vd^2
\left\vert\langle\bar\bk,+|\bar\bk',-\rangle\right\vert 
\frac{\left|\Im\Sigma_{\text{ee}}(\Omega)\right|}{\EF^2}.\label{eh1}
\eea

For Hubbard interaction, $\Im\Sigma_{\text{ee}}(\Omega)$ is given by Eq.~\eqref{imsigma3D}, while $|\bar\bk-\bar\bk'|\sim \kf$ and this
$\left\vert\langle\bar\bk,+|\bar\bk',-\rangle\right\vert\sim 1$. Then
\bea
\Re\sigma_{\text{eh1}}(\Omega)
\sim
e^2k_\text{F}(N\alpha_{\text{H}})^2\left(\frac{\Omega}{\EF}\right)^2,
\label{sigma_eh1_est}
\eea
which is of the same order as the \emph{ee} contribution, Eq.~\eqref{sigma_ee_est}. The actual calculation of the \emph{eh1} contribution gives
\bea
\Re\sigma_{\text{eh1}}(\Omega)=\frac{4}{945\pi}\frac{e^2
\kf}{\hbar 
} (N\alpha_{\text{H}})^2\left(\frac{\Omega}{\EF}\right)^2,\label{3D_sigma_eh1}
\eea
which matches the estimate Eq.~\eqref{sigma_eh1_est}.

The two-hole case is similar to the single-hole one,
except for 
now there are two matrix elements between electron and hole states, see Fig.~\ref{fig:1h2hdiags}\emph{c}.
Therefore, the \emph{eh2} contribution to the conductivity can be estimated
as
\bea
\Re\sigma_{\text{eh2}}(\Omega)\sim e^2 \mathcal{N}_{\fer,3} 
\vd^2
\left\vert\langle\bar\bk,+|\bar\bk',-\rangle\right\vert^2 \frac{\Im\Sigma_{\text{ee}}(\Omega)}{\EF^2}.\nn\\ 
\eea
For Hubbard interaction, the matrix element is on the order of unity, and
\bea
\Re\sigma_{\text{eh2}}(\Omega)\sim \Re\sigma_{\text{eh1}}(\Omega) \sim\Re\sigma_{\text{ee}}(\Omega),
\label{sigma_eh2_est}
\eea
whereas  the actual calculation gives 
\bea
\Re\sigma_{\text{eh2}}(\Omega)=\frac{2}{189\pi}\frac{e^2 \kf}{\hbar} (N\alpha_{\text{H}})^2\left(\frac{\Omega}{\EF}\right)^2.\label{3D_sigma_eh2}
\eea
The final result for the conductivity due to Hubbard interaction is the sum of the \emph{ee}, \emph{eh1}, and \emph{eh2} contributions, given by Eqs.~\eqref{3D_sigma_ee}, \eqref{3D_sigma_eh1}, and \eqref{3D_sigma_eh2}:
\bea
\Re\sigma(\Omega)=\frac{4}{175\pi}\frac{e^2 \kf}{\hbar} (N\alpha_{\text{H}})^2\left(\frac{\Omega}{\EF}\right)^2.\label{3D_sigma_Hubbard}
\eea
Note that Eq.~\eqref{3D_sigma_Hubbard} is valid for a gapless Dirac spectrum. For an almost parabolic spectrum, e.g., a gapped Dirac spectrum in the limit of $\EF\ll\Delta$, the \emph{ee} contribution is suppressed due to a small value of the non-parabolicity coefficient $w$ [cf. Eq.~\eqref{w}]. 
The \emph{eh1} and \emph{eh2} contributions are also suppressed because the eigenstates of the Hamiltonians \eqref{eq:3DKineticPartHamiltonian} and \eqref{eq:2DKineticPartHamiltonian} are either electron-like or hole-like and, therefore, the matrix elements $\langle \bk,+|\bk',-\rangle$ are small. In addition, there is also a partial cancellation between the diagrams in this case \cite{AGDM}. As a result, the total conductivity for an almost parabolic Dirac spectrum acquires a small factor of $(\EF/\Delta)^2\ll 1$. This is why these contributions were neglected in Refs.~ \cite{gavoret_optical_1969,ruckenstein_many-body_1987,Sham:1990,Hawrylak:1991,Pimenov:2017}.

For Coulomb interaction, $|\bar\bk-\bar\bk'|\sim \kappa_{3}\ll \kf$ and, therefore, the  matrix element between almost orthogonal electron and hole states is small: $\left\vert\langle\bar\bk,+|\bar\bk',-\rangle\right\vert\sim \kappa_3/\kf\ll 1$. Therefore, the \emph{eh1} and \emph{eh2} contributions to the conductivity are smaller than the \emph{ee} one in Eq.~\eqref{Cest1} by a factor of $\kappa_3/\kf$ and $(\kappa_3/\kf)^2$, respectively.
Thus, these contributions can be neglected, and the leading contribution to the conductivity for the Coulomb case is still given by Eqs.~\eqref{Cee1} and \eqref{Cee2}.

\begin{figure}
    \includegraphics[width=\linewidth
                ,left]
        {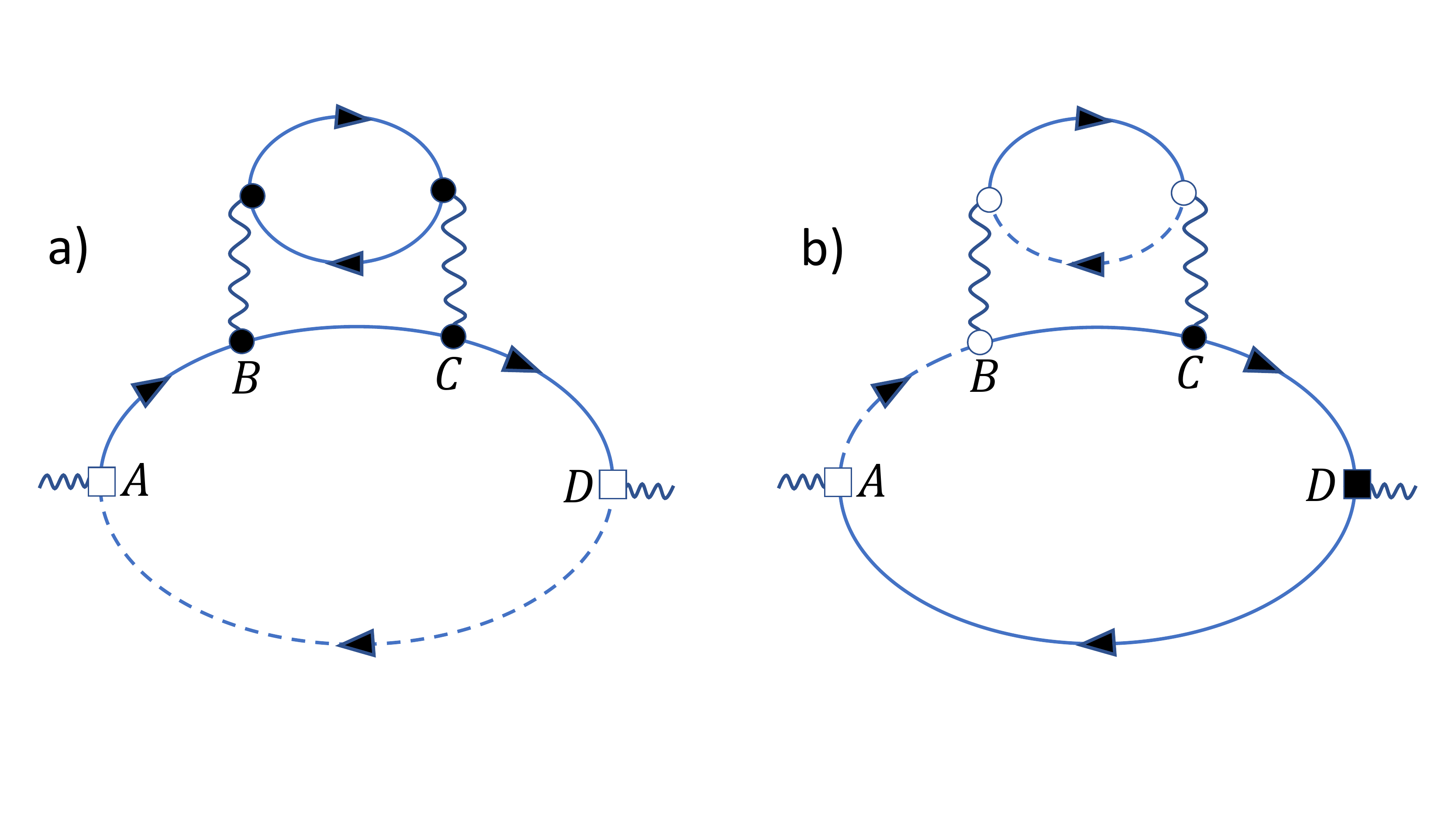}

        \caption[Examples of AM-like process]{Examples of diagrams describing Auger-Meitner scattering processes. Diagram \emph{a} corresponds to the case of $s_3=-1,s_4=+1$, when absorption occurs as depicted in Fig.~\ref{fig:AugerProcessCartoon}\emph{a}.
                Diagram \emph{b} corresponds to the case of $s_3=+1,s_4=-1$, when absorption occurs as depicted in  Fig.~\ref{fig:AugerProcessCartoon}\emph{b}. 
        The filled and blank circles denote matrix elements of $ee$ and $eh$ interactions, respectively, while the filled and blank squares denote the intra- and inter-band current vertices, respectively. The lines connecting vertices A and B, and C and D can, in general, be of either helicity; the diagrams shown in the figure correspond to specific choices of those helicities.
                \label{fig:Augerdiags}}
\end{figure}
\subsection{
Intermediate frequencies: $\oi\leq\Omega\leq\od$
 \label{sec:IF3D}}
In the intermediate frequency regime, there are eight possible terms contributing for each type of diagrams. These terms are specified by the helicities $s_3=-s_4$ and four possibilities of $s_1,s_2=\pm 1$ therein. 
As shown in Sec.~\ref{sec:IFregime}), these terms  start to contribute only for $\Omega$ above the indirect threshold, $\omega_{\text{I}}=\EF
$, which is below the direct (Pauli) threshold $\omega_\text{D}=2\EF$. 
Previous work by Gavoret et al.~\cite{gavoret_optical_1969} and others after them~\cite{ruckenstein_many-body_1987,Pimenov:2017} has studied only the  
diagrams allowed by the Hamiltonian \eqref{prev}.
For the Dirac spectrum, all diagrams are allowed and we analyze the leading-order ones, either within the large-$N$ or RPA approximations.
\subsubsection{Threshold behavior for $\Omega\gtrapprox\oi$}
\label{sec:3Dthreshold}
Analytic results in the intermediate frequency regime can be obtained only for frequencies just above $\oi$; for the rest of this regime, we will have to defer to numerical computation, discussed in Sec.~\ref{sec:num3D}.

To simplify analysis, we note that  the contributions for the $s_3=+1,s_4=-1$ case can be mapped onto the $s_3=-1,s_4=+1$ one just by relabelling the helicities.
    Thus, we need to consider only the $s_4=+1,s_3=-1$ case.  
    The sum 
over $\mathcal{S}_B=\{s_1,s_2\}=\{\pm 1,\pm 1\}$ in Eq.~\eqref{sigmaSE}
contains terms of three types: $t_1\sim 1/\Omega^2$, $t_2\sim 1/\Omega\min\{\Omega,\ve_\bk\}$, and  $t_3\sim 1/\min\{\Omega^2,\ve^2_\bk\}$. Near the threshold, $\Omega\gtrapprox \oi=\EF
$ while $\ve_\bk\approx \EF
$. Therefore, $t_1\sim t_2\sim t_3\sim 1/\EF^2$.
Next, the 
current vertex is $\sim\vd$,
and the conductivity is estimated as
    \bea
\Re\sigma_{\text{IF}}(\Omega)\sim\frac{e^2}{\oi}
\frac{v_\text{D}^2}{E_\text{F}^2}\int_{\mathbf{k}}
\theta(\oi+\delta\Omega+\xi^{-}_\mathbf{k})\nn\\
\times|\Im\Sigma_{\text{ee
}}(\bk,\oi+\delta\Omega+\xi^{-}_{\mathbf{k}})|,
\label{sigmaSE2_Auger}
\eea
where $\delta\Omega\equiv \Omega-\oi\ll \EF$ and $\Im\Sigma_{\text{ee}}(\bk,\omega)$ is given by Eq.~\eqref{Sigma_full}.  
The theta function imposes a constraint $\oi+\delta\Omega+\xi_\bk^->0$ or $
\epsilon_\bk< 
\delta\Omega$.
Therefore, the integral over $k$ in Eq.~\eqref{sigmaSE2_Auger} is 
confined
to a narrow region near the Dirac point
\bea 
k\lesssim k_0\equiv \frac{
\delta\Omega
}{\vd}\ll\kf.
\label{krange}
\eea
Under this condition, $\Im\Sigma_{\text{ee}}$ for Hubbard interaction is still of the FL form, but with $\Omega$ replaced by $\delta\Omega$, i.e., $\Im\Sigma_{\text{ee}}\sim(\delta\Omega)^2$.
For Hubbard interaction, the self-energy is given by
Eq.~\eqref{sigma_ee_est} with $\Omega$ replaced by $
\delta\Omega$. Collecting all the estimates together, we obtain
\bea
\Re\sigma_{\text{IF}}(\Omega)\sim e^2\kf 
(N\alpha_{\text{H}})^2
\left(\frac{\delta\Omega}{\EF}\right)^{5},\nn\\
\label{sigma_Auger_gapped_est0}
\eea
while the actual calculation gives
\be
\Re\sigma^{\text{IF}}
(\Omega)=\frac{71}{3240\pi}\frac{e^2k_\text{F}}{\hbar}(N\alpha_{\text{H}})^2\theta(\delta\Omega)\left(\frac{\delta\Omega}{E_\text{F}}\right)^5.\label{3D:IFT}
\ee
For Coulomb interaction, we have
\be
\Re\sigma^{\text{IF,C}}
(\Omega)
=\frac{71}{3240\pi} \frac{e^2k_\text{F}}{\hbar}\alpha_{\text{C}}^4\theta(\delta\Omega)\left(\frac{\delta\Omega}{E_\text{F}}\right)^5\label{3D:IFTC}
\ee
for $
\delta\Omega\ll\opthree\ll\EF$. 
The results for the Hubbard and Coulomb cases are identical, up to a different coupling constant, because, close to the 
indirect
threshold, the Coulomb interaction is effectively a constant, equal to $4\pi e^2/\kf^2$.

In fact, the results in Eqs.~\eqref{3D:IFT} and \eqref{3D:IFTC}  can be readily generalized for an arbitrary dimensionality and spectrum. Indeed, the dependence on $\delta\Omega$ comes from the $(\delta\Omega)^2$-scaling of the self-energy, which does not depend on dimensionality (as long as $d\geq 2$), 
and the factor of $k_0^d$, whose dependence on $\delta\Omega$ is determined both by the dimensionality and  the energy spectrum. In particular, for $\epsilon_\bk\propto k^a$, we obtain 
\be\beta_{\text{A}}=d/a+2.\label{betaA}
\ee For $d=3$ and $a=1$ this gives $\beta_{\text{A}}=5$, in agreement with Eq.~\eqref{3D:IFT}, while for $d=3$ and $a=2$ we obtain $\beta_{\text{A}}=7/2$, in agreement with Ref.~\cite{gavoret_optical_1969}.

Note that the threshold singularities occur in the presence of slowly varying contributions from the \emph{ee}, \emph{eh1}, and \emph{eh2} processes, which were discussed in Sec.~\ref{sec:lowOmega}. Certainly, the asymptotic forms of these contributions, Eqs.~\eqref{3D_sigma_Hubbard} and \eqref{Cee2},  
are no longer valid for $\Omega\sim \oi=\EF$. However, if we naively extrapolate these expressions  to the region of $\Omega\gtrapprox \oi$, we would find that the threshold singularities are completely masked by slowly varying contributions, unless, of course, one differentiates the total conductivity with respect to $\Omega$ an appropriate number of times.\footnote{We need to use Eq.~\eqref{Cee2} for the Coulomb case because we are in the interval $\opthree\ll \EF\lessapprox \Omega$.} This result is confirmed by numerical calculations presented in  Secs.~\ref{sec:num3D} and \ref{sec:num2D}. 
Only if the spectrum is gapped and almost parabolic, i.e., $\EF\ll\Delta$, can the threshold singularities be detected against the background of other contributions [see the discussion after Eq.~\eqref{3D_sigma_Hubbard}].

\subsubsection{Generic frequencies in the interval $\oi\lesssim\Omega\lesssim\od$}
\label{sec:3Dgeneric}
For a generic frequency above $\oi=\EF$ but below $\od=2\EF$ and away from both thresholds, we can obtain only an estimate for the conductivity, by replacing $\delta\Omega$ in Eqs.~\eqref{3D:IFT} and~\eqref{3D:IFTC} with $\EF$. This yields 
\bea
\Re\sigma_{\text{IF}}(\Omega)\sim \frac{e^2\kf}{\hbar}\left\{
\begin{array}{cc}
  (N\alpha_{\text{H}})^2,   &  \\
   \alpha_{\text{C}}^4,  & 
\end{array}
\right.
\eea
for the Hubbard and Coulomb cases, respectively. Extrapolating the asymptotic results for the electron-electron and electron-hole contributions by putting $\Omega\sim \EF$ in Eqs.~\eqref{3D_sigma_Hubbard} and \eqref{Cee2},
we see that all the contributions are comparable to each other in this range. 
The numerical results in this range are discussed in Sec.~\ref{sec:num3D}.

\subsection{High frequencies: \texorpdfstring{$\Omega>
\od$}{Omega>omegaD}}
At the level of non-interacting electrons, the optical conductivity of undoped and gapless 3D Dirac metal scales linearly with frequency [see Eq.~(\ref{3Dfree0})].
In the doped case, the onset of the linear scaling is shifted to $\od$:
\bea
\Re\sigma_\text{NI3}(\Omega)=\frac{Ne^2\Omega}{24\pi\hbar \vd}\theta(\Omega-\omega_\text{D}).
\label{3Dfree}
\eea

To the best of our knowledge, effects of electron-electron interaction in 3D Dirac systems were studied only for the undoped case.
In this case, 
the Coulomb interaction is marginally irrelevant  and, consequently, the Dirac velocity acquires an upward logarithmic renormalization while the coupling constant is  renormalized downward \cite{Abrikosov:1971,Rosenstein:2013}. The optical conductivity also experiences a logarithmic renormalization and, at $\Omega\to 0$, the slope of the linear scaling approaches a universal limit of $1+1/(N+1)$ \cite{Roy:2018}. By analogy with the 2D case, however (see Sec.~\ref{sec:Highfreq}), we expect the optical conductivity to exhibit a logarithmic singularity at $\Omega=\oi$ both for Coulomb and Hubbard interactions.
Renormalization of the optical conductivity is the first-order interaction effect, while the absorption processes studied in this paper are second-order ones. Therefore, the latter should be subleading to the former for $\Omega> \od$. Due to the lack of known first-order results for the doped case in this range, we will model the optical conductivity by its non-interacting value in Eq.~\eqref{3Dfree}.

\subsection{Numerical results in 3D}
\label{sec:num3D}
\begin{figure}
                                                                    \includegraphics[width=\linewidth,
                left]{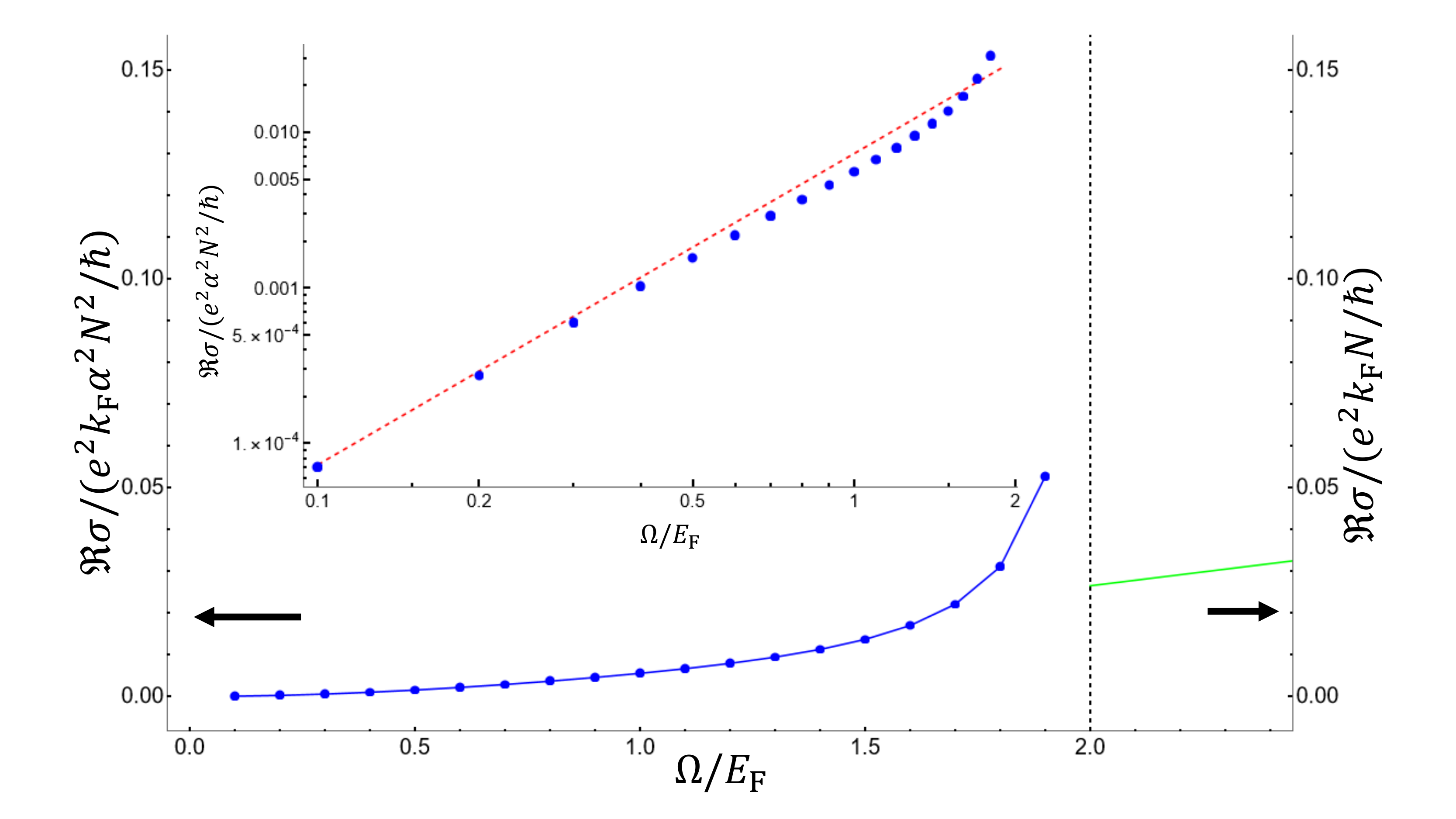}
        \caption{
        Numerical results for the optical conductivity,         $\Re\sigma(\Omega)$, as a function of $\Omega$ (in units of $\EF$) for a gapless 3D Dirac metal with a Hubbard-like interaction.         The left vertical axis is in  units of $(e^2\kf/\hbar) N^2\alpha_{\text{H}}^2$, where $N$ is the total degeneracy, 
                e.g., the number of distinct Dirac points, and $\alpha_{\text{H}}$ is the dimensionless coupling constant of Hubbard interaction.
                        The blue dots are the numerically evaluated values of $\Re\sigma(\Omega)$, while the continuous blue curve is a guide to the eye.
                The green line is the non-interacting result, Eq.~(\ref{3Dfree}), plotted along the right vertical axis in units of $(e^2\kf/\hbar)N$.
                The dashed vertical line demarcates the direct (Pauli) threshold at $\od=2\ef$.         Inset: The conductivity in the range of $0\leq \Omega <2\EF$ on a log-log scale (blue dots). The red dashed line is the analytic result for $\Omega\ll \EF$, Eq.~\eqref{3D_sigma_Hubbard},   which is extrapolated
        beyond the nominal range of its validity.       \label{fig:3DConductivityPlot}}
    \end{figure}

We evaluate Eq.~\eqref{Eq:generalzeroTexpression} numerically for each diagram,
for frequencies up to $\od=2\EF$ assuming Hubbard interaction.  
[To treat the Coulomb case for $\Omega$ comparable to $\EF$, we would need to use the exact dynamic interaction in Eq.~\eqref{eq:VdynRPA}, which is very expensive computationally.] Then we sum the results according to Eqs.~\eqref{RJ} and~\eqref{eq:cuur-curr-corr-sum-J} to obtain the 
total 
Eq.~\eqref{eq:ReConducDefi}. The conductivity in units $e^2\kf\alpha_{\text{H}}^2N^2/\hbar$ for $\Omega<\od$ is plotted in the main panel of
Fig.~\ref{fig:3DConductivityPlot}, left axis.  For the region $\Omega>\od$, where, at least in the weak-coupling limit,  absorption by non-interacting Dirac electrons 
dominates over interaction-induced absorption,  we plot the non-interacting result,  Eq.~\eqref{3Dfree}, normalized by 
$e^2\kf N/\hbar$ 
(right vertical axis). It is worth pointing out
that the rescaled conductivity 
is numerically small for almost the entire range of $\Omega<\od$, except for a narrow window near $2\od$, where the weak-coupling approximation breaks down (see a more detailed discussion at the end of this section). This implies that the interaction effects are numerically weaker than they can be expected to be. For example, for $\Omega\sim \EF$ an order of magnitude estimate for the rescaled conductivity is a number of order one. Instead, the actual result at, for example, $\Omega/\EF=1.1$, is equal to $0.00667$. This feature is in agreement with the asymptotic results for $\Omega\ll \EF$ in Table \ref{table:ResultsSummary}, all of which have small numerical coefficients. 

\begin{figure}
\vspace{0.5cm}
        \includegraphics[width=\linewidth,
                left]{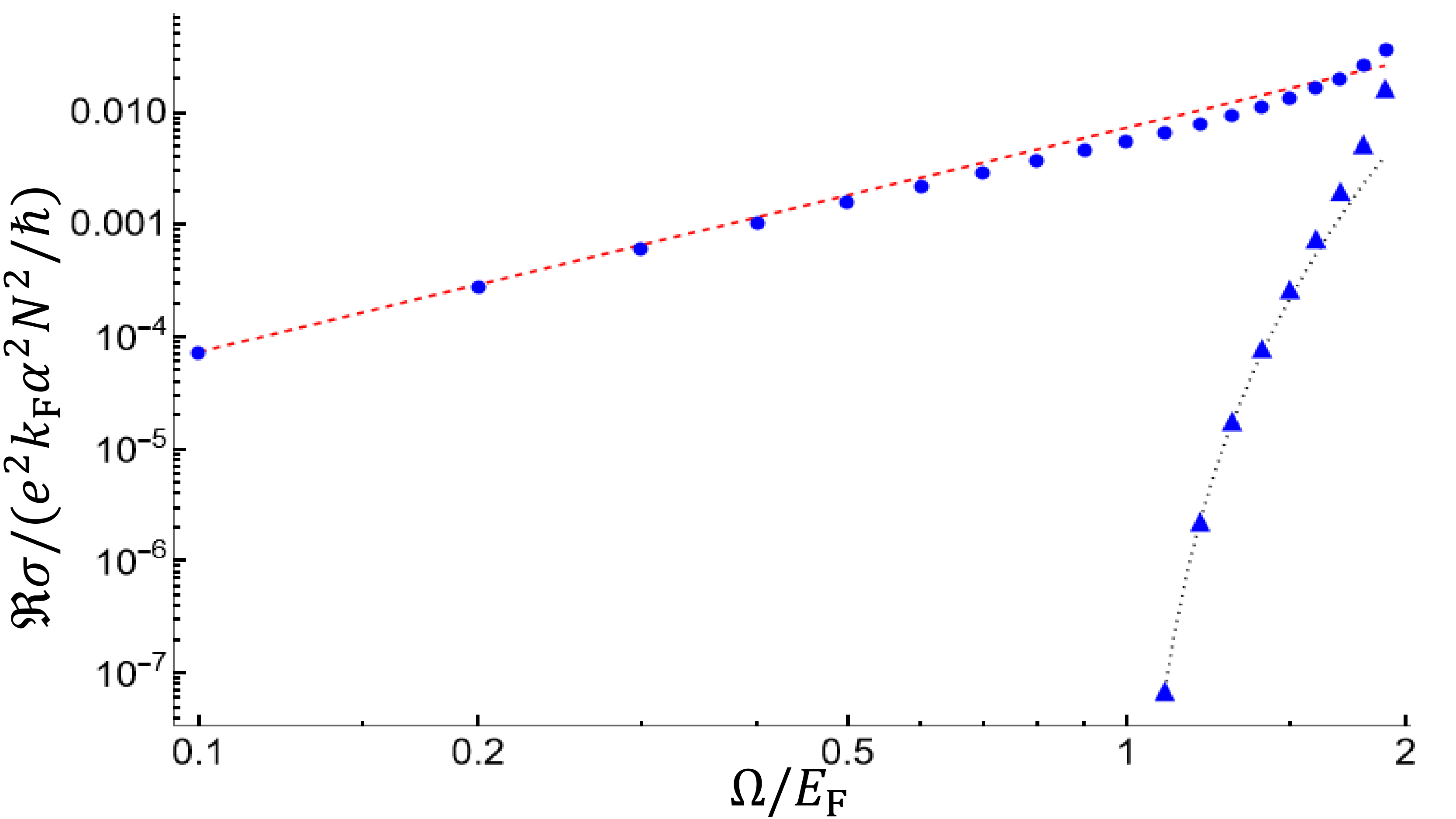}
        \caption{
                Numerically evaluated all-frequencies contribution,  as defined in Sec.~\ref{sec:LFregime},
                (blue dots) and the intermediate frequency contribution, as defined in Sec.~\ref{sec:IFregime},
                (blue triangles) to the optical conductivity 
                as a function of $\Omega$ (in units of $\EF$) for a gapless 3D Dirac metal with Hubbard interaction.
        The left vertical axis is in the units of $(e^2\kf/\hbar) N^2\alpha_{\text{H}}^2$, where $N$ is the total degeneracy, 
        e.g., the number of distinct Dirac points, and $\alpha_{\text{H}}$ is the dimensionless coupling constant of Hubbard interaction [Eq.~\eqref{eq:alphadef}].
        Also plotted is  
                the low-frequency asymptotic result for the all-frequency contribution, Eq.~\eqref{3D_sigma_Hubbard}, (red dashed curve) and the asymptotic result for the AM contribution near $\EF$,   Eq.~\eqref{3D:IFT}, (black dotted curve). 
                \label{fig:3DAllfreqvsAuger}}
\end{figure}

Also, as expected (see Sec.~\ref{sec:3Dthreshold}), the threshold singularity due to AM processes at $\Omega=\oi=\EF$ is completely masked by the \emph{ee} and \emph{eh} contributions due to the non-parabolicity of the Dirac spectrum:
there is no trace of the AM singularity in the main panel of Fig.~\ref{fig:3DConductivityPlot}. We illustrate this point further in Fig.~\ref{fig:3DAllfreqvsAuger}, in which the contributions to the conductivity from all but AM processes and from AM processes
are plotted separately.
As can be seen from the figure,  the AM contribution is smaller by orders of magnitude than the sum of other contributions near the indirect threshold, and becomes comparable to the latter only near the direct threshold of $2\EF$. While these plots are for a model Hubbard interaction, we expect a similar behavior 
for a more realistic Coulomb case, because the threshold singularity is not sensitive to the type of interaction.
The inset in Fig.~\ref{fig:3DConductivityPlot} shows the numerical results (blue dots) plotted versus the low-frequency analytic result, Eq.~\eqref{3D_sigma_Hubbard},
on a log-log scale.
As we see, 
the analytic result still works well up to $\Omega\approx \EF$.

Lastly, we see an upturn in $\Re\sigma(\Omega)$ as $\Omega$ approaches $2\EF$
from below. This indicates that
our perturbative approach, in which the Green's functions in all diagrams of Fig.~\ref{fig:AllDiagrams} are replaced by the free ones, breaks down near the direct threshold, more precisely, when $0<\od-\Omega\lesssim \alpha_{\text{H}}^2\EF$ for the Hubbard case and for $0<\od-\Omega\lesssim \alpha_{\text{C}}^4\EF$ for the Coulomb case.
This breakdown can be seen from, e.g., 
Eq.~\eqref{sigmaSE}. Indeed, substituting $s_1=s_2=+1,s_3=-1$ into the denominators in Eq.~\eqref{sigmaSE}, we see that the product of two fractions becomes 
equal to $1/(\Omega-2\epsilon_\bk)^2$. 
Now, from the paragraph above Eq.~\eqref{krange}, we know that $\epsilon_\bk<\delta\Omega=\Omega-\EF$ in the intermediate-frequency regime, i.e., for $\EF\leq\Omega<2\EF$. As $\Omega$ approaches $2\EF$, the maximum value of $2\epsilon_\bk$ also approaches $2\EF$, and the integral over $\bk$ diverges.
In principle, this singularity 
should be mitigated by re-summation of the perturbation theory,
which is beyond the scope of this work.

\bwt

\ewt

\section{Optical conductivity of a 2D Dirac metal\label{sec:2Dsystems}}

Just as in 3D, we first discuss the lowest frequency regime for 2D, and then the intermediate and high-frequency regimes.

\subsection{Lowest frequencies: \texorpdfstring{$\Omega\ll E_\text{F}$}{Omega<<EF}}
As in the 3D case, this regime corresponds to $s_3=s_4=+1$. The conductivity can be split into two contributions: a purely electron one and a contribution from processes that involve up to two holes. 

\subsubsection
{Intra-band absorption due to electron-electron interaction\label{sec:ee2D}}

The reasoning about partial cancellation of diagrams for an isotropic spectrum follows the same lines as  for the 3D case, see Sec.~\ref{sec:ee3D}. We thus have exactly the same expressions for the conductivity as in Eqs.~\eqref{sigmaSE2} and~\eqref{Sigma_full}, but now with the momentum integrals being 2D rather than 3D. 
The self-energy in 2D has an extra logarithmic factor; however, this factor cancels between different diagrams. Nevertheless, the $q$ integrand in Eq.~\eqref{Sigma_full} has an extra factor of $q$ in the denominator, which does lead to a logarithmic enhancement of the conductivity compared to the 3D case, cf. Ref.~\cite{Sharma:2021}.

For Hubbard interaction, 
we estimate the conductivity as
\bea
\Re\sigma_{\text{ee
}}(\Omega)
\sim e^2(N\alpha_{\text{H}})^2\left(\frac{\Omega}{
E_\text{F}
}\right)^2\ln\left(\frac{E_\text{F}}{\Omega}\right),\nn\\
\label{sigma_ee_2D_est}
\eea
whereas the actual calculation gives 
    \bea
   \Re\sigma_{\text{ee}}(\Omega)=\frac{e^2}{\hbar}(N\alpha_{\text{H}})^2\left(\frac{
   1} {80\pi^2}\ln\frac{E_\text{F}}{\Omega}+
   \frac{30\ln 2-1}{1200\pi^2}
      \right)
              \left(\frac{\Omega}{\EF}\right)^2.\nn\\
  \label{Hee2}
  \eea
  Note that Eq.~\eqref{Hee2} contains not only the leading logarithmic term but also a subleading one.
  Keeping the subleading term is necessary for comparison with the \emph{eh1} and \emph{eh2} contributions, which do not have a logarithmic enhancement. 

As in 3D, the case of Coulomb interaction in the region $\Omega\ll
\optwo
\ll\EF$ is similar to the Hubbard one. Explicit calculation shows that 
\bea
\Re\sigma^{\text{C}}_{\text{ee1}}(\Omega)=\frac{e^2}{\hbar}
\left( 
    \frac{1} {80\pi^2}\ln\frac{
        \optwo}{\Omega}+
    \frac{5\ln 2-2}{16\pi^2}
  \right)\left(\frac{\Omega}{\ef}\right)^2,\nn\\\label{Cee2D}
\eea
where $
\optwo$ is defined in Eq.~\eqref{oplasma}.
The 
leading logarithmic term in the last equation
coincides with the result of Ref.~\cite{Sharma:2021}. Note that, in contrast to the 3D case, the conductivity depends on the coupling constant of the Coulomb interaction only via the cutoff of the logarithmic term.

In the range of frequencies $\optwo\ll\Omega\ll\EF$, the estimate for the self-energy in Eq.~\eqref{ImS1Om} is modified as
\bea
\Im\Sigma_{\text{ee}}(\Omega)&\sim&-
\frac{\kappa^2_2}{\mathcal{N}_{\fer,2}\vf^2}\int^\Omega_0 \dee\nu \nu \int^{\infty}_{\max\{\Omega,\Omega-\nu\}/\vf} \frac{\dee q}{q^3}\nn\\
&\sim&-\omega_{\text{p}2}.\label{ImS1Om2d}
\eea
In contrast to the 3D case, the self-energy remains constant in this frequency interval and, according to Eq.~\eqref{sigma_ee_est}, the same is true for the conductivity. The actual calculation gives
\bea
\Re\sigma^{\text{C}}_{\text{ee}2}(\Omega)=\frac{5}{576\pi^2}\frac{e^2}{\hbar}\frac{e^2}{\vd}.
\label{Cee2_2D}
\eea

\subsubsection{ Absorption processes involving up to two holes 
\label{sec:eh2D}}

Now we analyze the scattering processes which involve up two holes. Again, the general reasoning here  is exactly the same as the 3D eh case. Namely, there are two types of such processes: with one hole ({\it eh1}, $s_1=-s_2=\pm 1$) 
and with two holes ({\it eh2}, $s_1=s_2= -1$), with the same corresponding conditions on the helicities as in the 3D case.

The estimates for the \emph{eh1} and \emph{eh2} contributions to the conductivity are the same as in the 3D case, i.e., they are given by Eqs.~\eqref{sigma_eh1_est} and \eqref{sigma_eh2_est},
modulo a replacement $e^2\kf\to e^2$.

  Therefore, the estimates for the \emph{eh1} and \emph{eh2} contributions in 2D 
read
\be
\Re\sigma_{\text{eh1}}(\Omega)\sim\Re\sigma_{\text{eh2}}(\Omega)\sim
e^2\alpha_{\text{H}}^2
\left(\frac{\Omega}{
E_\text{F}
}\right)^2,
\label{sigma_eh1_2D_est}\label{sigma_eh2_2D_est}
\ee
whereas the actual calculation  shows that 
\bea
\Re\sigma_{\text{eh1
}}&=&\Re\sigma_{\text{eh2
}}= \frac{1}{96\pi^2}\frac{e^2}{\hbar}(N\alpha_{\text{H}})^2
\left(\frac{\Omega}{\ef}\right)^2.\label{eh12}
\eea
Adding up  the \emph{eh1} and \emph{eh2} contributions, we have
\be
\Re\sigma_{\text{eh}}(\Omega)&=\Re\sigma_{\text{eh}1}(\Omega)+\Re\sigma_{\text{eh}2}(\Omega)
\nn\\
&=\frac{1}{48\pi^2}\frac{e^2}{\hbar}(N\alpha_{\text{H}})^2
\left(\frac{\Omega}{\ef}\right)^2.\label{eq:net2Deh}
\ee
While the combined \emph{eh} contribution is smaller than the leading logarithmic term in \emph{ee} contribution [cf. Eq.~\eqref{Hee2}], it is of the same order as the next-to-leading \emph{ee} term. The total conductivity is then a sum of Eq.~\eqref{Hee2} and Eq.~\eqref{eq:net2Deh}:
\be
\Re\sigma(\Omega)=\frac{e^2}{\hbar}(N\alpha_{\text{H}})^2\left(\frac{
   1} {80\pi^2}\ln\frac{E_\text{F}}{\Omega}+
   \frac{5\ln 2+4}{200\pi^2}
   \right)
  \left(\frac{\Omega}{\EF}\right)^2.\nn
  \\\label{2D_sigma_Hubbard}
\ee
In terms of numbers, the logarithmic term becomes the leading one for $\Omega/\EF<0.05$.

As in the 3D case, the \emph{eh} contribution for Coulomb interaction is smaller than the \emph{ee} one by a factor of $
\alpha_{\text{C}}$, and thus Eqs.~\eqref{Cee2D} and \eqref{Cee2_2D}  are the leading contributions to the conductivity in the corresponding frequency intervals. Note that the logarithmic term in Eq.~\eqref{Cee2D} becomes the leading one only at very low frequencies: $\Omega/\optwo<6.6\times 10^{-4}$. 

\subsection{Intermediate frequencies: $\oi\leq\Omega<\od$
}

The optical conductivity of a 2D Dirac metal in the intermediate frequency regime is completely analogous to the 3D case, discussed in Sec.~\ref{sec:IF3D}. As in 3D, the analytic results are attainable only for $\Omega\gtrapprox \oi$. In fact, in Sec.~\ref{sec:3Dthreshold} we have already derived a general expression for the scaling exponent $\beta_{\text{A}}$, see Eq.~\eqref{betaA}. For the case of $d=2$ and $a=2$, we obtain $\beta_{\text{A}}=3$, in agreement with Ref.~\cite{Pimenov:2017}. For our case of $d=2$ and $a=1$, this equation gives $\beta_{\text{A}}=4$. 
Without repeating the same steps as in 3D, we just present the results for the Hubbard case 
\be
\Re\sigma^{\text{IF}}=\frac{5}{108\sqrt{3}\pi}\frac{e^2}{\hbar}(N\alpha_{\text{H}})^2
\theta(\delta\Omega)\left(\frac{\delta\Omega}{
    E_\fer}\right)^{4}\label{2D:IFT}
\ee
and for the Coulomb case
\be
\Re\sigma^{\text{IF,C}}=\frac{5}{108\sqrt{3}\pi}\frac{e^2}{\hbar}\alpha_{\text{C}}^2
\theta(\delta\Omega)\left(\frac{\delta\Omega}{E_\fer}\right)^{4}.\label{2D:IFTC}
\ee

As in the 
3D case, the results for the Hubbard and Coulomb cases are identical for the reason explained in Sec.~\ref{sec:3Dthreshold}.
Also, as in 3D, the estimates for the conductivity for a generic frequency within the interval $\{\oi,\od\}$ and away from either of the thresholds, 
can be obtained by replacing $\Omega$ with $\EF$ in Eq.~\eqref{Hee2} and assuming that Eq.~\eqref{Cee2_2D} continues to be valid within an order of magnitude for $\Omega\sim E_F$. This gives
\bea
\Re\sigma_{\text{IF}}(\Omega)\sim \frac{e^2}{\hbar}\left\{
\begin{array}{cc}
  (N\alpha_{\text{H}})^2,   &  \\
   \alpha_{\text{C}}^2,  & 
\end{array}
\right.
\eea
for the Hubbard and Coulomb cases, respectively. 

\subsection{High frequencies: \texorpdfstring{$\Omega>\omega_{\text{D}}$}{Omega>omegaD}\label{sec:Highfreq}}
The optical response of 2D Dirac metals, e.g., graphene, has been studied extensively; see, e.g., reviews \cite{review_Peres2010,DasSarma:2011,kotov:2012} and references therein. At the non-interacting level, the optical conductivity has a universal form, given by Eq.~\eqref{2Dfree0}.
At finite doping, this result is modified to
\be
\Re\sigma_\text{NI2}(\Omega)=\frac{e^2N}{16\hbar}\theta(\Omega-\od).
\label{2Dfree}
\ee

As in 3D, the Coulomb interaction is also marginally irrelevant in 2D, which leads to an upward logarithmic renormalization of the Dirac velocity and, consequently, to the downward renormalization of the coupling constant. 
On the other hand, Hubbard interaction is irrelevant and can be neglected for frequencies below the ultraviolet cutoff of the model.

The optical conductivity of doped graphene 
was studied by Abedinpour et al.~\cite{abedinpour_drude_2011} to first order in both Coulomb and Hubbard interaction. 
As expected, the results reduce to those for the undoped case in the limit of $\EF\ll\Omega$.
Near the direct threshold $\od=2\EF$, the conductivity is logarithmically enhanced compared to the non-interacting value for both Coulomb and Hubbard cases.
Because the absorption processes studied in this paper occur to second order in the interaction, they are subleading to those studied in Ref.~\cite{abedinpour_drude_2011} and, therefore, we will not extend our results to the region $\Omega> \od$.

\subsection{Numerical results in 2D}\label{sec:num2D}
{We evaluated the optical conductivity numerically for Hubbard  interaction in a way similar to the  3D case,  The results are shown  in Fig.~\ref{fig:grapheneConductivityPlot}.
The conductivity in units of $(e^2/\hbar)\alpha_{\text{H}}^2N^2$ in the range of $\Omega<\od=2\EF$ is plotted on the left axis 
of the main panel. The inset shows the same data on the log-log scale (blue dots) and the low-frequency analytic result from Eq.~\eqref{Hee2} (red dashed line). On the right axis of the main panel, we plot the conductivity in units of $e^2 N/16\hbar$ for the non-interacting case, given by Eq.~\eqref{2Dfree}, (green solid line) and the analytic result to first order in Hubbard interaction from Ref.~\cite{abedinpour_drude_2011} for $\alpha_{\text{H}}=0.045$ (red solid curve). 

As in the 3D case, the rescaled conductivity is small compared to unity even for $\Omega\sim \EF$. 
Also, as in 3D, the threshold AM singularity 
from the on-set of AM processes 
at $\Omega=\omega_\text{I}$ is washed out, see Fig.~\ref{fig:2DAllfreqvsAuger}.
\bwt

\begin{figure}
                                                                    \includegraphics[width=\linewidth,
                left]{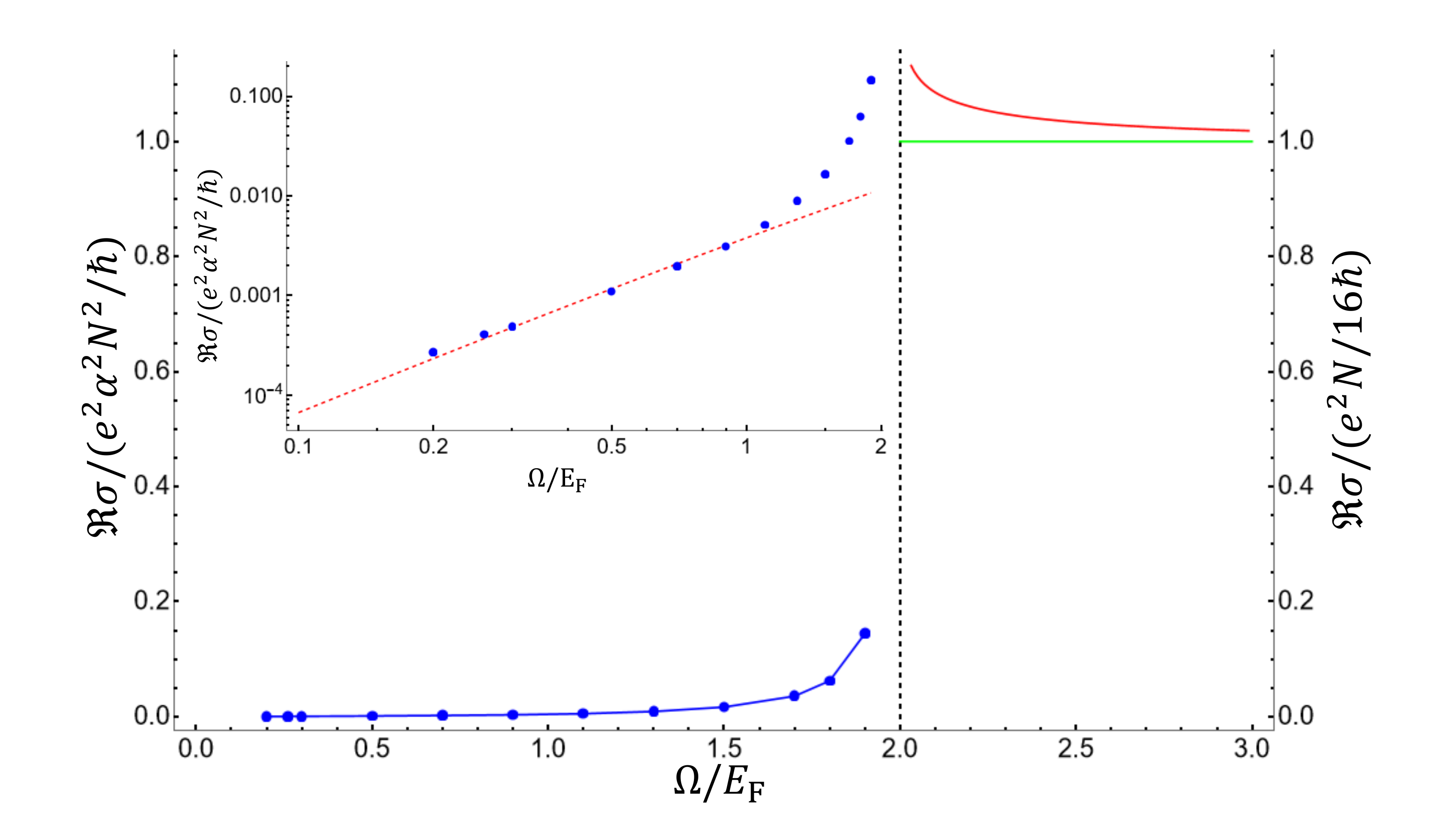}
        \caption{
        Numerical results for the optical conductivity,         $\Re\sigma(\Omega)$, as a function of $\Omega$ (in units of $\EF$) for a gapless 2D Dirac metal with a Hubbard-like interaction.         The left vertical axis is in  units of $(e^2/\hbar) N^2\alpha_{\text{H}}^2$, where $N$ is the total degeneracy, 
                e.g., the number of distinct Dirac points, and $\alpha_{\text{H}}$ is the dimensionless coupling constant of Hubbard interaction.
        The blue dots are the numerically evaluated values of $\Re\sigma(\Omega)$, while the continuous blue curve is a guide to the eye. The dashed vertical line demarcates the direct (Pauli) threshold at $\od=2\ef$. 
        The green solid line is the non-interacting result, Eq.~(\ref{2Dfree}), plotted along the right vertical axis in units of $e^2N/16\hbar$.
        The red solid line is the analytic result from Ref.~\cite{abedinpour_drude_2011}
         to first order in Hubbard interaction
        for $\alpha_{\text{H}}=0.045$.
        Inset: The conductivity in the range of $0< \Omega <2\EF$ on a log-log scale (blue dots). The red dashed line is the analytic result for $\Omega\ll \EF$, Eq.~\eqref{Hee2},   which is extrapolated
        beyond the nominal range of its validity.
              \label{fig:grapheneConductivityPlot}}
    \end{figure}

\ewt

\begin{figure}
\vspace{0.5cm}
        \includegraphics[width=\linewidth,
                left]{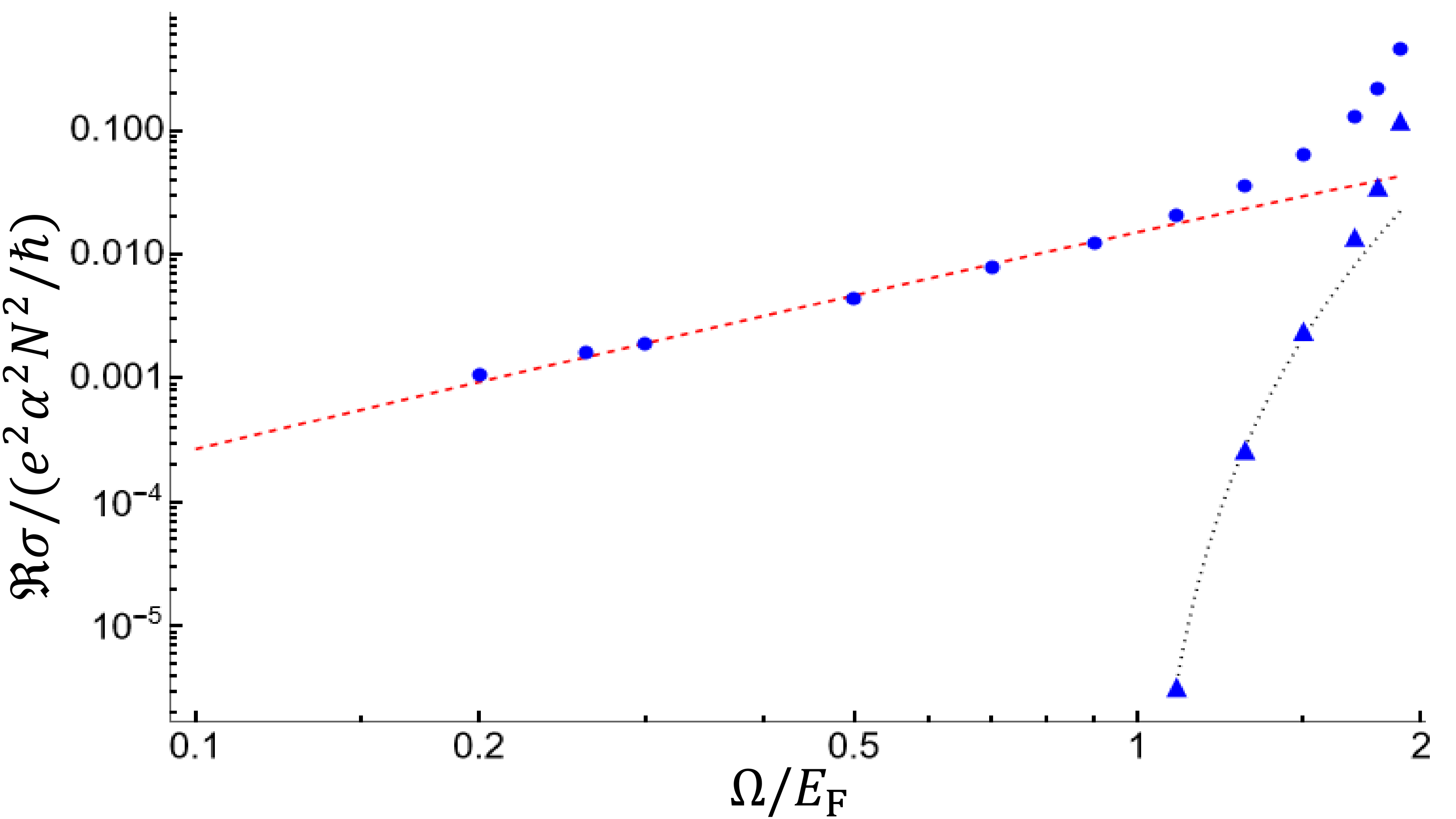}
        \caption{
                Numerically evaluated all-frequencies contribution,  as defined in Sec.~\ref{sec:LFregime},
                (blue dots) and the intermediate frequency contribution, as defined in Sec.~\ref{sec:IFregime},
                (blue triangles) to the optical conductivity 
                as a function of $\Omega$ (in units of $\EF$) for a gapless 2D Dirac metal with Hubbard interaction.
        The left vertical axis is in the units of $(e^2/\hbar) \alpha_{\text{H}}^2N^2$, where $N$ is the total degeneracy
                        and $\alpha_{\text{H}}$ is the dimensionless coupling constant of Hubbard interaction [Eq.~\eqref{eq:alphadef}].
        Also plotted are
                the low frequency analytical result for the all-frequency contribution, Eq.~\eqref{2D_sigma_Hubbard} (red dashed curve) and the analytical result for the AM contribution near $\EF$,   Eq.~\eqref{2D:IFT} (black dotted curve). 
                \label{fig:2DAllfreqvsAuger}}
\end{figure}

\section{Conclusions\label{sec:Conclusions}}
We studied optical absorption is 2D and 3D Dirac metals due to electron-electron (\emph{ee}) and electron-hole (\emph{eh}) interactions. The latter were described by two models: a Hubbard-like interaction, with a radius shorter than the Fermi wavelength but longer than the lattice spacing, and a dynamically screened Coulomb potential. To keep the perturbation theory under control, both types of interactions were assumed to be weak. The optical conductivity, $\Re\sigma(\Omega)$, was obtained by computing the leading diagrams for the current-current correlation functions, in the large-$N$ approximation for the Hubbard case and in the random-phase approximation for the Coulomb case. The main focus of this paper is the behavior of $\Re\sigma(\Omega)$ in the range of frequencies $0<\Omega<\od=2\EF$, where absorption is blocked by the Pauli principle in the single-particle picture. This range is further split into two ranges: $0<\Omega<\oi=E_F$ (I) and $\oi<\Omega<\od$ (II).

In range I, absorption starts at the lowest frequencies. The conductivity in this range comes from purely \emph{ee} scattering, which is allowed to contribute due to broken Galilean invariance, and from certain \emph{eh} scattering processes, which involve up to two holes. For $\Omega\ll \EF$,
we derived the analytic results for the conductivity, which are presented in Tables \ref{table:ResultsSummary} and \ref{table:ResultsSummaryCoulomb}, for the Hubbard and Coulomb cases respectively. In both cases, $\Re\sigma(\Omega)$
scales as $\Omega^2\ln\Omega$ in 2D and as $\Omega^2$ in 3D. In other words, the effective current relaxation rate, $1/\tau_j\equiv(\kf/\vd ne^2)\Omega^2\Re\sigma(\Omega)$ scales as $\Omega^4\ln\Omega$ in 2D and as $\Omega^4$ in 3D. (Here, $n$ is the carrier number density, $\kf$ is the Fermi momentum, and $\vd$ is the Dirac velocity.)
The \emph{ee} contribution to $\Re\sigma(\Omega)$ has been derived in Ref.~\cite{Sharma:2021} for the Coulomb case by a different method, via the Heisenberg equations of motion for the current operator, and our results for this contribution agree with those of Ref.~\cite{Sharma:2021} (modulo a discrepancy in the numerical coefficient in 3D). 
Remarkably, the \emph{eh} contribution, studied in this paper, is comparable to the \emph{ee} one in 3D and  subleading to the \emph{ee} in 2D only in the leading logarithmic sense. For the rest of range I, $\Re\sigma(\Omega)$ was calculated numerically.

In range II, another type \emph{eh} scattering processes, similar the Auger-Meitner (AM) processes in atomic physics \cite{meitner:1922,Auger:1923,PhysicsToday:2019},  start to contribute to the conductivity. These processes have been studied extensively in the context of doped semiconductors (see, e.g., Refs.~\cite{gavoret_optical_1969,ruckenstein_many-body_1987,Sham:1990,Hawrylak:1991,Pimenov:2017}), but only in the model of parabolic bands, within  which absorption in range I is absent, and the onset of absorption due to AM processes at $\Omega=\oi$ is manifested by a threshold singularity in $\Re\sigma(\Omega)$. We showed that a similar singularity also exists for Dirac metals.  However, in contrast to the parabolic-bands case, the AM singularity occurs at the background of absorption due to \emph{ee} and other \emph{eh} processes, which start to contribute in region I, but continue to contribute in region II as well.
 Our numerical calculations show that the AM threshold singularity is completely masked by these other processes.

In the range of $\Omega\sim \EF$ (but not in the immediate vicinity of either $\EF$ and $2\EF$), all \emph{ee} and \emph{eh} scattering processes give comparable contributions to $\Re\sigma(\Omega)$. As $\EF$ is the only energy scale in this regime, the effective current relaxation rate is of order $g\EF$, where $g$ is the dimensionless coupling constant for either type of interaction. However, our analytic and numerical results show that the numerical coefficient $C$ in the relation $1/\tau_j=Cg\EF$ is anomalously small, on the order of $10^{-3}$, i.e., in reality $1/\tau_j\ll \EF$ even at $g=1$. This may explain the observation of well-resolved collective modes below $2\EF$ in the helical surface state of a doped 3D topological insulator \cite{Kung2017PRL} (see Ref.~\cite{Maslov:2022} for more details).

As mentioned in Sec.~\ref{sec:Intro},  experiments on monolayer graphene find significant optical absorption at frequencies above the Drude tail but below $2\EF$ \cite{li_dirac_2008,Mak:2008,Horng:2011,Mak:2012,Drew:2016} and also significant Raman response in the same frequency range \cite{riccardi_gate-dependent_2016}. In real materials, absorption in this frequency range is not only due to \emph{ee} and \emph{eh} interactions, but also due to electron-impurity and electron-phonon scattering. Moreover, it was argued in Ref.~\cite{peres:2008} that the data of Ref.~\cite{li_dirac_2008} can be well explained by taking into account only electron-impurity and electron-phonon scattering (with an addition of excitonic effects \cite{peres:2010PRL}). We hope that future experiments on samples with higher mobilities will be able to resolve  intrinsic, \emph{ee} and \emph{eh} contributions to absorption.

\acknowledgements
This paper is dedicated to the memory of Konstantin B. Efetov, an outstanding physicist and a kind human being. We thank D. Basov, A. Jahin, A. Kumar, S. Maiti,  and I.  Michaloliakos for stimulating discussions. This work was supported by the US  National Science Foundation under Grants No. DMR-1720816 and No. DMR-2224000. 
\bibliography{graphenenozierespaperreferences}

\onecolumngrid
\appendix
\section{General structure of the diagrams for the optical conductivity}
\label{app:genstructure}
In this Appendix, we derive the general forms of the contributions from particular diagrams for the current-current correlation function in Fig.~\ref{fig:AllDiagrams}. These forms are valid 
for Hubbard interaction for all frequencies of interest, $\Omega<2\EF$, and for Coulomb interaction for $\Omega\ll E_F$.
We discuss only the self-energy and vertex diagrams; the analysis of the Aslamazov-Larkin diagrams follows the same lines.
\subsection{Some common properties}
As explained in Sec.~\ref{sec:listofdiagrams}, the imaginary parts of the contributions from the self-energy (SE) and vertex (V) diagrams to the current-current correlation function contain a dynamic polarization bubble 
$\pi_{\text{dyn}}(\bq,i\nu)=\pi_0(\bq,i\nu)-\pi_0(\bq,0)$. For practical purposes, however, it is more convenient to replace $\pi_{\text{dyn}}(\bq,i\nu)$ by the full bubble, $\pi_0(\bq,i\nu)$, with the understanding that its static part will drop out on taking the imaginary part. Then the self-energy and vertex diagrams can be written as
\be
\Pi^{\text{SE,V}}(i\Omega)=\frac{1}{d}
\int_\mathbf{q}T\sum\limits_{\nu}\pi^{\text{SE,V}}_1(\mathbf{q},i\nu;i\Omega)
V^2_{\text{st}}(\bq)\pi_{0}(\bq,i\nu),
\ee
where SE stands for both SE$_1$ and SE$_2$ parts of the SE diagram, $V_{\text{st}}(\bq)$ is either a (momentum-independent) Hubbard interaction or statically screened Coulomb potential,
\be
\pi_0(\mathbf{q},i\nu)=-\frac{1}{4}\sum\limits_{s'_4,
s'_6
=\pm 1}\int_\mathbf{p}\tr\left(\hat{M}_\mathbf{p}^{s'_6
}\hat{M}_{\mathbf{p+q}}^{s'_4
}
\right)\frac{\left[n_\text{F}(\xi^{s'_{6}
}
_\mathbf{p})-n_\text{F}(\xi^{s'_4
}_\mathbf{p+q})\right]}{i\nu-\xi^{s'_4
}_\mathbf{p+q}+\xi^{s'_{6}
}_\mathbf{p}}
\label{pi0}
\ee
is the polarization bubble,  and $\pi_1^{\text{SE,V}}(\bq,i\nu;i\Omega)$ are given by
\be
\pi_1^{\text{SE}_1}(\mathbf{q},i\nu;i\Omega)=-\int_\mathcal{K}\tr\left(\hat{\mathbf{v}}\hat{G}(\mathcal{K+W})\hat{G}(\mathcal{K+Q+W})\hat{G}(\mathcal{K+W})\cdot\hat{\mathbf{v}}\hat{G}(\mathcal{K})\right)\nn\\
\pi_1^{\text{SE}_2}(\mathbf{q},i\nu;i\Omega)=-\int_\mathcal{K}\tr\left(\hat{\mathbf{v}}\hat{G}(\mathcal{K+W})\cdot\hat{\mathbf{v}}\hat{G}(\mathcal{K})\hat{G}(\mathcal{K+Q})\hat{G}(\mathcal{K})\right)\nn\\
\pi_1^{\text{V}}(\mathbf{q},i\nu;i\Omega)=-\int_\mathcal{K}\tr\left(\hat{\mathbf{v}}\hat{G}(\mathcal{K+W})\hat{G}(\mathcal{K+Q+W})\cdot\hat{\mathbf{v}}\hat{G}(\mathcal{K+Q})\hat{G}(\mathcal{K})\right).
\ee
It will be convenient to introduce a quantity 
 \be
\tilde{\Sigma}_{s}(\mathbf{k},\mathbf{q},i\omega)=T\sum\limits_{\nu}g_{s}(\mathbf{k},i\omega+i\nu)\pi_0(\mathbf{q},i\nu),\label{eq:TildeSigmaDefinition}
\ee
whose meaning is that $\tilde{\Sigma}_{s}(\mathbf{k}+\bq,\mathbf{q},i\omega)$ is the self-energy with fixed momentum transfer $\bq$ (but without the interaction potential).
Evaluating the Matsubara sum over $\nu$ and performing analytic continuation as in $i\omega\to \omega+i0^+$,
we obtain
\be
\tilde{\Sigma}_{s,\text{R}}(\mathbf{k},\mathbf{q},
\omega)
=\frac{1}{4}\sum\limits_{s'_4,s'_6=\pm 1}\int_\mathbf{p}\left[\tr(\hat{M}_\mathbf{p}^{s'_6}\hat{M}_\mathbf{p+q}^{s'_4})\right]\left[n_\text{F}(\xi^{s'_{6}}_\mathbf{p})-n_\text{F}(\xi^{s'_4}_\mathbf{p+q})\right]\frac{\left[n_\text{F}(\xi_\mathbf{k}^{s})+\bos(\xi^{s'_4}_\mathbf{p+q}-\xi^{s'_{6}}_\mathbf{p})\right]}{
\omega+\xi^{s'_4}_\mathbf{p+q}-\xi^{s'_{6}}_\mathbf{p}-\xi_\mathbf{k}^{s}+i0^+}.\label{eq:TildeSigmaForm}
\ee
At $T=0$
the last equation is reduced to
\be
\tilde{\Sigma}_{s,\text{R}}(\mathbf{k},\mathbf{q},
\omega)=\frac{1}{4}\sum\limits_{s'_4,s'_6=\pm 1}\int_\mathbf{p}\frac{\tr(\hat{M}_\mathbf{p}^{s'_6}\hat{M}_\mathbf{p+q}^{s'_4})}{
\omega+\xi^{s'_4}_\mathbf{p+q}-\xi^{s'_{6}}_\mathbf{p}-\xi_\mathbf{k}^{s}+i0^+}\left[
\theta(\xi^{s'_{6}}_\mathbf{p})\theta(-\xi^{s'_4}_\mathbf{p+q})\theta(\xi_\mathbf{k}^{s})+\theta(-\xi^{s'_{6}}_\mathbf{p})\theta(\xi^{s'_4}_\mathbf{p+q})\theta(-\xi_\mathbf{k}^{s})\right].\label{eq:TildeSigmaZeroTForm}
\ee

We also note a useful identity, namely, that for  any real functions $h_{1,2}(\omega)$, and functions $f_{1,2}(\omega)$ 
of the form
\be
f_{1,2}(\omega)=\frac{1}{\omega-x_{1,2}}+iy_{1,2}\delta(\omega-x_{1,2}),
\ee
we have
\be
\int \dee\omega\left[h_1(\omega)\Re f_1(\Omega+\omega)\Im f_2(\omega)+h_2(\omega)\Im f_1(\omega)\Re f_2(\Omega+\omega)\right]\nonumber\\
=\frac{y_2h_1(x_2)-y_1h_2(-\Omega+x_1)}{\Omega+x_2-x_1}.
\label{eq:ImReMixtureProperty}
\ee
This identity 
will appear often in our discussion.

\subsection{Self-energy diagrams}
\label{sec:SE}

\subsubsection{SE$_1$ diagram}
\label{SE1}
Explicitly, the first self-energy diagram (SE$_1$) in Fig.~\ref{fig:AllDiagrams} reads  
\bea
\Pi^{\text{SE}_1}(i\Omega)
=\frac{1}{16d}\sum\limits_{\mathcal{S}_{\text{SE}_1}}\int_\mathbf{k,q}\tr\left(\hat{\bv}
\hat{M}_\mathbf{k}^{s'_1}\hat{M}_{\mathbf{k+q}}^{s'_5}\hat{M}_{\mathbf{k}}^{s'_2}\cdot\hat{\bv}
\hat{M}_{\mathbf{k}}^{s'_3}\right)V^2_\text{st}(\mathbf{q})W_{\text{SE}_1}(i\Omega),\label{A7}
\eea
where $\mathcal{S}_{\text{SE}_1}=\{s'_1,s'_2,s'_3,s'_5\}$,
\be
W_{\text{SE}_1}(\bk,\bq,i\Omega)&= T\sum\limits_{\omega}
g_{s'_1}(\mathbf{k},i\omega+i\Omega)g_{s'_2}(\mathbf{k},i\omega+i\Omega)g_{s'_3}(\mathbf{k},i\omega)\tilde{\Sigma}_{s'_5}(\mathbf{k+q},\mathbf{q},i\omega+i\Omega)
\label{eq:WSE1definition}\\
&\equiv 
T\sum\limits_{\omega}A_{\text{SE}_1}(\bk,\bq,i\omega+i\Omega)g_{s'_3}(\mathbf{k},i\omega),\nn
\ee
\be
A_{\text{SE}_1}(\bk,\bq,i\omega+i\Omega)=g_{s'_1}(\mathbf{k},i\omega+i\Omega)g_{s'_2}(\mathbf{k},i\omega+i\Omega)\tilde{\Sigma}_{s'_5}(\mathbf{k+q},\mathbf{q},i\omega+i\Omega),
\ee
and $\tilde{\Sigma}_{s'_5}(\mathbf{k+q},\mathbf{q},i\omega+i\Omega)$ is defined in Eq.~\eqref{eq:TildeSigmaDefinition}.
We focus on evaluating $W_{\text{SE}_1}(i\Omega)$ first
and switch the primed helicity labels to unprimed ones at this point 
(cf. footnote~\ref{ftnt:PivsRexplanation} 
in the main text).
Upon summation over 
$\omega$ and analytic continuation as in 
$i\Omega\rightarrow\Omega+i0^+$, 
for obtain for the imaginary part:
\bea
\Im W_{\text{SE}_1,\text{R}}(\bk,\bq,\Omega)&=&-\frac{1}{\pi}\int^\infty_{-\infty} \dee\omega \left[n_\text{F}(\omega)-n_\text{F}(\omega+\Omega)\right] \Im g_{s_3,\text{R}}(\mathbf{k},\omega)\Im A_{\text{SE}_1,\text{R}}(\bk,\bq,\omega+\Omega),\nn\\
&=&-\frac{1}{\pi}\int^0_{-\Omega} \dee\omega\Im g_{s_3,\text{R}}(\mathbf{k},\omega)\Im A_{\text{SE}_1,\text{R}}(\bk,\bq,\omega+\Omega),
\label{A11}
\eea
where at the last step we implemented the conditions $T=0$ and $\Omega>0$, and
\bse
\be
\Im A_{\text{SE}_1}(\bk,\bq,\omega+\Omega)=\Im\left[g_{s_1,\text{R}}(\mathbf{k},\omega+\Omega)g_{s_2}(\mathbf{k},\omega+\Omega)\tilde{\Sigma}_{s_5,\text{R}}(\bk+\bq,\bq,\omega+\Omega)\right]
\\
=\Re\left[g_{s_1,\text{R}}(\mathbf{k},\omega+\Omega)g_{s_2,\text{R}}(\mathbf{k},\omega+\Omega)\right]\Im\tilde{\Sigma}_{s_5,\text{R}}(\bk,\bq,\omega+\Omega)+\Im\left[g_{s_1,\text{R}}(\mathbf{k},\omega+\Omega)g_{s_2,\text{R}}(\mathbf{k},\omega+\Omega)\right]\Re\tilde{\Sigma}_{s_5,\text{R}}(\bk+\bq,\bq,\omega+\Omega).\nn\\
\label{A12}
\ee
\ese

We will now show that the second term in Eq.~\eqref{A12}, proportional to $\Re\tilde{\Sigma}_{s_5}(\bk,\bq,\omega+\Omega)$, does not contribute to the real part of the conductivity for  $\Omega<\omega_{\text{D}}$. Indeed, 
the prefactor of $\Re\tilde{\Sigma}_{s_5}(\bk,\bq,\omega+\Omega)$ reads
\be
\Im\left[g_{s_1,\text{R}}(\mathbf{k},\omega+\Omega)g_{s_2,\text{R}}(\mathbf{k},\omega+\Omega)\right]=-\frac{\pi\delta(\Omega+\omega-\xi^{s_2}_\mathbf{k})}{\Omega+\omega-\xi^{s_1}_\mathbf{k}}+(s_1\leftrightarrow s_2).\label{A14}
\ee
Because $\Im A_{\text{SE}_1,\text{R}}(\bk,\bq,\omega+\Omega)$ in Eq.~\eqref{A11} is multiplied by 
\bea\Im g_{s_3}(\bk,\omega)=-\pi\delta(\omega-\xi^{s_3}_\mathbf{k}),\label{imgs3}\eea we can put $\omega=\xi^{s_3}_\mathbf{k}$ in Eq.~\eqref{A14}. 
Then the delta function in the first term  of Eq.~\eqref{A14} implies that $\Omega=(s_2-s_3)\epsilon_\mathbf{k}$. Since we chose $\Omega>0$,   the delta function is non-zero only  if $s_2=+1$ and $s_3=-1$. Therefore, $\Omega=2\epsilon_\mathbf{k}$ or $\epsilon_\mathbf{k}=\Omega/2$. However, the limits of integration over $\omega$ in Eq.~\eqref{A11} along with the condition $\omega=\xi_\bk^{s_3}$ imply that $-\Omega<\xi_k^{s_3}=-\epsilon_\mathbf{k}-\EF$. Thus, we have $-\Omega<-\Omega/2-\EF$ or $\Omega>2\EF=\od$.
The second term in Eq.~\eqref{A14} is analyzed in a similar way and with the same result.

Now we show that the first  term in Eq.~\eqref{A12}, proportional to $\Im\tilde{\Sigma}_{s_5}(\bk+\bq,\bq,\omega+\Omega)$, does contribute to the real part of the conductivity for $\Omega<\od$. 
The contribution of this term to $\Im W_{\text{SE}_1,\text{R}}$ (denoted by superscript 1) is given by
\be
\Im W^{(1)}_{\text{SE}_1,\text{R}}(\bk,\bq,
\Omega)=-\frac{1}{\pi}\Re\left[g_{s_1,\text{R}}(\mathbf{k},\Omega+\xi^{s_3}_\mathbf{k})g_{s_2,\text{R}}(\mathbf{k},\Omega+\xi^{s_3}_\mathbf{k})\right]\int^0_{-\Omega}\dee\omega
\Im\tilde{\Sigma}_{s_5,\text{R}}(\mathbf{k+q},\mathbf{q},\omega+\Omega)\Im g_{s_3,\text{R}}(\mathbf{k},\omega).\nn\\
\label{ImWSE1}
\ee
Note that, formally speaking, the real part of the product of two Green's function in the equation above contains a highly singular term: $\left[\delta\left(\Omega+\xi_\bk^{s_3}-\xi^{s_1}_\bk\right)\right]^2$. In fact, it can be shown that such a term only renormalizes the Drude weight but does not contribute to the regular part of the conductivity. Postponing the proof till  Sec.~\ref{sec:delta2}, we now proceed discarding this term.
From Eq.~\eqref{eq:TildeSigmaZeroTForm},
\be
\Im\tilde{\Sigma}_{s_5,\text{R}}(\mathbf{k+q},\mathbf{q},\omega+\Omega)=
-\frac{\pi}{4}\sum\limits_{s_4,s_6=\pm 1}&\int_\mathbf{p}\tr(\hat{M}_\mathbf{p}^{s_6}\hat{M}_\mathbf{p+q}^{s_4})
\times\delta(\omega+\Omega+\xi^{s_4}_\mathbf{p+q}-\xi^{s_{6}}_\mathbf{p}-\xi_\mathbf{k+q}^{s_5})\nn\\
&\times\left[\theta(\xi^{s_{6}}_\mathbf{p})\theta(-\xi^{s_4}_\mathbf{p+q})\theta(\xi_\mathbf{k+q}^{s_5})+\theta(-\xi^{s_{6}}_\mathbf{p})\theta(\xi^{s_4}_\mathbf{p+q})\theta(-\xi_\mathbf{k+q}^{s_5})\right].
\ee
Because $\Omega+\omega>0$, the delta function in the equation above implies that $\xi^{s_4}_\mathbf{p+q}-\xi^{s_{6}}_\mathbf{p}-\xi_\mathbf{k+q}^{s_5}<0$. Comparing this condition with the ones imposed by the theta functions, we see that only the first set of theta functions is non-zero
Integrating over $\omega$ with the help of Eq.~\eqref{imgs3}, we obtain
\bea
\Im W^{(1)}_{\text{SE}_1,\text{R}}(\bk,\bq,
\Omega)
=
-\frac{\pi}{4}\sum\limits_{s_4,s_6=\pm 1}\int_\mathbf{p}&&\tr(\hat{M}_\mathbf{p}^{s_6}\hat{M}_\mathbf{p+q}^{s_4})
\ag{\theta(\Omega+\xi^{s_3}_\mathbf{k})}
\theta(-\xi_\mathbf{k}^{s_3})\theta(\xi^{s_{6}}_\mathbf{p})\theta(-\xi^{s_4}_\mathbf{p+q})\theta(\xi_\mathbf{k+q}^{s_5})
\nonumber\\
&&\times\delta(\Omega+\xi_\mathbf{k}^{s_3}+\xi^{s_4}_\mathbf{p+q}-\xi^{s_{6}}_\mathbf{p}-\xi_\mathbf{k+q}^{s_5}).\IEEEeqnarraynumspace
\eea
Substituting the last result into 
$\Im\Pi^{\text{SE}_1}_\text{R}$, obtained by analytic continuation of Eq.~\eqref{A7}, 
yields
\be
\Im\Pi^{\text{SE}_1}_\text{R}&&(\Omega)=
-\frac{\pi}{64}
\sum\limits_{\mathcal{S}_{\text{SE}_1}}\int_\mathbf{k,q}\tr\left(\hat{\bv}
\hat{M}_\mathbf{k}^{s_1}\hat{M}_{\mathbf{k+q}}^{s_5}\hat{M}_{\mathbf{k}}^{s_2}\cdot\hat{\bv}
\hat{M}_{\mathbf{k}}^{s_3}\right)V^2_\text{st}(\mathbf{q})\times \frac{1}{\Omega-\xi^{s_1}_\mathbf{k}+\xi^{s_3}_\mathbf{k}}\frac{1}{\Omega-\xi^{s_2}_\mathbf{k}+\xi^{s_3}_\mathbf{k}}\times\nonumber\\
&&\times
\sum\limits_{s_4,s_6=\pm 1}\int_\mathbf{p}\tr(\hat{M}_\mathbf{p}^{s_6}\hat{M}_\mathbf{p+q}^{s_4})
\theta(\Omega+\xi^{s_3}_\mathbf{k})
\theta(-\xi_\mathbf{k}^{s_3})\theta(\xi^{s_{6}}_\mathbf{p})\theta(-\xi^{s_4}_\mathbf{p+q})\theta(\xi_\mathbf{k+q}^{s_5})\delta(\Omega+\xi_\mathbf{k}^{s_3}+\xi^{s_4}_\mathbf{p+q}-\xi^{s_{6}}_\mathbf{p}-\xi_\mathbf{k+q}^{s_5}).\IEEEeqnarraynumspace
\ee
Relabeling $\mathbf{p}\rightarrow\mathbf{p-q}$ and then $\mathbf{p}\rightarrow\mathbf{-p}$, and using that $\epsilon_\bk=\epsilon_{-\bk}$, we find
\bea
\Im\Pi^{\text{SE}_1}_\text{R}(\Omega)
=-\frac{\pi^2}{32} \sum\limits_{\mathcal{S}}&\int_{\mathbf{k,p,q},\nu}\tr\left(\hat{\bv}
\hat{M}_\mathbf{k}^{s_1}\hat{M}_{\mathbf{k+q}}^{s_5}\hat{M}_{\mathbf{k}}^{s_2}\cdot\hat{\bv}
\hat{M}_{\mathbf{k}}^{s_3}\right)\tr(\hat{M}_\mathbf{-p-q}^{s_6}\hat{M}_\mathbf{-p}^{s_4})V^2_\text{st}(\mathbf{q})\times \frac{1}{\Omega-\xi^{s_1}_\mathbf{k}+\xi^{s_3}_\mathbf{k}}\frac{1}{\Omega-\xi^{s_2}_\mathbf{k}+\xi^{s_3}_\mathbf{k}}\nonumber\\
&\times
\theta(\Omega+\xi^{s_3}_\mathbf{k})
\theta(-\xi_\mathbf{k}^{s_3})\theta(\xi^{s_{6}}_\mathbf{p+q})\theta(-\xi^{s_4}_\mathbf{p})\theta(\xi_\mathbf{k+q}^{s_5})\delta(\Omega+\nu+\xi_\mathbf{k}^{s_3}-\xi_\mathbf{k+q}^{s_5})\delta(\nu-\xi^{s_4}_\mathbf{p}+\xi^{s_{6}}_\mathbf{p+q}),\label{A18}
\eea
 where we have re-introduced 
 $\int_\nu=\int\dee\nu/2\pi$.
Thus,
\be
\Im\Pi^{\text{SE}_1}(\Omega)=-\frac{\pi^2}{32}\sum\limits_{\mathcal{S}}&\int_{\mathbf{k,p,q},\nu}V^2_\text{st}(\mathbf{q})\times\mathcal{T}^{\text{SE}_1}_{\mathcal{S}}\mathcal{G}^{\text{SE}_1}_{\mathcal{S}}\nonumber\\
&\times
\theta(\Omega+\xi^{s_3}_\mathbf{k})
\theta(-\xi_\mathbf{k}^{s_3})\theta(\xi^{s_{6}}_\mathbf{p+q})\theta(-\xi^{s_4}_\mathbf{p})\theta(\xi_\mathbf{k+q}^{s_5})\delta(\Omega+\nu+\xi_\mathbf{k}^{s_3}-\xi_\mathbf{k+q}^{s_5})\delta(\nu-\xi^{s_4}_\mathbf{p}+\xi^{s_{6}}_\mathbf{p+q}),
\label{ImSE1}
\ee
where
\bse
\bea
\mathcal{T}^{\text{SE}_1}_{\mathcal{S}}=\frac{1}{d}\tr\left(\hat{\bv}
\hat{M}_\mathbf{k}^{s_1}\hat{M}_{\mathbf{k+q}}^{s_5}\hat{M}_{\mathbf{k}}^{s_2}\cdot\hat{\bv}
\hat{M}_{\mathbf{k}}^{s_3}\right)\tr\left(\hat{M}_\mathbf{-p-q}^{s_6}\hat{M}_\mathbf{-p}^{s_4}\right)\\
\mathcal{G}^{\text{SE}_1}_{\mathcal{S}}=\frac{1}{\Omega-\xi^{s_1}_\mathbf{k}+\xi^{s_3}_\mathbf{k}}\frac{1}{\Omega-\xi^{s_2}_\mathbf{k}+\xi^{s_3}_\mathbf{k}}.
\eea
\ese
The last equation gives the SE$_1$ part of Eq.~\eqref{eq:firsteqofresultlist} in the main text.

\subsubsection{SE$_2$ diagram \label{SE2}}

On the Matsubara axis, the second self-energy diagram ($\text{SE}_2$) in Fig.~\ref{fig:AllDiagrams} is related  to the $\text{SE}_1$ one via
\be
\Pi^{\text{SE}_2}(i\Omega)=
\Pi^{\text{SE}_1}(-i\Omega),\IEEEeqnarraynumspace
\ee
which, upon analytic continuation, implies that
\bea
\Im\Pi^{\text{SE}_2}_\text{R}(\Omega)=\Im\Pi^{\text{SE}_1}_\text{A}(-\Omega)=-\Im\Pi^{\text{SE}_1}_\text{R}(-\Omega),\label{A23}
\eea
where $\Pi^{\text{SE}_1}_\text{A}(\ve)$ is the advanced correlation function. Note that although the total correlation function is a real function of time and, therefore, the imaginary part of its Fourier transform  is an odd function of frequency, the partial contributions to the total correlation function from different diagrams do not have definite parity. Therefore, Eq.~\eqref{A23} cannot be simplified further. In the previous section, we considered $\Im\Pi^{\text{SE}_1}_\text{R}(\Omega)$ for $\Omega>0$; all we have to do now is the extend this analysis for $\Omega<0$. Omitting the steps, which are almost identical to those in the previous section, 
we present only the final result:
\be
\Im\Pi^{\text{SE}_2}(\Omega)=
-\frac{\pi^2}{32}
\sum\limits_{\mathcal{S}}&\int_{\mathbf{k,p,q},\nu}V^2_\text{st}(\mathbf{q})\times\mathcal{T}^{\text{SE}_2}_{\mathcal{S}}\mathcal{G}^{\text{SE}_2}_{\mathcal{S}}\nonumber\\
&\times
\theta(\Omega+\xi^{s_3}_\mathbf{k})
\theta(-\xi_\mathbf{k}^{s_3})\theta(\xi^{s_{6}}_\mathbf{p+q})\theta(-\xi^{s_4}_\mathbf{p})\theta(\xi_\mathbf{k+q}^{s_5})\delta(\Omega+\nu+\xi_\mathbf{k}^{s_3}-\xi_\mathbf{k+q}^{s_5})\delta(\nu-\xi^{s_4}_\mathbf{p}+\xi^{s_{6}}_\mathbf{p+q}),
\label{ImSE2}
\ee
where
\bse
\be
\mathcal{T}^{\text{SE}_2}_{\mathcal{S}}=\tr\left(\hat{\bv}
\hat{M}_\mathbf{-k-q}^{s_1}\hat{M}_{\mathbf{-k}}^{s_3}\hat{M}_{\mathbf{-k-q}}^{s_2}\cdot\hat{\bv}
\hat{M}_{\mathbf{-k-q}}^{s_5}\right)\tr\left(\hat{M}_\mathbf{p}^{s_4}\hat{M}_\mathbf{p+q}^{s_6}\right)\\
\mathcal{G}^{\text{SE}_2}_{\mathcal{S}}=\frac{1}{\Omega-\xi^{s_5}_\mathbf{k+q}+\xi^{s_1}_\mathbf{k+q}}\frac{1}{\Omega-\xi^{s_5}_\mathbf{k+q}+\xi^{s_2}_\mathbf{k+q}}.
\ee
\ese
Thus we have derived the SE$_2$ part of Eq.~\eqref{eq:firsteqofresultlist} in the main text.
\subsubsection{Elimination of singular terms} \label{sec:delta2}
In Sec.~\ref{SE1}, we argued that the singularities of the $\delta^2(x)$ type, occurring in e.g., Eq.~\eqref{ImWSE1}, can be ignored. Here, we prove this statement. Because the singularity occurs already for the case when all helicities are the same, we put 
$s_1=s_2=s_3=s_5\equiv s$ and suppress the helicity index. Also, since the momentum-dependence of the self-energy is irrelevant for the present argument, we denote temporarily $\tilde S(i\omega)\equiv\tilde\Sigma_s(\bk+\bq,\bk,i\omega)$.
Adding  Eq.~\eqref{ImWSE1} to the corresponding contribution from diagram SE$_2$, we obtain
\bea
W_{\text{SE}}(i\Omega)=W_{\text{SE}1}(i\Omega)+W_{\text{SE}2}(i\Omega)
=
T\sum\limits_{\omega}
g^2(\mathbf{k},i\omega)\tilde S(i\omega)
\left[g(\mathbf{k},i\omega-i\Omega)+g(\mathbf{k},i\omega+i\Omega)\right],\label{WS}
\eea
where we also made a change  of variables  $i\omega\to i\omega+i\Omega$ in $W_{\text{SE}1}$. Applying the identity 
\bea
g_s(\bk,i\ve)g_{s'}(\bk,i\ve')=\frac{1}{i(\ve'-\ve)-\xi_\bk^{s'}+\xi_\bk^s}\left[g_s(\bk,i\ve)-g_{s'}(\bk,i\ve')\right]\label{ident}
\eea
with $s=s'$ to Eq.~\eqref{WS} twice, we obtain
\bea
W_{\text{SE}}(i\Omega)=\frac{1}{(i\Omega)^2} T\sum_\omega \left[g(\bk,i\omega+i\Omega)+g(\bk,i\omega-i\Omega)-2g(\bk,i\omega)\right] \tilde S(i\omega).
\eea
Summing over $\omega$ and performing analytic continuation, we get 
\bea
W_{\text{SE,R}}(\Omega)=\frac{\mathcal{Z}(\Omega)}{(\Omega+i0^+)^2}, 
\label{A30}\eea
where
\bea
\mathcal{Z}(\Omega)&=&\int\frac{d\omega}{\pi}n_\fer(\omega)\left\{g_{\text{R}}(\bk,\omega+\Omega)\Im \tilde S_{\text{R}}(\omega)+\Im g_{\text{R}}(\bk,\omega)\tilde S_\text{A}(\omega-\Omega)+g_{\text{A}}(\bk,\omega-\Omega)\Im \tilde S_\text{R}(\omega)+\Im g_{\text{R}}(\bk,\omega) \tilde S_\text{R}(\omega+\Omega)\right.\nn\\
&&\left.-2\Im\left[g_{\text{R}}(\bk,\omega)\tilde S_{\text{R}}(\omega)\right]\right\}.
\eea
Taking the imaginary part of Eq.~\eqref{A30}, we find
\bea
\Im W_{\text{SE,R}}(\Omega)=\frac{\Im\mathcal{Z}(\Omega)}{\Omega^2}+\Im\frac{1}{(\Omega+i0^+)^2}\Re\mathcal{Z}(\Omega).
\eea
It is the second term that contains an essential singularity. We re-write the singular part as
\bea
\Im\frac{1}{(\Omega+i0^+)^2}\Re\mathcal{Z}(\Omega)=-\Im\left(\frac{\partial}{\partial\Omega}\frac{1}{\Omega+i0^+}\right)\Re\mathcal{Z}(\Omega)=\pi\delta'(\Omega)\Re\mathcal{Z}(\Omega)=-\pi\delta(\Omega)\frac{\partial}{\partial\Omega}\Re\mathcal{Z}(\Omega)\Big\vert_{\Omega=0}.\label{A34}
\eea
The imaginary part of the retarded current-current correlator must be an odd function of $\Omega$, and the same is true for $\Im W_{\text{SE,R}}(\Omega)$ and 
$\Im\mathcal{Z}(\Omega)$.
But then $\Re\mathcal{Z}(\Omega)$ is odd in $\Omega$, and its derivative must vanish at $\Omega\to 0$ at least as $\Omega$, which means that left-hand side of Eq.~\eqref{A34} is proportional to $\Omega\delta(\Omega)$. Recalling also that $\Re\sigma(\Omega)=-\Im\Pi_\text{R}(\Omega)/\Omega$, we see that the singular part of Eq.~\eqref{A34} only renormalizes the weight of the delta function term in the conductivity. 
Such a term is rendered finite by taking momentum-relaxing scattering, e.g., scattering by impurities, into account, but is of no interest to our study and can be safely discarded.

\subsection{Vertex diagram\label{appen:VertexDetails}}
Explicitly, the vertex diagram (V) in Fig.~\ref{fig:AllDiagrams} reads 
\be
\Pi^\text{V}(i\Omega)&=\frac{1}{16d}\sum\limits_{\mathcal{S}_\text{V}}\int_\mathbf{k,q}\tr\left(\hat{\bv}
\hat{M}_\mathbf{k}^{s'_1}\hat{M}_{\mathbf{k+q}}^{s'_2}\cdot\hat{\bv}
\hat{M}_{\mathbf{k+q}}^{s'_3}\hat{M}_{\mathbf{k}}^{s'_5}\right)V^2_\text{st}(\mathbf{q})\nonumber\\
&\times T\sum\limits_{\omega}T\sum\limits_{\nu}g_{s'_1}(\mathbf{k},i\omega+i\Omega)g_{s'_2}(\mathbf{k+q},i\omega+i\nu+i\Omega)g_{s'_3}(\mathbf{k+q},i\omega+i\nu)g_{s'_5}(\mathbf{k},i\omega)\pi_0(\mathbf{q},i\nu),
\ee
where $\mathcal{S}_{\text{V}}=\{s'_1,s'_2,s'_3,s'_5\}$.
Applying the identity \eqref{ident}
to the pairs of the Green's functions with the same momenta, using Eq.~\eqref{pi0} for $\pi_0(\bq,i\nu)$, 
and summing over $\nu$, we obtain
\be
\Pi^\text{V}(i\Omega)&=\frac{1}{16d}
&
\sum\limits_{\mathcal{S}_\text{V}}\int_\mathbf{k,q}\tr\left(\hat{\bv}
\hat{M}_\mathbf{k}^{s'_1}\hat{M}_{\mathbf{k+q}}^{s'_2}\cdot\hat{\bv}
\hat{M}_{\mathbf{k+q}}^{s'_3}\hat{M}_{\mathbf{k}}^{s'_5}\right)V^2_\text{st}(\mathbf{q})\frac{1}{i\Omega-\xi^{s'_1}_\mathbf{k}+\xi^{s'_5}_\mathbf{k}}\frac{1}{i\Omega-\xi^{s'_2}_\mathbf{k+q}+\xi^{s'_3}_\mathbf{k+q}}\nn\\
&\times W_\text{V}(\bk,\bq,i\Omega),
\ee
where
\be
W_\text{V}(\bk,\bq,i\Omega)=T\sum\limits_{\omega}\left[g_{s'_5}(\mathbf{k},i\omega)-g_{s'_1}(\mathbf{k},i\omega+i\Omega)
\right]
\left[\tilde{\Sigma}_{s'_3}(\mathbf{k+q},\mathbf{q},i\omega)-\tilde{\Sigma}_{s'_2}(\mathbf{k+q},\mathbf{q},i\omega+i\Omega)\right],
\ee
and $\tilde{\Sigma}_{s}(\mathbf{k},\mathbf{q},i\omega)$ is defined in Eq.~\eqref{eq:TildeSigmaZeroTForm}.

Upon summation over $\omega$ and analytic continuation, we obtain for the real and imaginary parts of the retarded counterpart of $W_\text{V}$ (as for the self-energy diagram, we also switch to 
unprimed helicities
at this point, cf. footnote~\ref{ftnt:PivsRexplanation} in the main text):
\be
\Re W_{\text{V},\text{R}}(\bk,\bq,\Omega)=-\frac{1}{\pi}&\int d\omega n_\text{F}(\omega)\left[\Im\left(g_{s_5,\text{R}}(\mathbf{k},\omega)\tilde{\Sigma}_{s_3,\text{R}}(\mathbf{k+q},\mathbf{q},\omega)\right)+\Im\left(g_{s_1,\text{R}}(\mathbf{k},\omega)\tilde{\Sigma}_{s_2,\text{R}}(\mathbf{k+q},\mathbf{q},\omega)\right)\right.\nonumber\\
&-\left(\Re\tilde{\Sigma}_{s_2,\text{R}}(\mathbf{k+q},\mathbf{q},\omega+\Omega)\Im g_{s_5,\text{R}}(\mathbf{k},\omega)+\Re g_{s_5,\text{A}}(\mathbf{k},\omega-\Omega)\Im\tilde{\Sigma}_{s_2,\text{R}}(\mathbf{k+q},\mathbf{q},\omega)\right.\nonumber\\
&\left.\left.+\Re g_{s_1,\text{R}}(\mathbf{k},\omega+\Omega)\Im\tilde{\Sigma}_{s_3,\text{R}}(\mathbf{k+q},\mathbf{q},\omega)+\Re\tilde{\Sigma}_{s_3,\text{A}}(\mathbf{k+q},\mathbf{q},\omega-\Omega)\Im g_{s_1,\text{R}}(\mathbf{k},\omega)\right)\right],\label{ReWV}
\ee
\be
\Im W_{\text{V},\text{R}}(\bk,\bq,\Omega)=-\frac{1}{\pi}&\int d\omega\left[n_\text{F}(\omega+\Omega)-n_\text{F}(\omega)\right]\left[\Im\tilde{\Sigma}_{s_2,\text{R}}(\mathbf{k+q},\mathbf{q},\omega+\Omega)\Im g_{s_5,\text{R}}(\mathbf{k},\omega)
\right.\nonumber\\
&\left.
+\Im g_{s_1,\text{R}}(\mathbf{k},\omega+\Omega)\Im\tilde{\Sigma}_{s_3,\text{R}}(\mathbf{k+q},\mathbf{q},\omega) \right].\IEEEeqnarraynumspace\label{eq:WvertexImPart}
\ee
The imaginary part of the current-current correlator is then given by
\be
\Im\Pi^\text{V}_\text{R}
(\Omega)&=\frac{1}{16d}
&
\sum\limits_{\mathcal{S}_\text{V}}
\int_\mathbf{k,q}\tr\left(\hat{\bv}
\hat{M}_\mathbf{k}^{s_1}\hat{M}_{\mathbf{k+q}}^{s_2}\cdot\hat{\bv}
\hat{M}_{\mathbf{k+q}}^{s_3}\hat{M}_{\mathbf{k}}^{s_5}\right)V^2_\text{st}(\mathbf{q})\mathcal{B}(\bk,\bq,\Omega),
\ee
where
\bea\mathcal{B}(\bk,\bq,\Omega)&=&\frac{\Im W_\text{V,R}(\bk,\bq,\Omega)}{(\Omega-\xi^{s_1}_\mathbf{k}+\xi^{s_5}_\mathbf{k})(\Omega-\xi^{s_2}_\mathbf{k+q}+\xi^{s_3}_\mathbf{k+q})}-\pi\left(\frac{\delta(\Omega-\xi^{s_2}_\mathbf{k+q}+\xi^{s_3}_\mathbf{k+q})}{\Omega-\xi^{s_1}_\mathbf{k}+\xi^{s_5}_\mathbf{k}}+\frac{\delta(\Omega-\xi^{s_1}_\mathbf{k}+\xi^{s_5}_\mathbf{k})}{\Omega-\xi^{s_2}_\mathbf{k+q}+\xi^{s_3}_\mathbf{k+q}}\right)
\Re W_\text{V,R}(\bk,\bq,\Omega),
\nn\\
\label{calB}
\eea
and where we omitted the term with an essential singularity, which is similar to the singular term in the self-energy diagram discussed in Sec.~\ref{sec:delta2}.
We will now show that only the first term in 
Eq.~\eqref{calB} contributes to the real part of the conductivity 
for $0<\Omega<2\EF$.
The reasoning is similar to but more involved than the reasoning in Sec.~\ref{sec:SE}.

 Using the constraints imposed by the delta functions in front of $\Re W_\text{V}(\bk,\bq,\Omega)$, we obtain that, for $\Omega>0$,
only the sets 
\bea
s_2=+1,s_3=-1\; \text{and}\; s_1=+1,s_5=-1\label{2315}
\eea
are allowed for the first and second terms in the round brackets, respectively.
This implies that
$\epsilon_{\bk+\bq}=\Omega/2$ and $\epsilon_\bk=\Omega/2$
in the first and second terms, respectively.
Next, we will show that $\epsilon_\mathbf{k+q}>\EF$ and $\epsilon_\mathbf{k}>\EF$ in the first and second terms, respectively, which will imply that the corresponding contribution to $\Im\Pi^\text{V}_\text{R}(\Omega)$ is non-zero only if $\Omega>2\EF$. To see the constraints imposed on $\epsilon_{\bk+\bq}$ and $\epsilon_\bk$, we write down an explicit form of $\Re W_\text{V,R}(\bk,\bq,\Omega)$, using Eqs.~\eqref{ReWV}, 
\eqref{eq:ImReMixtureProperty}, and  \eqref{eq:TildeSigmaForm}.
After some re-arrangement, we get
\be
\Re & W_{\text{V},\text{R}}(\bk,\bq,\Omega)=
\frac 14\sum\limits_{s_4,s_6=\pm 1}\int_\mathbf{p}\left[\tr(\hat{M}_\mathbf{p}^{s_6}\hat{M}_\mathbf{p+q}^{s_4})\right]\left[n_\text{F}(\xi^{s_{6}}_\mathbf{p})-n_\text{F}(\xi^{s_4}_\mathbf{p+q})\right]\nonumber\\
&\times\left[
f_{s_3}
\frac{n_\text{F}(\xi_\mathbf{k}^{s_5})-n_\text{F}(-\xi^{s_4}_\mathbf{p+q}+\xi^{s_{6}}_\mathbf{p}+\xi_\mathbf{k+q}^{s_3})
}{\xi_\mathbf{k}^{s_5}+\xi^{s_4}_\mathbf{p+q}-\xi^{s_{6}}_\mathbf{p}-\xi_\mathbf{k+q}^{s_3}}-f_{s_2}\frac{
n_\text{F}(\xi_\mathbf{k}^{s_5})-n_\text{F}(-\xi^{s_4}_\mathbf{p+q}+\xi^{s_{6}}_\mathbf{p}+\xi_\mathbf{k+q}^{s_2})
}{\Omega+\xi_\mathbf{k}^{s_5}+\xi^{s_4}_\mathbf{p+q}-\xi^{s_{6}}_\mathbf{p}-\xi_\mathbf{k+q}^{s_2}}
\right.\nonumber\\
&+\left.
f_{s_2}\frac{
n_\text{F}(\xi_\mathbf{k}^{s_1})-n_\text{F}(-\xi^{s_4}_\mathbf{p+q}+\xi^{s_{6}}_\mathbf{p}+\xi_\mathbf{k+q}^{s_2})
}{\xi_\mathbf{k}^{s_1}+\xi^{s_4}_\mathbf{p+q}-\xi^{s_{6}}_\mathbf{p}-\xi_\mathbf{k+q}^{s_2}}-f_{s_3}\frac
{
n_\text{F}(-\xi^{s_4}_\mathbf{p+q}+\xi^{s_{6}}_\mathbf{p}+\xi_\mathbf{k+q}^{s_3})-n_\text{F}(\xi_\mathbf{k}^{s_1})
}{\Omega-\xi_\mathbf{k}^{s_1}-\xi^{s_4}_\mathbf{p+q}+\xi^{s_{6}}_\mathbf{p}+\xi_\mathbf{k+q}^{s_3}}
\right],\label{eq:ReWVertexsolved}
\ee
where $
f_s=n_\text{F}(\xi_\mathbf{k+q}^{s})+\bos(\xi^{s_4}_\mathbf{p+q}-\xi^{s_{6}}_\mathbf{p})$.
The difference of the Fermi functions in the first line of Eq.~\eqref{eq:ReWVertexsolved} is non-zero only if 
either $\xi^{s_{6}}_\mathbf{p}>0$ and $\xi^{s_4}_\mathbf{p+q}<0$, or vice versa. Now we analyze these two options one by one.  

For $\xi^{s_{6}}_\mathbf{p}>0$ and $\xi^{s_4}_\mathbf{p+q}<0$, we have $f_s=-\theta(\xi_\mathbf{k+q}^{s})$ and thus 
$\xi_{\bk+\bq}^{s_{2,3}}>0$. The combination of these conditions implies that the arguments of the Fermi functions containing three terms are positive, and thus the arguments of the Fermi functions containing a single term must be negative. With that, Eq.~\eqref{eq:ReWVertexsolved} is simplified to
\be
\Re W_{\text{V},\text{R}}(\Omega)=-\frac{1}{4}&\sum\limits_{s_4,s_6=\pm 1}\int_\mathbf{p}\tr(\hat{M}_\mathbf{p}^{s_6}\hat{M}_\mathbf{p+q}^{s_4})\theta(\xi^{s_{6}}_\mathbf{p})\theta(-\xi^{s_4}_\mathbf{p+q})\nonumber\\
&\times\left[\theta(-\xi_\mathbf{k}^{s_5})\left(\frac{\theta(\xi_\mathbf{k+q}^{s_{2}})}{\Omega+\xi_\mathbf{k}^{s_5}+\xi^{s_4}_\mathbf{p+q}-\xi^{s_{6}}_\mathbf{p}-\xi_\mathbf{k+q}^{s_2}}-\frac{\theta(\xi_\mathbf{k+q}^{s_{3}})}{\xi_\mathbf{k}^{s_5}+\xi^{s_4}_\mathbf{p+q}-\xi^{s_{6}}_\mathbf{p}-\xi_\mathbf{k+q}^{s_3}}\right)
\right.\nonumber\\
&\qquad\left.-\theta(-\xi_\mathbf{k}^{s_{1}})
\left(\frac{\theta(\xi_\mathbf{k+q}^{s_{2}})}{\xi_\mathbf{k}^{s_1}+\xi^{s_4}_\mathbf{p+q}-\xi^{s_{6}}_\mathbf{p}-\xi_\mathbf{k+q}^{s_2}}+\frac{\theta(\xi_\mathbf{k+q}^{s_{3}})}{\Omega-\xi_\mathbf{k}^{s_1}-\xi^{s_4}_\mathbf{p+q}+\xi^{s_{6}}_\mathbf{p}+\xi_\mathbf{k+q}^{s_3}}\right)
\right].\label{eq:vertWVintermediate}
\ee
Now, we focus on the first term in the round brackets in Eq.~\eqref{calB}. Using the constraint following from $\delta(\Omega-\xi^{s_2}_\mathbf{k+q}+\xi^{s_3}_\mathbf{k+q})$,
we arrive at
\be
\Re W_{\text{V},\text{R}}(
\xi^{s_2}_\mathbf{k+q}-\xi^{s_3}_\mathbf{k+q}
)=\frac{1}{4}
&
\sum\limits_{s_4,s_6=\pm 1}\int_\mathbf{p}\tr(\hat{M}_\mathbf{p}^{s_6}\hat{M}_\mathbf{p+q}^{s_4})\theta(\xi^{s_{6}}_\mathbf{p})\theta(-\xi^{s_4}_\mathbf{p+q})\left[\theta(\xi_\mathbf{k+q}^{s_{3}})-\theta(\xi_\mathbf{k+q}^{s_{2}})\right]\nonumber\\
&\times\left[
\frac{\theta(-\xi_\mathbf{k}^{s_5})}{
\xi_\mathbf{k}^{s_5}+\xi^{s_4}_\mathbf{p+q}-\xi^{s_{6}}_\mathbf{p}-\xi_\mathbf{k+q}^{s_3}
}
\right.
\left.
+
\frac{\theta(-\xi_\mathbf{k}^{s_{1}})}{
\xi_{\bk+\bq}^{s_2}-\xi_\mathbf{k}^{s_1}-\xi^{s_4}_\mathbf{p+q}+\xi^{s_{6}}_\mathbf{p}
}
\right].
\ee
The RHS of the last equation is non-zero only if the arguments of the theta functions in the first pair square brackets are of opposite signs. According to Eq.~\eqref{2315}, $s_3=-1$ while $s_2=1$.  Therefore, $\xi^{s_3}_{\bk+\bq}<0$ and $\xi^{s_2}_{\bk+\bq}=\epsilon_{\bk+\bq}-\EF>0$.
$\epsilon_\mathbf{k+q}>\EF$. But because $\epsilon_{\bk+\bq}=\Omega/2$ for the first term in the round brackets of Eq.~\eqref{calB}, we have the condition that $\Omega>2\EF$. 
Thus, this term only contributes for frequencies above $\omega_\text{D}$ and is not relevant for our analysis.

The second term in the round brackets of Eq.~\eqref{calB} is simplified using the condition following from 
$\delta(\Omega-\xi^{s_1}_\mathbf{k}+\xi^{s_5}_\mathbf{k})$:
{\be
\Re W_{\text{V},\text{R}}(\xi^{s_1}_\bk-\xi_\bk^{s_5})=
\frac{1}{4}
&
\sum\limits_{s_4,s_6=\pm 1}\int_\mathbf{p}\tr(\hat{M}_\mathbf{p}^{s_6}\hat{M}_\mathbf{p+q}^{s_4})\theta(\xi^{s_{6}}_\mathbf{p})\theta(-\xi^{s_4}_\mathbf{p+q})\left[\theta(-\xi_\mathbf{k}^{s_{5}})-\theta(-\xi_\mathbf{k}^{s_1})\right]\nn\\
&\times\left[
  \frac{\theta(\xi_\mathbf{k+q}^{s_{3}})}{\xi_\mathbf{k}^{s_5}+\xi^{s_4}_\mathbf{p+q}-\xi^{s_{6}}_\mathbf{p}-\xi_\mathbf{k+q}^{s_3}}
 +
 \frac{\theta(\xi_\mathbf{k+q}^{s_{2}})}{\xi_\mathbf{k+q}^{s_2}-\xi_\mathbf{k}^{s_1}-\xi^{s_4}_\mathbf{p+q}+\xi^{s_{6}}_\mathbf{p}}
 \right] 
.\ee
Using the constraints on $s_1$ and $s_5$ from Eq.~\eqref{2315} again, we conclude that 
$\epsilon_\mathbf{k}>\EF$. But because $\epsilon_\bk=\Omega/2$ for the second term in the round brackets of Eq.~\eqref{calB}, we again have the condition that $\Omega>2\EF$.

The case of $\xi^{s_{6}}_\mathbf{p}<0$ and $\xi^{s_4}_\mathbf{p+q}>0$ is analyzed in the same way and with the same result.
Therefore, the term  proportional to $\Re W_{\text{V},\text{R}}(\Omega)$ in Eq.~\eqref{calB} is absent for $\Omega<2\EF$.

Focusing now on the first term of Eq.~\eqref{calB}, we obtain, after some re-arrangements and re-labelling of momenta and helicities,
the following final expression for the vertex diagram 
for $\Omega<2\EF$:
\be
\Im\Pi^{\text{V}}_\text{R}(\Omega)=
\frac{\pi^2}{32}&\sum\limits_{\mathcal{S}}\int_{\mathbf{k,p,q},\nu}V^2_\text{st}(\mathbf{q})\left[\mathcal{T}^{\text{V}_1}_{\mathcal{S}}\mathcal{G}^{\text{V}_1}_{\mathcal{S}}+\mathcal{T}^{\text{V}_2}_{\mathcal{S}}\mathcal{G}^{\text{V}_2}_{\mathcal{S}}\right]\nonumber\\
&\times\theta(-\xi^{s_3}_\mathbf{k})\theta(\xi^{s_5}_\mathbf{k+q})\theta(-\xi^{s_4}_\mathbf{p})\theta(\xi^{s_6}_\mathbf{p+q})\delta(\Omega+\nu+\xi^{s_3}_\mathbf{k}-\xi^{s_5}_\mathbf{k+q})\delta(\nu+\xi^{s_6}_\mathbf{p+q}-\xi^{s_4}_\mathbf{p}),\label{ImV}
\ee
where, for the sake of convenience, we  re-introduced $\int_\nu=\int\dee\nu/2\pi$  and
\bse
\be
\mathcal{T}^{\text{V}_1}_{\mathcal{S}}=\frac{1}{d}\tr\left(\hat{\bv}
\hat{M}_\mathbf{k}^{s_1}\hat{M}_{\mathbf{k+q}}^{s_5}\cdot\hat{\bv}
\hat{M}_{\mathbf{k+q}}^{s_2}\hat{M}_{\mathbf{k}}^{s_3}\right)\text{Tr}\left(\hat{M}_{\mathbf{-p-q}}^{s_6}\hat{M}_{\mathbf{-p}}^{s_4}\right)\\
\mathcal{T}^{\text{V}_2}_{\mathcal{S}}=\frac{1}{d}\tr\left(\hat{\bv}
\hat{M}_{\mathbf{-k-q}}^{s_5}\hat{M}_{\mathbf{-k}}^{s_1}\cdot\hat{\bv}
\hat{M}_{\mathbf{-k}}^{s_3}\hat{M}_{\mathbf{-k-q}}^{s_2}\right)\text{Tr}\left(\hat{M}_\mathbf{p}^{s_4}\hat{M}_{\mathbf{p+q}}^{s_6}\right)\\
\mathcal{G}^{\text{V}_1}_{\mathcal{S}}=\mathcal{G}^{\text{V}_2}_{\mathcal{S}}=\frac{1}{\Omega-\xi^{s_1}_\mathbf{k}+\xi^{s_3}_\mathbf{k}}\frac{1}{\Omega-\xi^{s_5}_\mathbf{k+q}+\xi^{s_2}_\mathbf{k+q}}.
\ee
\ese
Thus we have derived Eq.~\eqref{eq:VertexTGresults} of the main text.

\subsection{Sum of the self-energy and vertex diagrams\label{appen:SEplusV}}
Adding up Eqs.~\eqref{ImSE1}, \eqref{ImSE2}, and \eqref{ImV}, we obtain the combined contribution of the self-energy and vertex diagram for 
$0<\Omega<\omega_{\text{D}}=2\EF$:
\bea
\Im\Pi^{\text{SE+V}}_\text{R}(\Omega)&=&\Im \Pi^{\text{SE}_1}_\text{R}(\Omega)+\Im \Pi^{\text{SE}_2}_\text{R}(\Omega)+\Im \Pi^\text{V}_\text{R}(\Omega)=
-\frac{\pi^2}{32}\sum\limits_{\mathcal{S}}\int_\mathbf{k,p,q}\int_\nu V^2_\text{st}(\mathbf{q})\times\theta(-\xi^{s_3}_\mathbf{k})\theta(\xi^{s_5}_\mathbf{k+q})\theta(-\xi^{s_4}_\mathbf{p})\theta(\xi^{s_6}_\mathbf{p+q})\nn\\
&&\times
\delta(\Omega+\nu+\xi^{s_3}_\mathbf{k}-\xi^{s_5}_\mathbf{k+q})\delta(\nu+\xi^{s_6}_\mathbf{p+q}-\xi^{s_4}_\mathbf{p})
\left[\mathcal{T}^{\text{SE}_1}_{\mathcal{S}}\mathcal{G}^{\text{SE}_1}_{\mathcal{S}}+\mathcal{T}^{\text{SE}_2}_{\mathcal{S}}\mathcal{G}^{\text{SE}_2}_{\mathcal{S}}-\mathcal{T}^{\text{V}_1}_{\mathcal{S}}\mathcal{G}^{\text{V}_1}_{\mathcal{S}}-\mathcal{T}^{\text{V}_2}_{\mathcal{S}}\mathcal{G}^{\text{V}_2}_{\mathcal{S}}\right].\label{eq:ImPiSEplusV}
\eea
Similar computations for PAL and CAL diagrams give Eqs.~\eqref{eq:PALTGresults} and~\eqref{eq:CALexactexpressions}, respectively.

\section{
Asymptotic expressions for the conductivity\label{appen:asymp_eval}}
As an example, we derive 
an asymptotic expression for the purely electron-electron contribution to the conductivity of a 3D Dirac metal with Hubbard interaction in the limit $\Omega\ll\EF$.
 We start with the general result  \eqref{Eq:generalzeroTexpression}, which we reproduce below for the reader's convenience:
\be
\mathcal{R}^{J_u
}_{\mathcal{S}}(\Omega)=K^{J_u}\int_{\mathbf{k,p,q}}\int_\nu &V^2_{\text{st}}(\mathbf{q})
\mathcal{T}^{J_u}_\mathcal{S}\left(\bk,\bp,\bq\right)\mathcal{G}^{J_u}_\mathcal{S}(\bk,\bp,\bq,\Omega)\nn\\
&\times \theta(-\xi^{s_3}_\mathbf{k})\theta(-\xi^{s_4}_\mathbf{p})\theta(\xi^{s_5}_\mathbf{k+q})\theta(\xi^{s_6}_\mathbf{p+q})\delta(\Omega+\nu+\xi^{s_3}_\mathbf{k}-\xi^{s_5}_\mathbf{k+q})\delta(\nu+\xi^{s_6}_\mathbf{p+q}-\xi^{s_4}_\mathbf{p}),
\ee
where
 $\mathcal{T}^{J_u}_\mathcal{S}(\mathbf{k,p,q})$  and $\mathcal{G}^{J_u}_\mathcal{S}(\mathbf{k,p,q})$ are defined in Eqs.~\eqref{eq:firsteqofresultlist}-\eqref{eq:CALexactexpressions}, respectively. 
We remind the reader that the purely electron-electron contribution corresponds to all helicities being positive, i.e., 
\bea
\mathcal{S}=\mathcal{S}_+\equiv \{s_1=1,s_2=1,s_3=1,s_4=1,s_5=1,s_6=1\}.\label{S+}
\eea
In a more explicit form,
\be
\mathcal{R}^{J_u}_{\mathcal{S}_+}(\Omega)&=\lambda_3^2K^{J_u}
\int^\infty_0\frac{dkk^{2}}{(2\pi)^3}
\int^\infty_0\frac{dpp^{2}}{(2\pi)^3}
\int^\infty_0\frac{dq q^{2}}{(2\pi)^3}\int^\infty_{-\infty}\frac{d\nu}{2\pi}
\nonumber\\
&\times
\int\limits_{0}^{\pi}d\theta_\mathbf{q\hat z}\sin\theta_\mathbf{q\hat z}\int\limits_{0}^{2\pi}d\phi_\mathbf{q,\hat z}
\int\limits_{0}^{\pi}d\theta_\mathbf{kq}\sin\theta_\mathbf{kq}\int\limits_{0}^{2\pi}d\phi_\mathbf{k, q}\int\limits_{0}^{\pi}d\theta_\mathbf{pq}\sin\theta_\mathbf{pq}\int\limits_{0}^{2\pi}d\phi_\mathbf{p,q}
\mathcal{T}^{J_u}_\mathcal{S_+}(\mathbf{k,p,q})\mathcal{G}^{J_u}_\mathcal{S_+}(\mathbf{k,p,q})\nn\\
&\times\theta(-\xi^{+}_\mathbf{k})\theta(-\xi^{+}_\mathbf{p})\theta(\xi^{+}_\mathbf{k+q})\theta(\xi^{+}_\mathbf{p+q})\delta(\Omega+\nu+\xi^{+}_\mathbf{k}-\xi^{+}_\mathbf{k+q})\delta(\nu+\xi^{+}_\mathbf{p+q}-\xi^{+}_\mathbf{p}),
 \label{eq:angularpartseparated}
\ee
where 
$\theta_{{\bf nn}'}$ is the angle between vectors ${\bf n}$ and ${\bf n'}$,
and 
$\phi_{{\bf n},{\bf n}'}$ 
is the 
azimuthal angle of vector ${\bf n}$ in a spherical system with vector ${\bf n}'$ taken as the polar axis.
Now we solve the integrals over $\theta_{\bk\bq}$ and $\theta_{\bp\bq}$, using the delta functions in in the expression above, which yields 
\bea
&&\int\limits_{0}^{\pi}d\theta_\mathbf{kq}\sin\theta_\mathbf{kq}\int\limits_{0}^{\pi}d\theta_\mathbf{pq}\sin\theta_\mathbf{pq}
\mathcal{T}^{J_u}_\mathcal{S_+}(\mathbf{k,p,q})\mathcal{G}^{J_u}_\mathcal{S_+}(\mathbf{k,p,q})
\delta(\Omega+\nu+\epsilon_\mathbf{k}-\epsilon_\mathbf{k+q})\delta(\nu+\epsilon_\mathbf{p+q}-\epsilon_\mathbf{p})\nn\\
&&=\frac{(\epsilon_\bk+\nu+\Omega)(\epsilon_\mathbf{p}-\nu)}{\epsilon_\bk\epsilon_\bp\vd^2 q^2}
\left[
{\mathcal{T}}^{J_u}_\mathcal{S_+}(\mathbf{k,p,q})
{\mathcal{G}}^{J_u}_\mathcal{S_+}(\mathbf{k,p,q})\right]\Big\vert_{\cos\theta_{\bk\bq}=\mu_\bk,\cos\theta_{\bp\bq}=\mu_\bp}
\Theta(\mu_\bk,\mu_\bp),
\eea
where $\Theta(\mu_\bk,\mu_\bp)$ 
encapsulates the constraints imposed by the delta functions in Eq.~\eqref{eq:angularpartseparated}, i.e.,
\be
 \qquad\,\,\,
\epsilon_\mathbf{k+q}=\Omega+\nu+\epsilon_\mathbf{k},\epsilon_\mathbf{p+q}=\epsilon_\mathbf{p}-\nu,\label{eq:deltafuncconstraint1}
\ee such that
\be
-1\leq \mu_\bk=\frac{(\Omega+\nu+\epsilon_\bk)^2-\epsilon_\bk^2-\vd^2q^2}{2\epsilon_\bk\vd q}\leq 1,\nn\\
-1\leq \mu_\bp=\frac{(-\nu+\epsilon_\mathbf{p})^2-\epsilon_\bp^2-\vd^2q^2
}{2\epsilon_\bp\vd q}\leq 1.\label{eq:cosineconstraints}
\ee
Note that the constraints~\eqref{eq:deltafuncconstraint1} are the same ones as in Eq.~\eqref{cb} of the main text for the special case of all helicities being positive. 
The last two equations are resolved in terms of $q$, which imposes constraints on the range of integration over $q$.
In addition, the first two theta functions in Eq.~\eqref{eq:angularpartseparated} imply that $\epsilon_\bk\leq\EF$ and $\epsilon_\bp\geq\EF$  for $s_3=s_4=1$, whereas the second two theta functions, in conjunction with Eq.~\eqref{eq:deltafuncconstraint1}, imply that $\Omega+\nu+\epsilon_\bk\geq\EF$
and $\epsilon_\bp-\nu\geq\EF$.

Next, we switch
from integration over $k$ and $p$ 
to integration over dispersions $\epsilon_\bk=\vd k$ and $\epsilon_\bp=\vd p$. Imposing explicitly all the constraints described above, we arrive at
\be
\mathcal{R}^{J_u}_{\mathcal{S_+}}(\Omega)=\int\limits_{\EF-\Omega}^{\EF}d\epsilon_\bk\int\limits_{\max\left\{
        0 ,2\EF-\Omega-\epsilon_\bk\right\}}^{\EF}
    d\epsilon_\bp\int\limits_{\EF-\Omega-\epsilon_\bk}^{\epsilon_\mathbf{p}-\EF}d\nu\int\limits_{\max\left\{(\Omega+\nu
    )
    ,-\nu
    \right\}/\vd}^{
    \min\left\{(\Omega+\nu
    +2\epsilon_\bk
    )
    ,(2\epsilon_\mathbf{p}-\nu)
    \right\}/\vd}dq \mathcal{L}^{J_u}_\mathcal{S_+}(\epsilon_\bk,\epsilon_\bp,q,\nu,\Omega),
\label{eq:3Dangularintegratedexpression}
\ee
where
\be
\mathcal{L}^{J_u}_\mathcal{S_+}(\epsilon_\bk,\epsilon_\bp,q,\nu,\Omega)=\frac{K^{J_u}\lambda_3^2}{(2\pi)^{10}\vd^8}\bar{\mathcal{Q}}^{J_u}_\mathcal{S_+}(\epsilon_\bk,\epsilon_\bp,q,\nu,\Omega)
\epsilon_\bk \epsilon_\bp(\epsilon_\bk+\Omega+\nu)(\epsilon_\mathbf{p}-\nu)
\ee
and
\be
\bar{\mathcal{Q}}^{J_u}_\mathcal{S_+}(\epsilon_\bk,\epsilon_\bp,q,\nu,\Omega)
=\int\limits_{0}^{\pi}d\theta_\mathbf{qz}
\sin\theta_\mathbf{qz}\int\limits_{0}^{2\pi}d\phi_\mathbf{qx}
\int\limits_{0}^{2\pi}d\phi_\mathbf{kq}\int\limits_{0}^{2\pi}d\phi_\mathbf{pq}
\left[\mathcal{T}^{J_u}_\mathcal{S_+}(\bk,\bp,\bq)\mathcal{G}^{J_u}_\mathcal{S_+}(\bk,\bp,\bq)\right]\Big\vert_{\cos\theta_{\bk\bq}=\mu_\bk,\cos\theta_{\bp\bq}=\mu_\bp}.\label{eq:restofangintegrals3D}
\ee

For the particular case of $J_u=\text{SE}_1$ and all helicities positive, we have $\mathcal{G}^{\text{SE}_1}_\mathcal{S_+}(\bk,\bp,\bq)=1/\Omega^2$, and thus the angular integrals in Eq.~\eqref{eq:restofangintegrals3D} involve only 
$\mathcal{T}^{\text{SE}_1}_\mathcal{S_+}(\bk,\bp,\bq)$ in  Eq.~\eqref{eq:firsteqofresultlist}, which is evaluated explicitly as 
\bea
\mathcal{T}^{\text{SE}_1}_\mathcal{S_+}(\bk,\bp,\bq)&=&-\frac{8 \vd^3q \sin\theta_{\bk\bq}}{\epsilon_{\bk+\bq}}\frac{\epsilon_\bp+\epsilon_\bpq+ q\cos\theta_{\mathbf{pq}}}{\epsilon_\bpq} 
\Big[8 \cos2 \theta_{\bk\bq} \cos\phi_{\mathbf{qx}} \sin\theta_{\bq\mathbf{z}} 
\left(
\cos\theta_{\bq\mathbf{z}} \cos\phi_{\mathbf{kq}} \cos\phi_{\mathbf{qx}}-\sin\phi_{\mathbf{kq}} \sin\phi_{\mathbf{qx}}\right)
\nn\\
&&+\frac{1}{8} \sin2 \theta_{\bk\bq}\left(
4+12 \cos2 \theta_{\bq\mathbf{z}}
+4 \cos2 \theta_{\bq\mathbf{z}}\cos2\phi_{\mathbf{kq}}
-4 \cos2 \phi_{\mathbf{kq}}
+12\cos2 \theta_{\bq\mathbf{z}}\cos2\phi_{\mathbf{qx}}
\right.\nn\\
&&\left.+4\cos2\phi_{\mathbf{qx}}\cos2\theta_{\bq\mathbf{z}}
\cos2\phi_{\mathbf{kq}}
+12\cos2 \phi_{\bk\mathbf{q}}\cos2\phi_{\mathbf{qx}}
-12 \cos2 \phi_{\mathbf{qx}}
-16 \cos\theta_{\bq\mathbf{z}} \sin2 \phi_{\mathbf{kq}} \sin2 \phi_{\mathbf{qx}}
\right)\Big].
\nn\\
\label{trace}
\eea
Integrating Eq.~\eqref{trace} over the angles and imposing the constraints from Eq.~\eqref{eq:deltafuncconstraint1}, we obtain
\be
\mathcal{L}^{\text{SE}_1}_{\mathcal{S}_+}(\epsilon_\bk,\epsilon_\bp,q,\nu,\Omega)=-\frac{
\lambda_3^2}{
384\pi^5
\vd^6 \Omega^2 }
\left
(\left(2\epsilon_\bk+\Omega+\nu\right)^2-\vd^2q^2\right)\left(\left(2\epsilon_\bp-\nu\right)^2-\vd^2q^2\right) 
.
\IEEEeqnarraynumspace
\label{D9}
\ee
where we used $K^{\text{SE}_1}=-\pi^2/32$ as defined in Eq.~\eqref{KJu} of the main text.

In this particular case, the integrand of Eq.~\eqref{eq:3Dangularintegratedexpression} is a purely polynomial function, and thus it is possible to obtain an exact result for all frequencies $\Omega<2\EF$. However, this is not the case for other diagrams (except for SE$_2$) and for the 2D case. For 
consistency with these other cases,
we will only evaluate the integral \eqref{eq:3Dangularintegratedexpression} for $\Omega\ll\EF$. To this end, we note that for $\Omega<\EF$ the limits of the integral for $\epsilon_\bp$ in Eq.~\eqref{eq:3Dangularintegratedexpression} become
\be
2\EF-\epsilon_\bk-\Omega<\epsilon_\bp<\EF.
\ee
Thus, for $\Omega\ll\EF$ it is convenient to define new dimensionless
variables 
    \be
    x=(\epsilon_\bk-\EF+\Omega)/\Omega;\,y=(\epsilon_\mathbf{p}-\EF)/\Omega+x;\,z=\nu/\Omega+x,\nn
        \,
            \\
         x\in(0,
      1),y\in(0,x),z\in(0,y).
               \label{eq:xyzdefintionlastline}
    \ee
Note that $x,y$ and $z$ are all of order $1$.
Also, the upper and lower limits of the integral over $q$ in Eq.~\eqref{eq:3Dangularintegratedexpression} can be replaced by $2k_F$ and $0$, respectively.
Now 
we
substitute $\epsilon_\mathbf{k},\epsilon_\mathbf{p},\nu$ in terms of $x,y,z$ into Eq.~\eqref{D9}, define $r=q/2k_F$,
and expand the result in $x,y,z$  and $\Omega$ for $\Omega\ll\EF$.
After the expansion, Eq.~\eqref{D9} is reduced to
\bea
\mathcal{L}
^{\text{SE}_1}_{\mathcal{S}_+}(x,y,z,
r,\Omega)&=&
-\frac{\lambda_3^2}{
384\pi^5
\vd^6}(2\EF)^2\Big[
\left(\frac{2\EF}{\Omega}\right)^2
(r^2-1)^2
+\frac{4\EF}{\Omega}r^2(r^2-1)(1-2y)
\nonumber\\
&&+ 
4r^4\left(1-x+x^2
-2 y-xy+3y^2-yz+z^2\right)
\nonumber\\
&&-
r^2\left(5-10x+10x^2-12xy+12y^2-2z+4xz-12yz+10z^2\right)
\nn\\
&&
+\left(2 x^2-4 xz-2x+2 z^2+2 z+1\right)
+\mathcal{O}(\Omega/\EF)
\Big].
\eea
Performing the four-dimensional integral
\be
\mathcal{R}^{\text{SE}_1}_{\mathcal{S}_+}(\Omega)=\Omega^3(2\kf)\int\limits_0^1\dee x\int\limits_0^{x}\dee y\int\limits_0^{y}\dee z\int\limits_0^{
1}\dee
r
\mathcal{L
}^{\text{SE}_1}_{\mathcal{S}_+}(x,y,z,
r,\Omega),
\ee
we obtain
\bse
\be
\mathcal{R}^{\text{SE}_1}_{\mathcal{S}_+}(\Omega)=
-\frac{\lambda_3^2}{135\pi^5} \frac{\kf^6}{\vd}
\left
[
\frac{\Omega}{\EF}
+
\frac{9}{80}\left(\frac{\Omega}{\EF}\right)^3\right
]
.
\label{D15a}
\ee
Similar expansions are 
performed for other contributions.
Below we just list the results:
\be
\mathcal{R}^{\text{SE}_2}_{\mathcal{S}_+}(\Omega)=
-\frac{\lambda_3^2}{135\pi^5} \frac{\kf^6}{\vd}
\left[
\frac{\Omega}{\EF}+
\frac{9}{80}\left(\frac{\Omega}{\EF}\right)^3\right],
\\
\mathcal{R}^{\text{V}_1}_{\mathcal{S}_+}(\Omega)+\mathcal{R}^{\text{V}_2}_{\mathcal{S}_+}(\Omega)=
-\frac{\lambda_3^2}{135\pi^5} \frac{\kf^6}{\vd}
\left[-
\frac{10}{7}
\frac{\Omega}{\EF}
-
\frac{129}{280}\left(\frac{\Omega}{\EF}\right)^3\right],
\\
\mathcal{R}^{\text{PAL}_1}_{\mathcal{S}_+}(\Omega)+\mathcal{R}^{\text{PAL}_2}_{\mathcal{S}_+}(\Omega)+\mathcal{R}^{\text{CAL}_1}_{\mathcal{S}_+}(\Omega)+\mathcal{R}^{\text{CAL}_2}_{\mathcal{S}_+}(\Omega)=
-\frac{\lambda_3^2}{135\pi^5} \frac{\kf^6}{\vd}
\left[-
\frac{4}{7}
\frac{\Omega}{\EF}
+
\frac{71}{140}
\left(\frac{\Omega}{\EF}\right)^3\right].\IEEEeqnarraynumspace\label{D15d}
\ee
\ese
The conductivity is obtained by multiplying the sum of all $\mathcal{R}^{J_u}_{\mathcal{S}_+}$ by $-1/\Omega$. The leading, linear-in-$\Omega$ terms in Eqs.~\eqref{D15a}-\eqref{D15d}, would then produce an $\Omega$-independent conductivity, which would be the case for a Fermi liquid with Galilean invariance broken completely by, e.g., umklapp scattering. However, in our case Galilean invariance is broken only partially, and the conductivity is suppressed compared to the case with fully-broken Galilean invariance. Indeed, adding up 
Eqs.~\eqref{D15a}-\eqref{D15d}, we see that linear-in-$\Omega$ terms cancel out, while the sum the subleading, cubic terms 
reproduces Eq.~\eqref{3D_sigma_ee} of the main text.

\section{Combining contributions from all diagrams for intra-band absorption due to electron-electron interaction
\label{appen:CombingLFeediagrams}}
In this Appendix, we demonstrate how the contributions from all diagrams are combined together for the case of electron-electron interaction, when all the helicities are positive: $s_i=+1$ for $i=1\dots 6$.
We introduce the following definitions \bse
\bea
\ket{s_\mathbf{k}}&\equiv&\ket{\mathbf{k},s}\equiv\ket{\psi_{\mathbf{k},s}},\\
\hat{M}_\mathbf{k}^{s}&=&\ket{s_\mathbf{k}}\bra{s_\mathbf{k}},\\
\Phi^{s,s'}_\mathbf{k,
\bk'}&=&\braket{
s_\bk|s'_{\bk'}
},\\
\mathbf{v}^{s,s'}_{\mathbf{k}}&=&\bra{
s_\bk}\hat{\mathbf{v}}\ket{
s'_\bk},
\eea
\ese
and list the properties that will be used in this section:
\bse
\bea
\braket{s_\mathbf{k}|s'_\mathbf{k'}}&=&\braket{s_\mathbf{-k}|s'_\mathbf{-k'}},\\
\braket{s_\mathbf{k}|s_\mathbf{k}}&=&1,\\
\braket{s_\mathbf{k}|-s_\mathbf{k}}&=&0,\\
\mathbf{v}^{+,+}_{\mathbf{-k}}&=&-\mathbf{v}^{+,+}_{\mathbf{k}}.\\
\eea
\ese
We remind the reader that an algebraic expression for each diagram is the sum of two terms, labeled as $J_1$ and $J_2$, where $J=$SE, V, etc. denotes the type of a diagram.\footnote{The two terms for the self-energy diagram corresponds to two distinct diagrams, labeled as SE$_1$ and SE$_2$ in Fig.~\ref{fig:AllDiagrams}. For the rest of the diagrams, the two terms appear only in the algebraic expressions rather as distinct diagrams.} We first combine the $J_1$ terms together, and then do the same for the $J_2$ terms. 
Note that, according to Eqs.~\eqref{eq:firsteqofresultlist}-\eqref{eq:CALexactexpressions}, $\mathcal{G}^{J_1}_{\mathcal{S}_+}=\mathcal{G}^{J_2}
_{\mathcal{S}_+}
=1/\Omega^2$,
where $\mathcal{S}_+$ is defined in Eq.~\eqref{S+}.
Therefore, we need to combine only the trace parts.  
For $\mathcal{T}^{\text{SE}_1}_{\mathcal{S}_+}$ we obtain
\be
\mathcal{T}^{\text{SE}_1}_{\mathcal{S}_+}
=\text{Tr}\left(\hat{\mathbf{v}}\hat{M}_\mathbf{k}^{+}\hat{M}_{\mathbf{k+q}}^{+}\hat{M}
_{\mathbf{k}}^{+}\cdot\hat{\mathbf{v}}\hat{M}_{\mathbf{k}}^{+}\right)\text{Tr}(\hat{M}_\mathbf{-p-q}^{+}\hat{M}_\mathbf{-p}^{+})
\nonumber\\
=\bra{
+_\bk}\hat{\mathbf{v}}\ket{
+_\bk}\cdot\braket{+_\mathbf{k}|+_\mathbf{k+q}}\braket{+_\mathbf{k+q}|+_\mathbf{k}}\bra{
+_\bk}\hat{\mathbf{v}}\ket{
+_\bk}\braket{+_\mathbf{k}|+_\mathbf{k}}\braket{+_\mathbf{-p}|+_\mathbf{-p-q}}\braket{+_\mathbf{-p-q}|+_\mathbf{-p}}\braket{+_\mathbf{-p}|+_\mathbf{-p}}
\nonumber\\
=\mathbf{v}^{+,+}_{\mathbf{k}}\cdot\mathbf{v}^{+,+}_{\mathbf{k}}\braket{+_\mathbf{k}|+_\mathbf{k+q}}\braket{+_\mathbf{k+q}|+_\mathbf{k}}\left|\Phi^{+,+}_\mathbf{p,p+q}\right|^2
\nonumber\\
=\mathbf{v}^{+,+}_{\mathbf{k}}\cdot\mathbf{v}^{+,+}_{\mathbf{k}}\left|\Phi^{+,+}_\mathbf{k,k+q}\right|^2\left|\Phi^{+,+}_\mathbf{p,p+q}\right|^2.
\ee
Performing the same steps for the V$_1$, PAL$_1$ and CAL$_1$ contributions and adding them up, we obtain 
\be
\mathcal{T}_1\equiv
\sum_{J_1=\text{SE}_1,\text{V}_1,\text{PAL}_1,\text{CAL}_1}K^{J_1}\mathcal{T}^{J_1}_{\mathcal{S}_+}
=-\frac{\pi^2}{32}\mathbf{v}^{+,+}_{\mathbf{k}}\cdot
\Delta\mathbf{v}
\left|\Phi^{+,+}_\mathbf{k,k+q}\right|^2\left|\Phi^{+,+}_\mathbf{p,p+q}\right|^2,
\label{T1}
\ee 
where $K_{J_u}$ is defined in Eq.~\eqref{KJu} of the main text, and 
\be
\Delta\mathbf{v}
=
\mathbf{v}^{+,+}_{\mathbf{k}}+\mathbf{v}^{+,+}_{\mathbf{p+q}}-\mathbf{v}^{+,+}_{\mathbf{k+q}}-\mathbf{v}^{+,+}_{\mathbf{p}}
.
\ee
In the same way, we re-write the $J_2$ terms and combine them together. The only difference compared to the $J_1$ case is that,  when combining the PAL$_2$ and CAL$_2$ contributions, we 
need 
 to re-label the momenta as $\mathbf{k}\leftrightarrow\mathbf{p}$ in the integrand of Eq.~\eqref{Eq:generalzeroTexpression}. This can be done for the present case, 
when all the helicities are positive, and thus the delta and theta functions in Eq.~\eqref{Eq:generalzeroTexpression} can be reduced back to their original forms  by replacing first $\nu\rightarrow\nu-\Omega$ and then $\nu\rightarrow-\nu$. After these manipulations, we obtain for the sum of the $J_2$  terms
\be
\mathcal{T}_2\equiv\sum_{J=\text{SE}_2,\text{V}_2,\text{PAL}_2,\text{CAL}_2}K^{J_2}\mathcal{T}^{J_2}_{\mathcal{S}_+}=\frac{\pi^2}{32}\mathbf{v}^{+,+}_{\mathbf{k+q}}\cdot\Delta\mathbf{v}\left|\Phi^{+,+}_\mathbf{k,k+q}\right|^2\left|\Phi^{+,+}_\mathbf{p,p+q}\right|^2.\label{T2}
\ee
Adding up Eqs.~\eqref{T1} and \eqref{T2},
we find
\be
\mathcal{T}_{\mathcal{S}_+}\equiv  \mathcal{T}_1+\mathcal{T}_2=
-\frac{\pi^2}{32}\Delta\mathbf{v}
\cdot\left(
\mathbf{v}^{+,+}_{\mathbf{k}}
-\mathbf{v}^{+,+}_{\mathbf{k+q}}\right)\left|\Phi^{+,+}_\mathbf{k,k+q}\right|^2\left|\Phi^{+,+}_\mathbf{p,p+q}\right|^2.\label{sumT}
\ee
Using the same reasoning 
as for the PAL$_2$ case above, we relabel $\mathbf{k}\leftrightarrow\mathbf{p}$ in Eq.~\eqref{sumT} and rewrite it as 
\be
\mathcal{T}_{\mathcal{S}_+}
=-\frac{\pi^2}{32}\Delta\mathbf{v}\cdot
\left(
\mathbf{v}^{+,+}_{\mathbf{p}+\bq}-
\mathbf{v}^{+,+}_{\mathbf{p}}\right)
\left|\Phi^{+,+}_\mathbf{p,p+q}\right|^2\left|\Phi^{+,+}_\mathbf{k,k+q}\right|^2,
\label{sumT2}
\ee
where we used that $\Delta\mathbf{v}
\rightarrow-\Delta\mathbf{v}
$ 
on $\mathbf{k}\leftrightarrow\mathbf{p}$. Adding Eqs.~\eqref{sumT} and \eqref{sumT2}, we obtain a symmetrized form of $\mathcal{T}_{\mathcal{S}_+}$:
\be
\mathcal{T}_{\mathcal{S}_+}=-\frac{\pi^2}{64}
\left(\Delta\mathbf{v}\right)^2
\left|\Phi^{+,+}_\mathbf{p,p+q}\right|^2\left|\Phi^{+,+}_\mathbf{k,k+q}\right|^2,
\ee
which is reproduced in Eq.~\eqref{sumTmain} of the main text.
\end{document}